\documentclass[12pt]{scrartcl} %{article}
\usepackage{listings, xcolor}
\usepackage{amsmath}
\usepackage{amssymb}
\usepackage{graphicx}
\usepackage{caption}
\usepackage{subcaption}
\usepackage[T1]{fontenc}
\usepackage[latin1]{inputenc}
\usepackage[english]{babel}
\usepackage[latin1]{inputenc}
\usepackage{lmodern}
\usepackage{bbm}
\usepackage{float} 
\usepackage{epstopdf}
\usepackage{subcaption}

\usepackage{threeparttable}
\usepackage{setspace}
\usepackage[round]{natbib}
\usepackage{babelbib}
\usepackage{url}
\usepackage{rotating}
\usepackage{dcolumn}
\newcolumntype{d}[1]{D{.}{.}{#1}}

\usepackage{booktabs}
\usepackage{tabularx}
\usepackage{indentfirst}
\usepackage{multirow}
\usepackage{framed}
\usepackage{enumitem}
\usepackage{lscape}
\usepackage{longtable}
\usepackage{tabulary}  
\usepackage{rotating}

\usepackage{chngcntr}
\counterwithin{figure}{section}
\counterwithin{table}{section}
\counterwithin{equation}{section}
\usepackage{setspace}

\usepackage{colortbl}

\usepackage[left=2.54cm,right=2.54cm,top=2.54cm,bottom=2.54cm,marginparwidth=1.75cm]{geometry}
\usepackage{scrlayer-scrpage}
\usepackage{ragged2e}
\usepackage[overload]{empheq}
\ofoot[\pagemark]{\pagemark}
\usepackage{tikz}
\usetikzlibrary{shapes,decorations,arrows,calc,arrows.meta,fit,positioning}
\tikzset{
	-Latex,auto,node distance =1 cm and 1 cm,semithick,
	state/.style ={ellipse, draw, minimum width = 0.7 cm},
	point/.style = {circle, draw, inner sep=0.04cm,fill,node contents={}},
	bidirected/.style={Latex-Latex,dashed},
	el/.style = {inner sep=2pt, align=left, sloped}
}
\tikzset{
	vertex/.style = {
		circle,
		fill            = black,
		outer sep = 2pt,
		inner sep = 1pt,
	}
}

\tikzstyle{line} = [draw, -latex']
\usetikzlibrary{arrows,decorations.markings,patterns,calc}
\usetikzlibrary{shadows}
\tikzset{shadow scale=1, shadow xshift=-.5ex, shadow yshift=-.5ex,
	opacity=.5, fill=black!50, every shadow}
\color{black}
\usepackage[titletoc,title]{appendix}

\usepackage[hidelinks,breaklinks=true]{hyperref}

\newcommand{\indep}{\rotatebox[origin=c]{90}{$\models$}}

\newtheorem{ass}{Assumption}[section]

\newtheorem{thm}{Theorem}[section]

\newcommand{\IQ}[1]{\mathbbm{1}_{\{Y < q_1(#1(X),X)\}}}
\newcommand{\IQg}[1]{\mathbbm{1}_{\{Y > q_1(#1(X),X)\}}}

\newcommand{\IQo}[1]{\mathbbm{1}_{\{Y < q_0(#1(X),X)\}}}
\newcommand{\IQog}[1]{\mathbbm{1}_{\{Y > q_0(#1(X),X)\}}}

\newcommand{\ip}{\mathbbm{1}^+}
\newcommand{\im}{\mathbbm{1}^-}

\usepackage{dcolumn}
\newcolumntype{d}[1]{D{.}{.}{#1} }
\newcolumntype{Y}{>{\centering\arraybackslash}X}
\newcolumntype{Z}{>{\flushleft\arraybackslash}X}

\newcommand{\titleinfo}{\vspace{-3cm} \LARGE  Heterogeneous Treatment Effect Bounds under Sample Selection with an Application to the Effects of Social Media on Political Polarization}
\title{\titleinfo}
\def\authora{Phillip Heiler}
 
\def\emaila{\href{mailto:pheiler@econ.au.dk}{pheiler@econ.au.dk}}

\begin{document}
	\begin{titlepage}
		\title{\titleinfo \thanks{ \scriptsize I would like to thank Laurent Davezies, Xavier d'Hautfoeuille, Patrik Guggenberger, Julius Ilciukas, Giovanni Mellace, Tomasz Olma, Pedro Sant'Anna, Vira Semenova, Jörg Stoye, Nadja van't Hoff, the participants of the Econometric Society European Winter Meeting 2022, the CREST Microeconometrics Seminar, the Penn State Econometrics Seminar, and the SEA 91st Annual Meeting for fruitful discussions and comments that helped to greatly improve the paper. A special thank you is also sent to Ro'ee Levy for providing access to the data and insightful discussions. All remaining errors are mine.}}
		\author{\authora\thanks{{\scriptsize Aarhus University. Department of Economics and Business Economics, TrygFonden's Centre for Child Research, Fuglesangs All\'e 4, 8210 Aarhus V, Denmark, email: \emaila. Part of this research was conducted during main affiliation with Harvard University, Department of Economics. 02138 Cambridge MA, United States.}} }
	
		%\date{\underline{First version}: September 09, 2022 \\[-1ex]
		%	\underline{This version}: \today}
		\maketitle
		\thispagestyle{empty}
		
		\begin{abstract} \singlespacing	\small
			We propose a method for estimation and inference
			for bounds for heterogeneous causal effect parameters in general sample selection
			models where the treatment can affect whether an outcome
			is observed and no exclusion restrictions are available. 
			The method provides conditional effect
			bounds as functions of policy relevant pre-treatment variables. It allows for
			conducting valid statistical inference on the unidentified conditional effects.
			We use a flexible debiased/double machine learning approach that can
			accommodate non-linear functional forms and high-dimensional confounders. Easily verifiable high-level conditions for estimation, misspecification robust confidence intervals, and uniform confidence bands are provided as well. 
			We re-analyze data from a large scale field experiment on Facebook on counter-attitudinal news subscription with attrition. Our method yields substantially tighter effect bounds compared to conventional methods and suggests depolarization effects for younger users. 
		\end{abstract}
		%\vfill
	\end{titlepage}

\setlength\abovedisplayskip{3pt}
\setlength{\belowdisplayskip}{3pt}

\newpage \pagenumbering{arabic}

%\vspace{-12pt}
\section{Introduction}
In this paper, we propose a novel method for estimation and inference for bounds of heterogeneous causal effects when outcome data is only selectively observed and no exclusion restrictions or instruments are available. In particular, we are concerned with the case when the treatment of interest itself can affect the selection process and when effects are heterogeneous along both observable and unobservable dimensions. The bounds are derived from a \textit{conditional monotonicity} assumption in the selection equation. %\citep{semenova2023generalized}.
They can be used to study the effects of interventions on \textit{always-taker} units, e.g.~the effects of active labor market policies on earnings on the population that is working regardless of whether they were subject to the intervention or not.\footnote{Note that in contrast to the typical setup and nomenclature in the instrumental variables literature, the principal strata are defined with respect to potential selection state caused by treatment, not potential treatment states caused by an instrument. The always-taker stratum is also sometimes referred to as \textit{inframarginal} or \textit{always-observed}.} %\citep{lee2009training}. 
They can also be applied to obtain credible bounds in experimental studies where the original treatment can affect selection. %This is common in (field-)experiments where the willingness of participants to reply to e.g~follow-up surveys can heavily depend on the initial treatment investment \citep{hjortskov2018encouraging}.

When applying established partial identification approaches for similar sample selection problems in practice, unconditional or subgroup specific effect bounds \citep{horowitz2000nonparametric,zhang2003estimation,lee2009training,semenova2023generalized} are often wide and thus too uninformative for assisting policy. Narrower bounds that also exploit covariate information can be used for a better targeting of interventions under weaker, i.e.~more credible, conditions compared to restrictive point-identified methods that require exclusion restrictions and/or distributional assumptions. The nonparametric heterogeneity based approach in this paper helps to tighten bounds along policy relevant pre-treatment variables. This is due to the fact that the severity of the identification problem, i.e.~the width of the identified set, can vary substantially along the confounding dimensions most associated with the heterogeneity variables of interest. In addition, the procedure has significant advantages over calculating bounds within discrete partitions of the data: It can accommodate continuous variables and, as it extracts signals for the bounds \textit{before} conditioning on heterogeneity variables, exploits larger samples as well as potential group patterns/restrictions for modeling selection probabilities and other relevant nuisance functions. 

The method can also incorporate a high-dimensional number of confounders building on debiased machine learning (DML) methodology \citep{chernozhukov2018double,semenova2021debiased}. %\citep{chernozhukov2018double}. 
%It exploits recent advances using approximately unbiased moments for nonparametric regression analysis of heterogeneous effect parameters \citep{semenova2021debiased}. 
We derive explicit high-level conditions regarding the quality of the nuisance quantity estimators that can be verified in a variety of settings for popular nonparametric or machine learning estimators such as high-dimensional sparse regression, deep neural networks, or random forests. 
We provide analytical confidence intervals for heterogeneous effects that are robust against different types of model misspecification as well as uniform confidence bands using a multiplier bootstrap. %exploiting recent advances in the literature on robust inference in partially identified models \citep{andrews2019inference,stoye2020simple}.   
\vspace{-8pt}
\begin{figure}[!h] \caption{Partially Identified Effects, Confidence Intervals and Band} \label{fig_example1_both} \centering 
	\includegraphics[width=0.49\textwidth, trim= 0 120 0 100, clip]{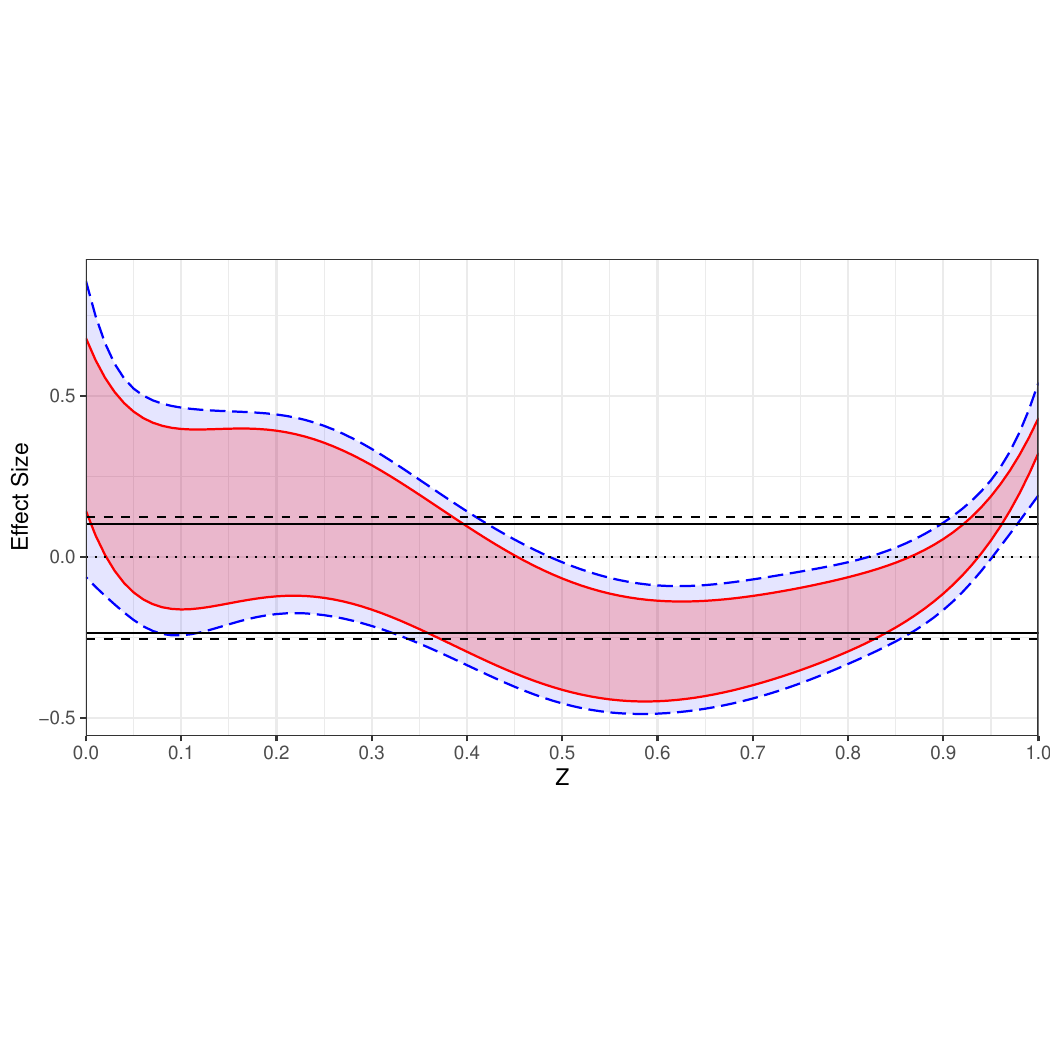}
	\includegraphics[width=0.49\textwidth, trim= 0 120 0 100, clip]{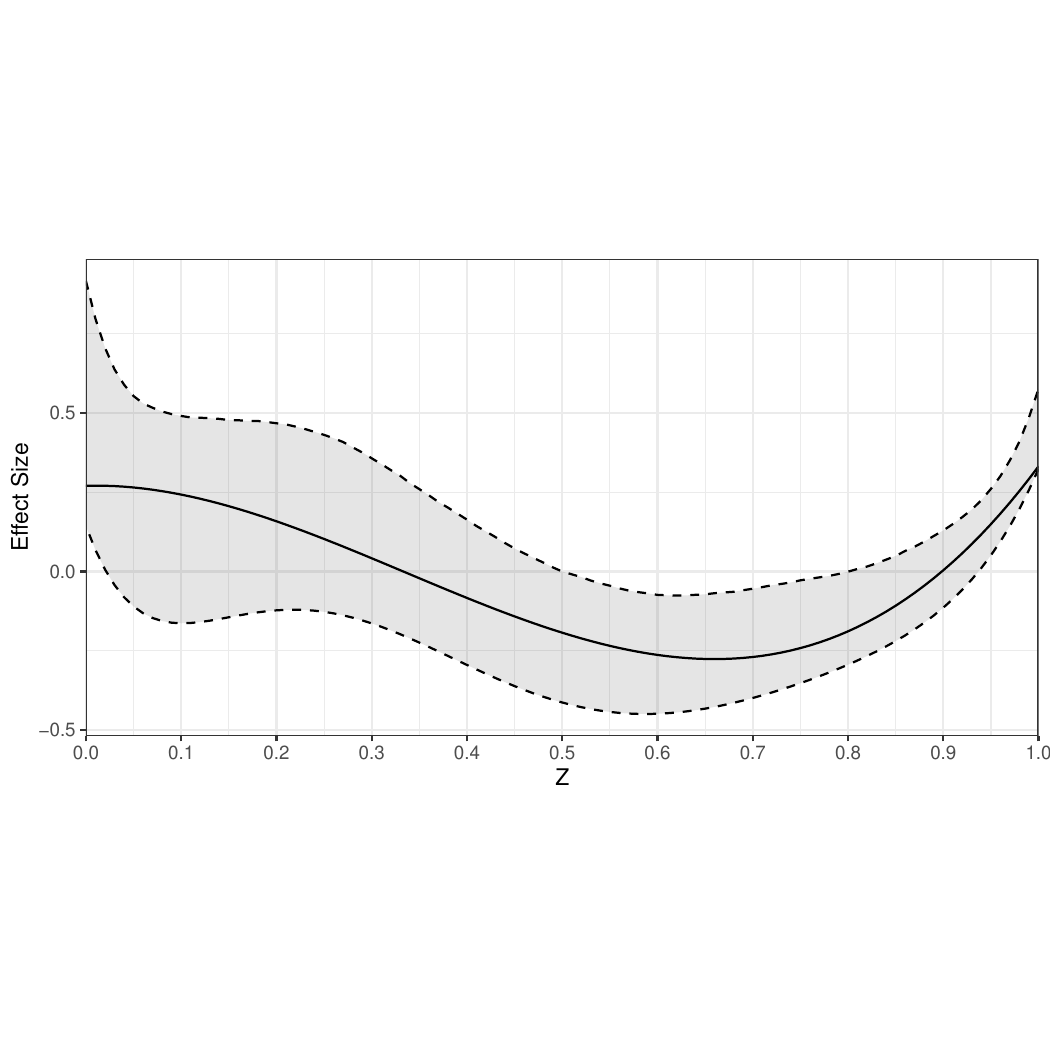}	\vspace{-12pt}
	\begin{justify} \footnotesize
		The black lines (left plot) are generalized Lee bounds \citep{semenova2023generalized} with pointwise 95\%-confidence intervals (dashed). The red area is the identification region using the method from this paper. The dashed blue lines are the pointwise 95\%-confidence intervals for the heterogeneous effects. The black line (right plot) is the true conditional average treatment effect curve. The black area is the 95\%-confidence band. Nuisance parameters are estimated via honest generalized random forests and 10-fold cross-fitting with sample size $n = 5000$ and $p = 100$ regressors. Heterogeneous bounds use basis-splines. For more details consider Section \ref{sec_MC1}.
	\end{justify} \vspace{-14pt}
\end{figure}

Figure \ref{fig_example1_both} illustrates the proposed method for a one-dimensional heterogeneity analysis. It plots identified sets, confidence intervals, and confidence band for the causal effect in dependence of a pre-treatment variable $z$. These could e.g.~be bounds on the effect of job training on earnings as a function of pre-training earnings. 
%The black lines (left plot) correspond to state-of-the-art unconditional ``generalized Lee bounds'' \citep{semenova2023generalized} with 95\%-confidence intervals (dashed). The shaded red area is the identified region using the method from this paper. The blue area between the dashed blue lines is the 95\%-confidence region for the heterogeneous effect. The black line (right plot) depicts the true conditional average treatment effect of the design.
We can see that unconditional analysis cannot rule out a zero effect while the confidence intervals of the heterogeneous bounds clearly suggest significant negative ($0.5 < z < 0.8$) and positive effects ($z > 0.95$) leading to different policy recommendations. 
Note that these conclusions are achieved not just by \textit{heterogeneous locations} of the bounds but also by \textit{narrower widths} for certain $z$ values. % compared. % to the unconditional case. 
Thus, heterogeneous bounds can reduce uncertainty stemming from weaker identification assumptions for certain sub-populations. Monte Carlo simulations suggest that the presented confidence intervals perform well in finite samples. %We also provide a simple bootstrap method for uniform inference on the whole effect curve that control asymptotic size under correct specification.

Our application is concerned with the effect of social media news consumption on political polarization. In 2022, over 70\% of US adults consumed news on social media.\footnote{Pew Research Center, Survey of U.S. adults, July 18-Aug 21, 2022.}
The consequences of social media and online news consumption on political polarization are of major importance: High partisan attitudes threaten the functioning of society and democracy as well as trust in public and private institutions \citep{phillips2022affective}. Many democratic societies have been experiencing significant changes in partisan attitudes since the broad roll-out of social networks such as Facebook and Twitter (now X). 
However, the exact contributions are under debate (\cite{haidtONGOINGsocialmediaPolitical}, ongoing). % In particular, research findings are often heterogeneous or contradicting depending on context, time, country, social media platform/algorithm, political ideology, demographics, and outcome measures 

We study the effect of Facebook news subscription on \textit{affective polarization}. Facebook is the most dominant social media site for news consumption among US adults (31\% regularly get news on this site). Affective polarization measures relative attitudes toward opposing partisans in terms of (dis)like and (dis)trust. We re-examine data collected by \cite{levy2021social} who employs a large scale field experiment on Facebook where units are nudged towards subscribing to popular media outlets with clear partisan ideology such as \textit{FoxNews} or \textit{MSNBC}. In particular, we measure the effects of the counter-attitudinal treatment in terms of political leaning on affective polarization after two months. 
The outcome measure suffers from large differential attrition rates between treatment and control groups and within political ideology (over 50\% total). %Moreover, there is heterogeneity in the direction of the effect of treatment. 

Overall, the findings do not contradict \cite{levy2021social} who suggests a decrease in affective polarization by $-0.06$ standard deviations not corrected for attrition. The relative size of the identified set benefits from the inclusion of covariates even under the randomized treatment assignment. Independently of nuisance parameter specifications, our identified sets are around $[-0.08,0.01]$ and 58\% tighter compared to conventional monotonicity bounds that cannot rule out a moderate increase in affective polarization ($[-0.16,0.06]$). 
Looking at subgroup heterogeneity, we are getting closer to point identification for some groups. For example, for conservatives and 18-year olds, the identified sets are $[-0.07,-0.04]$ and $[-0.09,-0.04]$ respectively. Additionally accounting for the statistical uncertainty, there is weak evidence in favor of depolarization effect for the younger users.

%This is an ideal setup for the method developed in this paper. 
%We find weak evidence against an overall null-effect after correcting for attrition. Our identified set using random forest-based methods of $[-0.156, -0.016]$ (standard deviations) suggest a decrease in affective polarization and covers the original point estimate of $-0.055$ by \cite{levy2021social} not corrected for attrition.  We also detect significant heterogeneity: The counter-attitudinal treatment seems to significantly reduce affective polarization for moderates and conservatives. The corresponding identified sets cover parameters of 2-3 times the magnitude of the unconditional estimate by \cite{levy2021social}. The identified set for liberals also excludes zero but is statistically insignificant.
%Moreover, there are significant negative effects for Facebook users below the age of 44. 
%The identified sets for younger users again suggest 2-3 times larger effects compared to the uncorrected unconditional point estimate. 

The paper is structured as follows: Section \ref{sec_literature1} discusses the methodological literature. Section \ref{sec_methodology1} contains model, estimator, and confidence intervals. Section \ref{sec_LargeSample1} provides technical assumptions, confidence bands, and large sample properties. Section \ref{sec_MC1} presents Monte Carlo simulations. Section \ref{sec_application} contains the empirical study. 
Section \ref{sec_conclusion1} concludes. All proofs and extensions are in the Appendix. The \verb|R| package \verb|HeterogeneousBounds| and replication notebook can be found on the author's personal web-page. 
 
\section{Methodological Literature}\label{sec_literature1}
%\subsection{Methodological Literature} 
%Estimating causal effects and sample selectivity is a long-standing problem in economic research \citep{heckman1979sample}. 
Bounds for causal effects under weak assumptions have been considered in a series of papers by Charles Manski and others, see \cite{molinari2020microeconometrics} for a comprehensive overview. \cite{horowitz2000nonparametric} develop nonparametric bounds for treatment effects in selected samples. \cite{zhang2003estimation} consider bounds for always-taker units under a monotonicity assumption regarding the effect of treatment on selection, commonly imposed in generic sample selection models, and/or stochastic dominance assumptions on the potential outcomes. \cite{imai2008sharp} demonstrates sharpness of these bounds. \cite{huber2015sharp} consider similar sharp bounds for other principal strata. \cite{lee2009training} provides asymptotic theory %and an application to the evaluation of a large-scale job training program 
for \cite{zhang2003estimation} bounds using (conditional) monotonicity. Without further assumptions, these bounds are only applicable unconditionally or for low-dimensional discrete partitions of the covariate space and now commonly referred to as ``Lee bounds''. 
\cite{semenova2023generalized} provides ``generalized Lee bounds'' under a conditional monotonicity assumption. Our paper uses similar identification assumptions. \cite{semenova2023generalized} also allows for high-dimensional and continuous confounders and generalizes the approach to multiple outcomes but does neither consider heterogeneity analysis nor misspecification-robust inference. 
\cite{bartalotti2021identifying} also propose identification of bounds for always-takers within a marginal treatment effect framework. 
Using monotonicity and stochastic dominance, they tighten effect bounds based on underlying treatment propensities. However, they do neither address heterogeneity beyond the propensity score, asymptotic properties, inference, nor potential misspecification. Moreover, their method is not suitable for many confounding variables without imposing additional parametric assumptions.

Our work is also directly related to the literature on robust or (Neyman-)orthogonal moment functions and DML. 
%Orthogonal moment functions are a key element to cope with flexible nonparametric or machine learning estimators for complex functional relationships and/or a large number of confounding variables.  
For point-identified parameters, there are now many approaches that use machine learning and orthogonal moments in both experimental and observational studies, see e.g.~\cite{belloni2014inference}, \cite{farrell2015robust}, \cite{chernozhukov2018double}, and \cite{wager2018estimation}. \cite{chernozhukov2018double} develop a canonical framework that can be used for inference on low-dimensional target parameters such as the average treatment effect.
In the context of heterogeneity analysis, orthogonal moments have been exploited in point-identified problems by using them as pseudo-outcomes in (nonparametric) regression models to obtain predictive causal summary parameters, see e.g.~\cite{lee2017doubly}, \cite{fan2020estimation}, \cite{semenova2021debiased}, and \cite{heiler2021effect} or \cite{knaus2022double} for an overview. %We follow the approach by \cite{semenova2021debiased} who use nonparametric series regression \citep{belloni2015some} together with machine learning estimators for estimating predictive causal heterogeneity parameters. 
Our localization approach is closest to \cite{semenova2021debiased} who estimate heterogeneity parameters via nonparametric projections using least squares series methods \citep{belloni2015some}. Employing a series approach is particularly useful for studying potential misspecification since the heterogeneity step can then be reformulated as a relaxed moment inequality problem with deviation parameters that are proportional to the difference between effect bounds and their linear predictors. It also allows for construction of uniform confidence bands under correct specification. 

\cite{semenova2023generalized} constructs DML estimators for unconditional Lee-type bounds. \cite{semenova2023debiased} considers partially identified parameters for linear moment functions with DML. A crucial point in both of these papers is that, while the effect of interest might not be point-identified, bounds themselves are characterized by well-understood convex moment problems or the corresponding support function. The same applies to the heterogeneous bounds considered in this paper. 
We derive the asymptotic distribution of the nonparametric heterogeneous DML based estimator for the identified set. % instead of a single parameter or functional as in by \cite{semenova2021debiased}. 
This nests the univariate generalized Lee bounds by \cite{semenova2023generalized} as a special case. In addition, we also provide the bounds and inference theory for separate subgroups of always-takers defined by their monotonicity type, i.e.~their effect sign of treatment on selection.
In contrast to \cite{semenova2023debiased}, the moment functions are nonlinear in the outcome.
The use of machine learning (random forests) for Lee-type bounds has also been heuristically discussed by \cite{cornelisz2020addressing}. They do, however, not provide any formal theory for estimation and inference.  %\cite{maria2022adaptivePI} also considers heterogeneous partially identified parameters using random forests under different identification assumptions.   
\cite{olma2021nonparametric} also discusses nonparametric estimation of generic truncated conditional expectations based on similar conditional moments as the ones used in this paper. He suggests kernel estimation for both stages and provides pointwise linearization and distribution results. In contrast, our approach can handle generic first-stage learners that also work in setups with high-dimensional confounding. Moreover, \cite{olma2021nonparametric} does neither consider local power improvements, misspecification, nor uniform inference.   

Inference for partially identified parameters in sample selection models is a non-trivial task. %In this paper, we are concerned with constructing two-sided confidence regions and associated tests for the heterogeneous causal effects. 
Relying on quantiles of large sample distributions of the \textit{bounds} can be overly conservative for the actual \textit{effect} of interest. The relevant uncertainty for the latter depends on the actual width of the identified set. If small, then deviations from a null value are likely to occur in both positive and negative direction, i.e.~the problem is effectively two-sided. If large, then uncertainty in one direction dominates, rendering the testing problem close to one-sided. \cite{imbens2004confidence} consider related confidence intervals for partially identified parameters. % and provide a method for correction.
Their method is not uniformly valid with regards to width of the underlying identified set due to an implicit superefficiency assumption.
%Their method relies on an implicit superefficiency assumption that does not apply in many setups \citep{stoye2009more}. In these cases, the \cite{imbens2004confidence} bounds are not uniformly valid with regards to the width of the underlying identified set. 
\cite{stoye2009more} suggests to artificially impose superefficiency via shrinkage. \cite{andrews2010inference} provide a more general approach within a moment inequality framework. \cite{andrews2013inference} and \cite{andrews2017inference} study inference based on (many) conditional moment inequalities in the presence of nuisance parameters. While the latter can be infinite dimensional as in this paper, their inference target is essentially parametric. Our approach is closest in spirit to \cite{andrews2014nonparametric} who consider nonparametric estimation based on conditional moment inequalities. Their method can be adjusted to conduct pointwise inference on conditional effect bounds. However, as they study a relatively general setup, they do neither address local power improvements for effects, estimated nuisances, misspecification, nor uniform inference. \cite{chernozhukov2019inference} also consider inference using generic moment inequalities in possibly high dimensions allowing for nonparametric sieve-type parameters. In contrast, we exploit the specific shape of the identified set as well as the restriction to low-dimensional heterogeneity variables to obtain stronger asymptotic results that can be used for power improvements.\footnote{\cite{chernozhukov2019inference}, Appendix B also heuristically discusses approximate moment inequalities/misspecification arising from estimation using for parametric nuisance models.}   

When analyzing heterogeneity, we are concerned with effect bounds at potentially many points and thus there is an additional risk of \textit{local misspecification} compared to unconditional effects. For instance, even when the conditional moments are correctly specified and ordered, their linear predictors might intersect. In addition, there can be \textit{global misspecification}, e.g.~due to misspecified selection probabilities, that can lead to a reversed ordering of the orthogonalized local bounds in both the population and in finite samples. Under such misspecification, aforementioned bounds and moment inequality methods can produce empty or very narrow confidence regions suggesting spuriously precise inference \citep{andrews2019inference}. \cite{andrews2019inference} propose inference methods that extend the notion of coverage to pseudo-true parameter sets in general moment inequality frameworks, see also \cite{stoye2020simple} for a power improvement for generic regular parametric bound estimators. This paper adapts the inference method of \cite{stoye2020simple} to nonparametric heterogeneous bounds with machine learning in the first stage.

\section{Methodology} \label{sec_methodology1}
\subsection{Model and Identification Assumptions}
In this section, we introduce the sample selection model and the main identification assumptions followed by a brief review of the construction of effect bounds. We then show how to exploit the latter to construct, estimate, and conduct inference on the heterogeneous partially identified effect parameters.  

\begin{figure}[!h] \caption{Nonparametric Sample Selection Model} \centering \label{fig_dag1}
	\begin{tikzpicture}
		\tikzset{line width=1.5pt}

		\node[ellipse,draw,line width = 1.2pt,dashed, drop shadow, fill = white] (-1) at (8,0) {\scriptsize Unobservables $U$};
		\node[ellipse,draw,line width = 1.2pt, drop shadow, fill = white] (0) at (4,1.2) {\scriptsize Covariates $X$};
		\node[ellipse,draw,line width = 1.2pt, drop shadow, fill = white] (1) at (0,0) {\scriptsize Treatment $D$};
		\node[ellipse,draw,line width = 1.2pt, drop shadow, fill = white] (2)  at (5.5,-1.2) {\scriptsize Selection $S$};
		\node[ellipse,draw,line width = 1.2pt, drop shadow, fill = lightgray] (3) at (2,-1.2) {\scriptsize Outcome $Y$ (latent)};
		\node[ellipse,draw,line width = 1.2pt, drop shadow, fill = white] (4) at (4,-2.4) {\scriptsize Observed Outcome $Y\times S$};
		
		\path (1) edge (3);
		\path (2) edge (4);
		\path (3) edge (4);

		\path (0) edge[out=west] (1);	
		\path (0) edge (2);	
		\path (0) edge (3);	
		
		\path[<->] (0) edge[out=east,dashed]  (-1);	
		
		\path (-1) edge (2);	
		\path (-1) edge (3);	
		\path (1) edge (2);	
		
	\end{tikzpicture} \footnotesize \begin{justify}
	This is a graphical representation of a nonparametric sample selection model. Nodes denote variables and edges are structural relationships, i.e.~a missing arrow from one node to another is an exclusion restriction. Unobserved independent components at each node are omitted. Unobserved variables are in dashed nodes. %The Markov assumption for the graph holds, i.e.~it is a directed acyclical graph where variables are independent of all their non-descendants given their direct parental nodes \citep{tian2002ageneral}. For inclusion of missing variables and their use in graphical models see \cite{mohan2021graphical}. %Note that the graph is not a directed acylclical graph (DAG) due to failure of the Markov assumption: Conditional on their parental nodes, outcome and selection here are not independent allowing for endogeneity as in the classic sample selection framework by \cite{heckman1979sample}.\footnote{}
	\end{justify} \vspace{-12pt}
\end{figure}
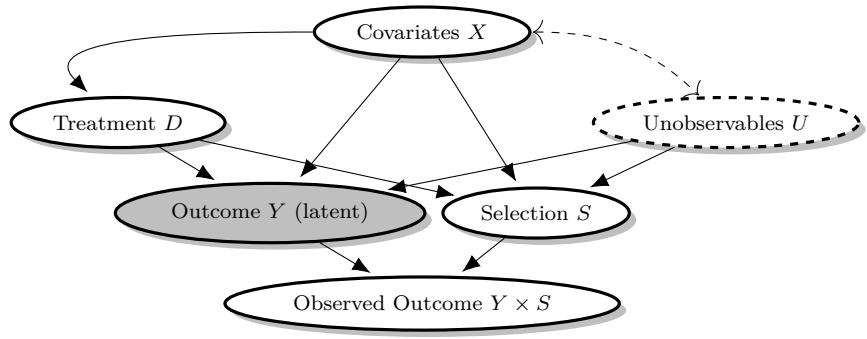

Assume for $i=1,\dots,n$ we observe iid data $W_i = (X_i',D_i,S_i,Y_iS_i)$ where $X_i$ is a vector of predetermined covariates supported on $\mathcal{X} \subseteq \mathbb{R}^d$, $D_i$ is a treatment indicator, $S_i$ indicates whether the outcome is observed, i.e.~we only observe $Y_iS_i$ and not $Y_i$. We would like to evaluate the average causal effect of $D_i$ on outcome $Y_i$ for the units that are selected under both treatment or control condition. We focus on the case of known treatment propensities $e(x) = P(D_i=1|X_i=x)$ during the exposition, but the methodology can also be extended to observational studies where they are unknown and have to be estimated. The prototypical setup is depicted by the graph in Figure \ref{fig_dag1} derived from the nonparametric structural equation model.

The model nests the classic sample selection model \citep{heckman1979sample} where the treatment of interest enters in selection step, see \cite{lee2009training} for a more restrictive parametric sample selection model representation. It allows for the treatment $D$ to affect the selection $S$. 
The model does not restrict the relationship between selection and the partially unobserved outcome of interest. In particular, there can be unobservables $U$ related to both selection and potential outcomes. 
In addition, there are no variables available which only affect selection that could be used to identify causal effects on the partially unobserved outcome via instrumental variable based methods. 
%The model

Based on this model, we define unit potential outcomes $Y_i(1)$ and $Y_i(0)$ and potential selection indicators $S_i(1)$ and $S_i(0)$ for units $i=1,\dots,n$. They correspond to a unit's value in outcome or selection when the treatment is exogeneously set to $D_i = 1$ or $D_i=0$ respectively under the model in Figure \ref{fig_dag1}.
Note that potential selection and potential outcomes can still be dependent. This implies that, even conditional on $X_i$, causal effects can differ for different sub-types defined by their respective $S_i(1)$ and $S_i(0)$ variables. The target is to evaluate the causal effect of $D_i$ for the %\textit{always-takers} or \textit{inframarginal} 
units that are selected in both control and treatment state, i.e.~units for which $S_i(1)=S_i(0)=1$.
As $D$ affects $S$, a comparison of treated and control units for selected units does not yield a valid causal comparison for the sub-population $S_i(1) = S_i(0) = 1$.  
However, the model implies the following conditional independence relationships in Rubin-Neyman potential outcome notation\footnote{These are the marginal versions of the joint independence assumption $Y_i(0), Y_i(1), S_i(0), S_i(1) \indep\ D_i | X_i$ considered in \cite{lee2009training} that yield the same identification results.}

\begin{align}
	Y_i(d), S_i(d)\ \indep\ D_i\ |\ X_i \ \text{ for } \ d=0,1. \label{eq_CIA1}
\end{align}

Condition \eqref{eq_CIA1} cannot be exploited in a standard selection on observables strategy for point identification as, even conditional on $X_i$, we only observe the selected subset of potential outcomes for which $S_i = 1$.
%However, conditional on $X_i$, the treatment is exogenous with respect to potential selection. to obtain bounds for the causal effect in the $S_i(1) = S_i(0) = 1$ population, i.e.~the effect is \textit{partially identified} under the model. For that, it is not necessary to impose any restrictions on the sign of the effect of $D$ on $S$. 
We now impose a monotonicity assumption on the selection equation. In particular, treatment can affect selection positively or negatively. The sign, however, must be uniquely determined by the vector of covariates $X_i$. This is a \textit{weak} or \textit{conditional monotonicity} assumption based on ex-ante unknown partitions: \begin{ass}[Weak/Conditional Monotonicity]
	There exist partitions of the covariate space $\mathcal{X} = \mathcal{X}_+ \cup \mathcal{X}_-$ such that $P(S_i(1)\geq S_i(0)|X_i\in\mathcal{X}_+) = P(S_i(1)\leq S_i(0)|X_i\in\mathcal{X}_-) = 1$.
\end{ass} This assumptions is consistent with additive separability of observables and unobservables in an otherwise unrestricted selection equation. We also assume that, with positive probability, there are comparable units between selected treated and selected non-treated as well as between treated and non-treated overall: \begin{ass}[Multiple Overlap]
	For all $x\in\mathcal{X}$ and $d \in \{0,1\}$ we have that
	$0 < P(S_i=1|X_i=x,D_i=d) < 1$ and $0 < P(D_i=d|X_i=x) < 1$.
\end{ass}
We also rule out the case where treatment does not affect selection as in this case effects are point-identified:
\begin{ass}[Margin]
	There is a constant $c > 0$ and a set $\mathcal{\bar{X}} \subset \mathcal{X}$ with $P(\mathcal{X}\backslash\mathcal{\bar{X}}) = 0$ such that $\inf_{x\in\mathcal{\bar{X}}}\ |P(S_i=1|X_i=x,D_i=1)-P(S_i=1|X_i=x,D_i=0)| > c$.
\end{ass}
This assumption is most plausible when there are discrete variables and/or continuous variables with bounded support. For example, a standard binary response model for selection where $D_i$ enters with a fixed, non-zero parameter then obeys this condition. Assumption 3.3 could be weakened to a margin condition that restricts the behavior of the treatment on selection effect distribution around zero with some technical modifications \citep{semenova2023generalized}. We abstract from these issues for simplicity. 

At last, we need a standard continuity condition in each treatment arm: \begin{ass}[Continuity]
	The conditional outcome distributions $P(Y_i\leq y|D_i=1,S_i=1,X_i=x)$ and $P(Y_i\leq y|D_i=0,S_i=1,X_i=x)$ are continuous almost everywhere. %on $\mathcal{X^+}$ and $\mathcal{X^-}$ respectively. 
\end{ass}

For identification of the sharp effect bounds, we first outline the case of \textit{strong} or \textit{unconditional} monotonicity as considered in \cite{zhang2003estimation} and \cite{lee2009training}, i.e.~when $\mathcal{X}^- = \{ \emptyset \}$ or $S_i(1)\geq S_i(0)$ with probability one. In this case, selected units within the control group must be \textit{always-takers} or \textit{inframarginal} in the sense of $S_i(1) = S_i(0) = 1$ while within the treated group there is a mixture of always-takers and \textit{compliers} or \textit{marginal} units who are induced to be selected by the treatment. Let the conditional causal effect of the always-takers be defined as \begin{align}
	\theta_{AT}(x) = E[Y_i(1)-Y_i(0)|S_i(0)=S_i(1)=1,X_i=x].
\end{align}
This is the expected causal effect for a unit with covariates $x$ whose outcome would be observed independently of its treatment status. Conditional on $X_i=x$, let $q_1(u,x)$ be the $u$-th quantile of the treated selected units, $s(d,x) = P(S_i=1|X_i=x,D_i=d)$ the selection probability, and $p_0(x) = {s(0,x)}/{s(1,x)}$ the share of always-takers relative to always-takers and compliers. \cite{zhang2003estimation} show that the sharp bounds are given by \begin{align} 
	\theta_L(x) \leq \theta_{AT}(x) \leq \theta_U(x), \label{eq_boundsX}
\end{align}
where \begin{align}
	\theta_L(x) &= E[Y_i|D_i=1,S_i=1,Y_i\leq q_1(p_0(x),x),X_i=x] - E[Y_i|D_i=0,S_i=1,X_i=x], \notag \\
	\theta_U(x) &= E[Y_i|D_i=1,S_i=1,Y_i\geq q_1(1-p_0(x),x),X_i=x] - E[Y_i|D_i=0,S_i=1,X_i=x]. \notag \\[-7ex] \label{eq_boundsX_DEFINITION}
\end{align} 
%If $p_0(x) = 1$, the treatment would not affect the response and thus, conditional on observables, causal effects are point-identified by comparing the outcome of selected treated to selected control units and $\theta_L(x) = \theta_U(x)$. 

While these $x$-specific or ``personalized bounds'' can provide some insights, they are not very useful in applications with continuous variables and/or many discrete cells to evaluate. Moreover, in such cases it is often no longer possible to consistently estimate the conditional bounds and to construct asymptotically valid confidence intervals without further restrictions. This is equivalent to estimation of personalized treatment effects under unconfoundedness in high dimensions \citep{chernozhukov2018generic}. Therefore, we propose to instead consider heterogeneous effect bounds conditional on a smaller, pre-specified (policy relevant) variables $Z_i$ supported on $\mathcal{Z}$. These can be (functions of) any observable variable possibly affecting treatment, selection, and outcome. Thus, without loss of generality, we can treat $Z_i = f(X_i)$ where $f(\cdot)$ is a measurable, possibly multi-valued function. As $\mathcal{Z}$ is of low dimension, this allows for conducting asymptotically valid inference for heterogeneous causal effects even if the original confounding dimension of $\mathcal{X}$ is large. The main parameter of interest is then given by \begin{align}
	\theta_{AT}(z) &= E[Y_i(1)-Y_i(0)|S_i(0)=S_i(1)=1,Z_i=z] \notag \\
	&= E[\theta_{AT}(X_i)|S_i(0)=S_i(1)=1,Z_i=z]. \label{eq_theta_ATz}
\end{align}
This can be interpreted as expected causal effect for the always-takers at ``summary group'' $f(X_i) = Z_i = z$.

\subsection{Heterogeneous Effect Bounds} 
We now demonstrate how to obtain sharp bounds and moment estimators for \eqref{eq_theta_ATz} using (strong) monotonicity. Using the sharp upper bounds $\theta_U(x)$ from \cite{zhang2003estimation} or \cite{lee2009training} yields \begin{align}
	\theta_{AT}(z) &\leq E[\theta_U(X_i)|S_i(0)=S_i(1)=1,Z=z] \notag \\
	&= \frac{E[\theta_U(X_i)S_i(0)|Z_i=z]}{E[S_i(0)|Z_i=z]} \notag \\
	&= \frac{E[\theta_U(X_i)s(0,X_i)|Z_i=z]}{E[s(0,X_i)|Z_i=z]},
\end{align}
by monotonicity and the law of iterated expectations. The same steps apply analogously to the lower bound. Thus, for identification, it is sufficient to have two (non-centered) moment functions: (i) $\psi_B^+(W_i,\eta)$ that identifies the conditional always-taker bound scaled by the density of always-takers and (ii) $\psi_{S_0}(W_i,\eta)$ that identifies the conditional always-taker share itself. $\eta$ here is a vector of nuisance quantities, see Section \ref{sec_weakmon2}, Table \ref{tab_momentsAll1} for a complete definition. In particular, we need that these are valid conditional on $X_i$ at the true nuisance parameters $\eta = \eta_0$, i.e.~that \begin{align}
	E[\psi_B^+(W_i,\eta_0)|X_i=x] &= \theta_{B}(x)s(0,x), \notag \\
	E[\psi_{S_0}(W_i,\eta_0)|X_i=x] &= s(0,x).
\end{align}
% In general, we need two moment functions for each bound $B \in \{L,U\}$: i) $\psi_B^+(W_i,\eta)$ that identifies the conditional always-taker bound scaled by the density of always-takers and ii) $\psi_{S_0}(W_i,\eta)$ that identifies the conditional always-taker share itself. $\eta$ here is a vector of potential nuisance quantities, see Section \ref{sec_weakmon2}, Table \ref{tab_momentsAll1} for a complete definition. For identification, it is sufficient that these functions are valid conditional on $X_i$ at the true nuisance parameters $\eta = \eta_0$, i.e.~that \begin{align}
%	E[\psi_B^+(W_i,\eta_0)|X_i=x] &= \theta_{B}(x)s(0,x) \notag \\
%	E[\psi_{S_0}(W_i,\eta_0)|X_i=x] &= s(0,x)
%\end{align}
%and equivalently conditional on $Z_i=z$.
 In principle, many moment functions have these properties. For estimation and inference, however, we focus on the moment functions $\psi_B^+(W_i,\eta)$ as developed by \cite{semenova2023generalized} in combination with the well-known doubly robust/augmented inverse probability weighting moment for the expected potential selection \citep{robins1994estimation}. This will be important to control the large sample properties with unknown nuisance parameters, see Section \ref{sec_LargeSample1}.  %ollects all the required nuisance functions, see Section [REF] for a complete definition. 
%Moreover, $\psi_{S_0}$ provides and unbiased signal for the density $E[s(0,X)|Z=z] = E[\psi_{S_0}(W_i,\eta_0)|Z=z]$. 
We can then express the sharp heterogeneous bounds as \begin{align}
	\frac{E[	\psi_L^+(W_i,\eta_0) | Z=z]}{E[	\psi_{S_0}(W_i,\eta_0) | Z=z]}	= \theta_L(z) \leq \theta_{AT}(z) \leq \theta_U(z) =  \frac{E[	\psi_U^+(W_i,\eta_0) | Z=z]}{E[	\psi_{S_0}(W_i,\eta_0) | Z=z]}.
\end{align}
The corresponding estimators can be obtained by replacing all the conditional expectations with their (nonparametric) projections, see Section \ref{sec_estimation1}.
\subsection{Weak Monotonicity} \label{sec_weakmon2}
We now extend the identification to the weak/conditional monotonicity case. We consider the separate effects for both positive and negative monotonicity partitions as well as the conventional aggregate always-taker effect in the presence of both. Denote $\ip(x) = \mathbbm{1}(p_0(x) < 1)$ and $\im(x) = \mathbbm{1}(p_0(x) > 1)$. Given $X_i=x$ the partitions are identified. We can write the conditional always-takers effect as a piece-wise combination 
\begin{align*}
	\theta_{AT}(x) &= \theta_{AT}(x)(\ip(x) + \im(x)).
\end{align*}
We can correspondingly define partition-specific, truncated estimands as \begin{align} 
	\theta_{AT}^+(x) = \begin{cases}
		 \theta_{AT}(x) & \text{ if } p_0(x) < 1 \\	 
		 0 & \text{ else} 
	\end{cases}, \qquad 
	\theta_{AT}^-(x) = \begin{cases}
		\theta_{AT}(x) & \text{ if } p_0(x) > 1 \\
		0 & \text{ else} 
	\end{cases}, 
\end{align}
and equivalently for $\theta_B^+(x)$ and $\theta_B^-(x)$. As in the previous section, we can now apply monotonicity within the different partitions to bound $\theta_{AT}(z)$. In particular \begin{align}
	\theta_{AT}(z) 
	&\leq E[(\ip(X_i) + \im(X_i))\theta_U(X_i)|S_i(0)=S_i(1)=1,Z=z] \notag \\
	&= \frac{E[\ip(X_i)\theta_U(X_i)S_i(0) + \im(X_i)\theta_U(X_i)S_i(1)|Z_i=z]}{E[\ip(X_i)S_i(0) + \im(X_i)S_i(1)|Z_i=z]} \notag \\
	&= \frac{E[\ip(X_i)\theta_U(X_i)s(0,X_i) + \im(X_i)\theta_U(X_i)s(1,X_i)|Z_i=z]}{E[\ip(X_i)s(0,X_i) + \im(X_i)s(1,X_i)|Z_i=z]},
\end{align}
and analogously for the lower bound. Thus, for identification, it is sufficient to have four moment functions that again identify the within partition always-taker shares and partition specific effect bounds scaled with the corresponding density, i.e. \begin{align}
	\ip(x) E[\psi_B^+(W_i,\eta_0)|X_i=x] &= \theta_{B}^+(x)s(0,x), \notag \\
		\ip(x)E[\psi_{S_0}(W_i,\eta_0)|X_i=x] &= 	\ip(x)s(0,x),  \notag  \\
	\im(x) E[\psi_B^-(W_i,\eta_0)|X_i=x] &= \theta_{B}^-(x)s(1,x), \notag  \\
	\im(x) E[\psi_{S_1}(W_i,\eta_0)|X_i=x] &= \im(x) s(1,x).
\end{align}
%
%
% The low-dimensional summary effects are then obtained equivalently to the previous section as \begin{align}
%	\theta_{AT}(z) = E[\theta_{AT}(X_i)|S_i(0)=S_i(1)=1,Z_i=z]
%\end{align}
%and 
%\begin{align*}
%	\theta_{AT}^+(z) &= \begin{cases*}
%		\frac{E[\theta_{AT}(X_i)|S_i(0)=S_i(1)=1,Z_i=z,\ip]}{E[\ip|S_i(0)=S_i(1)=1,Z_i=z]} &\text{ if } P(\ip = 1|S_i(0)=S_i(1)=1,Z_i=z) > 0 	\\	 
%		0 &\text{ else} 
%	\end{cases*}
%\end{align*}
%and analogously for $\theta_{AT}^+(z)$ and $\theta_{AT}^-(z)$. We now again exploit the set of moment functions for always-taker times density as well as the potential selection means together. %For $B \in \{L,U\}$ let \begin{align*}
%%	\psi_B(W_i,\eta_0) = \psi_B^+(W_i,\eta_0)\ip + \psi^-_B(W_i,\eta_0)\im
%%\end{align*}
%We use the same moments as before within the $\ip(x)$ partition plus two functions, $\psi_B^-(W_i,\eta)$ within $\im(x)$ and $\psi_{S_1}(W_i,\eta)$ such that
%\begin{align}
%	\ip(x) E[\psi_B^+(W_i,\eta_0)|X_i=x] &= \theta_{B}^+(x)s(0,x) \notag \\
%	E[\psi_{S_0}(W_i,\eta_0)|X_i=x] &= s(0,x)  \notag  \\
%	\im(x) E[\psi_B^-(W_i,\eta_0)|X_i=x] &= \theta_{B}^-(x)s(1,x)  \notag  \\
%	E[\psi_{S_1}(W_i,\eta_0)|X_i=x] &= s(1,x)
%\end{align}
%%and equivalently for $Z_i = z$.
%These are the respective moment functions for the scaled always-taker bounds of both types scaled with a factor proportional to the corresponding always-taker shares. Using the properties of conditional expectations and of truncated densities then yields 

Overall, we obtain that
\begin{align}
		\theta_L(z) \leq \theta_{AT}(z) \leq \theta_{U}(z),  \label{eq_bounds_ATz}
\end{align}
with \begin{align}
		\theta_L(z)	&=	\frac{E[\ip(X_i) \psi_L^+(W_i,\eta_0) + \im(X_i)\psi_L^-(W_i,\eta_0)|Z=z]}{E[\ip(X_i)\psi_{S_0}(W_i,\eta_0) + \im(X_i)\psi_{S_1}(W_i,\eta_0)|Z=z]}, \notag \\
		\theta_{U}(z)		&= 	\frac{E[\ip(X_i) \psi_U^+(W_i,\eta_0) + \im(X_i)\psi_U^-(W_i,\eta_0)|Z=z]}{E[\ip(X_i)\psi_{S_0}(W_i,\eta_0) + \im(X_i)\psi_{S_1}(W_i,\eta_0)|Z=z]}. 
\end{align}
For the different monotonicity types, separate bounds are given by 
 \begin{align}
	\theta_L^+(z) \leq \theta_{AT}^+(z) \leq \theta_{U}^+(z), \notag \\
	\theta_L^-(z) \leq \theta_{AT}^-(z) \leq \theta_{U}^-(z), \label{eq_bounds_ATplmin1}
\end{align}
with \begin{align}
	\theta_L^+(z)	&=	\frac{E[\ip(X_i) \psi_L^+(W_i,\eta_0)|Z=z]}{E[\ip(X_i)\psi_{S_0}(W_i,\eta_0)|Z=z]}, \quad \theta_L^-(z) = \frac{E[\im(X_i) \psi_L^-(W_i,\eta_0)|Z=z]}{E[\im(X_i)\psi_{S_1}(W_i,\eta_0)|Z=z]}, \notag \\
	\theta_{U}^+(z)&= \frac{E[\ip(X_i) \psi_U^+(W_i,\eta_0)|Z=z]}{E[\ip(X_i) \psi_{S_0}(W_i,\eta_0)|Z=z]}, \quad \theta_U^-(z) = \frac{E[\im(X_i) \psi_U^-(W_i,\eta_0)|Z=z]}{E[\im(X_i)\psi_{S_1}(W_i,\eta_0)|Z=z]}.
\end{align}
The aggregate and partition-specific estimands are ratios. They differ by the denominator as, for the aggregate bounds, each partition-specific effect is weighted with its respective conditional probability. %One could write the aggregate effect as a weighted version of the type-specific effects above which would lead to six conditional expectations overall. Using Bayes' law reduces the problem to four conditional expectations only. 
These bounds are sharp. In the case of $Z_i = 1$, the aggregate bounds \eqref{eq_bounds_ATz} are identical to the one-dimensional generalized Lee bounds by \cite{semenova2023generalized}, i.e.~our theory covers this as a special case. 
 %In the case of $Z_i = 1$, the aggregate bounds are similar to the one-dimensional generalized Lee bounds by [REC] with a different estimator for the denominator. In the case of strong monotonicity, they are equivalent for $Z_i = 1$, i.e.~our theory covers this as a special case. 

\subsection{Estimation} \label{sec_estimation1}
We now focus on the particular moment functions with desirable properties. %For simplicity, we focus on the case where the treatment propensities $P(D=1|X=x)$ are known. 
For any required $B \in \{L^+,U^+,L^-,U^-,S_0,S_1,S_0^+,S_1^-\}$, let the (uncentered) moment function $\psi_B(W_i,\eta_0)$ be defined according to Table \ref{tab_momentsAll1}. %The moment functions for all parameters defined in the previous section and nuisance quantities are listed in Table \ref{tab_momentsAll1}. 
\begin{table}[!h] \footnotesize \centering \caption{Moment and Nuisance Functions} \label{tab_momentsAll1}
	\begin{tabular}{lc} \hline \hline \\[-.5ex]
		~ & Moment function 		\\ \hline \\[-.5ex]
		$\psi_{S_0}(W,\eta_0)$ & $\frac{(S-s(0,X))(1-D)}{1-e(X)} + s(0,X)$ \\[2ex]
		$\psi_{S_1}(W,\eta_0)$ & $\frac{(S-s(1,X))D}{e(X)} + s(1,X)$ \\[2ex]
		$\psi_L^+(W,\eta_0)$ & $\frac{D}{e(X)}SY\IQ{p_0} - \frac{(1-D)}{1-e(X)}SY + q_1(p_0(X),X)\big[\frac{(1-D)}{1-e(X)}(S-s(0,X))$ \\
		& $ - \frac{D}{e(X)}p_0(X)(S-s(1,X)) - \frac{D}{e(X)}S(\IQ{p_0} - p_0(X))\big]$ \\[2ex]
		$\psi_U^+(W,\eta_0)$ & $\frac{D}{e(X)}SY\IQg{1-p_0} - \frac{(1-D)}{1-e(X)}SY + q_1(1-p_0(X),X)\big[\frac{1-D}{1-e(X)}(S-s(0,X))$ \\
		& $ - \frac{D}{e(X)}p_0(X)(S-s(1,X)) + \frac{D}{e(X)}S(\IQ{1-p_0} - (1-p_0(X)))\big]$ \\[2ex]
		$\psi_L^-(W,\eta_0)$ & $\frac{D}{e(X)}SY - \frac{1-D}{1-e(X)}SY\IQog{1-1/p_0} -q_0(1-1/p_0(X),X)\big[ \frac{D}{e(X)}(S-s(1,X))$ \\
		&$ -\frac{(1-D)}{1-e(X)}\frac{1}{p_0(X)}(S-s(0,X)) + \frac{1-D}{1-e(X)}S(1/p_0(X) - \IQog{1-1/p_0})\big]$ \\[2ex]
		$\psi_U^-(W,\eta_0)$ & $\frac{D}{e(X)}SY - \frac{1-D}{1-e(X)}SY\IQo{1/p_0} - q_0(1/p(X),X)\big[ \frac{D}{e(X)}(S-s(1,X))$ \\
		&$ -\frac{(1-D)}{1-e(X)}\frac{1}{p_0(X)}(S-s(0,X)) - \frac{1-D}{1-e(X)}S(\IQo{1/p_0} - 1/p_0(X))\big]$ \\ \hline \\[-.5ex]
		\multicolumn{2}{c}{$\psi_{S_0^+} = \ip \psi_{S_0}$,\quad  $\psi_{S_1^-} = \im \psi_{S_1}$,\quad $\psi_{L^+} = \ip \psi_L^+$,\quad $\psi_{L^-} = \im \psi_L^-$ } \\ \hline \\[-.5ex]
		& Nuisance functions $\eta_0 = \{s(0,x),s(1,x),q_0(u,x),q_1(u,x)\}$  \\ \hline \\[-.5ex]
		$e(x)$ & $P(D=1|X=x)$ \\
		$s(d,x)$ & $P(S=1|D=d,X=x)$ \\
		%$p_0(x)$ & $s(d,x)/s(1,x)$ \\
		$q_d(u,x)$ & $\inf \{q : u \leq P(Y \leq q|D=d,S=1,X=x)\}$ \\
		\hline \hline 
		%		\multicolumn{2}{l}{$\psi_{S_0^+} = \ip \psi_{S_0}$ } \\
		%		\multicolumn{2}{l}{$\psi_{S_1^-} = \im \psi_{S_1}$ } \\
		%		\multicolumn{2}{l}{$\psi_{L^+} = \ip \psi_L^+$ } \\
		%		\multicolumn{2}{l}{$\psi_{L^-} = \im \psi_L^-$} \\ 
	\end{tabular}
\end{table}
Note that all bounds in \eqref{eq_bounds_ATz} and \eqref{eq_bounds_ATplmin1} are functions of various ratios of different conditional expectations $E[\psi_B(W_i,\eta_0)|Z=z]$. Thus, they can be obtained by separate regressions of the respective $\psi_B(W_i,\eta_0)$ onto the spaces spanned by their $k_B$-dimensional transformations of $Z_i$, $b_B(Z_i)$
\begin{align}
	\psi_B(W_i,\eta_0) &= b_B(Z_i)'\beta_{B,0} + r_{B}(Z_i) + \varepsilon_{i,B}, \label{eq_auxReg1}
\end{align}
with conditional mean errors $E[\varepsilon_{i,B}|Z_i] = 0$ and approximation errors \begin{align}
	r_{B}(Z_i) = E[\psi_B(W_i,\eta_0)|Z_i] - b_B(Z_i)'\beta_{B,0}.
\end{align} Based on this, we construct population estimands for the bounds $B=L,U$ using the linear predictors at each $Z=z$ as
\begin{align}
	{\theta}_B^{LP}(z) = \frac{b_{B^+}(z)'{\beta}_{B^+,0} + b_{B^-}(z)'{\beta}_{B^-,0}}{b_{S_0^+}(z)'{\beta}_{S_0^+,0} + b_{S_1^-}(z)'{\beta}_{S_1^-,0}} \label{eq_BLP}
\end{align}
and analogously for $\theta_{B}^{+,LP}$ and $\theta_{B}^{-,LP}$.
Replacing the unobserved true scores in \eqref{eq_auxReg1} by their sample counterparts yields estimators \begin{align}
	\hat{\beta}_B &= \bigg(\sum_{i=1}^nb_B(Z_i)b_B(Z_i)'\bigg)^{-1}\sum_{i=1}^nb_B(Z_i)\psi_B(W_i,\hat{\eta}), \label{eq_bhat1}
	%	&= \hat{Q}^{-1}E_n[b_i\psi_i(\hat{\eta},\hat{\pi})]
\end{align}
where the scores of the effect bounds with estimated nuisance quantities $\psi_B(W_i,\hat{\eta})$ serve as pseudo-outcomes in separate least squares regression on their respective basis functions.\footnote{In principle, the equations based on \eqref{eq_auxReg1} are only seemingly unrelated and could thus also be estimated using a system based approach that takes into account the correlation structure of the conditional mean errors to increase efficiency. While this is straightforward in the case of a finite-dimensional parametric mean functions, it introduces additional dependencies in the two-step estimation that might offset potential gains in efficiency in the nonparametric case. Thus, we leave an extension along this line for future work.} The estimated aggregate heterogeneous effect bounds can then be calculated by combining the point predictions of four models
\begin{align}
	\hat{\theta}_B(z) = \frac{b_{B^+}(z)'\hat{\beta}_{B^+} + b_{B^-}(z)'\hat{\beta}_{B^-} }{b_{S_0^+}(z)'\hat{\beta}_{S_0^+} + b_{S_1^-}(z)'\hat{\beta}_{S_1^-}}   \label{eq_theta_hat1}
\end{align}
and analogously for $\hat{\theta}_{B}^+(z)$ and $\hat{\theta}_{B}^-(z)$. In principle the two components in both numerator and denominator in \eqref{eq_theta_hat1} can also be obtained joint regressions that use $(\psi_{B^+}(W_i,\hat{\eta}) + \psi_{B^-}(W_i,\hat{\eta}))$ and $(\psi_{S_0^+}(W_i,\hat{\eta}) + \psi_{S_0^-}(W_i,\hat{\eta}))$ respectively as outcomes in \eqref{eq_auxReg1}. This requires overall fewer parameters and only two choices of basis functions which could be beneficial in finite samples but is overall less adaptive to different complexity and trade-offs in estimating the relevant conditional expectations. 

For the remainder, let $\theta(z) \in \{\theta_{AT}(z),\theta_{AT}^+(z),\theta_{AT}^-(z)\}$ denote a generic parameter of interest. The total sum of basis functions over all required regressions $k^*$ here will depend on this target parameter, see Table \ref{tab_k_all} for an overview using a separate basis for each component.%\footnote{Note that, for any of the final estimators, the nonparametric regressions are combined. Thus, the problem is less complex than running a single nonparametric regression with $k^*$ basis functions but rather behaves like a single nonparametric regression with $\sup_{B} k_B$ basis functions.}
\begin{table} \caption{Number of Basis Functions} \label{tab_k_all} \footnotesize
	\centering \begin{tabular}{ll} \hline \hline \\[-1.5ex]
		Parameter $\theta(z)$ & Total Number of Basis Functions $k^*$ \\ \hline \\[-1.5ex] %\\[-0.5ex]
		\multicolumn{2}{l}{Strong/Unconditional Monotonicity} \\[0.5ex]
		$\theta_{AT}(z) = \theta_{AT}^+(z)$ & $[k_U + k_L + 2k_{S_0}]$ \\[1ex]
		\multicolumn{2}{l}{Weak/Conditional Monotonicity} \\[0.5ex]
		$\theta_{AT}(z)$ & $[k_{U^+} + k_{L^+} + k_{U^-} + k_{L^-} + 2(k_{S_0^+} + k_{S_1^+})]$ \\
		$\theta_{AT}^+(z)$ & $[k_{U^+} + k_{L^+} + 2k_{S_0^+}]$ \\
		$\theta_{AT}^-(z)$ & $[k_{U^-} + k_{L^-} + 2k_{S_1^-}]$ \\ \hline \hline 
	\end{tabular} %\begin{justify}
	%$k^*$ is the total number of basis functions. 
	%\end{justify}
\end{table}
When all $b_B(Z_i)$ for the relevant $B$ consist only of a constant, the estimator is similar to one for the unconditional generalized Lee bounds in \cite{semenova2023generalized} with an augmented inverse probability weighting (AIPW) denominator.\footnote{This is possible due to the stronger margin assumption 3.3 compared to \cite{semenova2023generalized}. Under this strong margin condition, $\rho_N$ in \cite{semenova2023generalized}, Equation (5.7) can be set to $0$.}

Estimation of nuisance parameters $\hat{\eta}$ can be done via modern machine learning such as random forests, deep neural networks, high-dimensional sparse likelihood and regression models, or other non- and semiparametric estimation methods with good approximation qualities for the nuisance functions at hand. The influence of their learning bias/approximation on the functional bound estimator is limited by the Neyman-orthogonality of the chosen moment functions \citep{chernozhukov2018double,semenova2023generalized}. %, i.e.~we have that
%	\begin{align*}
	%		\partial_{r}E[\psi_B(W_i,\eta_0 + \tau(\eta-\eta_0))|X_i=x] =_{|\tau=0} o(n^{-1/2})
	%	\end{align*}
%	for all $x, \eta$ and $B \in \{L,U\}$. 
In particular, under typical basis choices and suitable regularity conditions, the estimation does not affect the large sample distribution if the $L_2$-approximation rates for all nuisance quantities $\hat{\eta} - \eta_0$ are of order $o((nk_B)^{-1/4})$ and the selection probabilities are $L_1$-consistent at rate $o({k_B}^{-1/2})$. The first condition is identical to standard DML estimation of conditional average treatment effects using Neyman-orthogonal moment functions for nonparametric projection \citep{semenova2021debiased}. When $z$ is one-dimensional, popular $k_B$ choices for many bases under weak smoothness assumptions are of rate $O(n^{1/5})$ leading to an overall RMSE convergence requirement for the nuisances of $o(n^{-3/10})$. This is a rate achievable by many nonparametric and machine learning estimators such as forests, deep neural networks, or high-dimensional sparse models under moderate complexity and/or dimensionality restrictions, see \cite{semenova2023generalized}, \cite{semenova2021debiased}, and \cite{heiler2021effect} for examples. The additional weak $L_1$ condition is required to control the variance of the trimming indicators. More details regarding the technical assumptions can be found in Section \ref{sec_LargeSample1}.

 We also require that all components in $\hat{\eta}$ are obtained via $K$-fold cross-fitting, see Definition 3.1 in \cite{chernozhukov2018double}.
%\begin{mydef}\textbf{K-fold cross-fitting} (see Definition 3.1 in \cite{chernozhukov2018double})
%	Take a K-fold random partition $(I_f)_{f=1}^K$ of observation indices $[K] = \{1,\dots,n\}$ with each fold size $n_f = n/K$. For each $f \in [K] = \{1,\dots,K\}$, define $I_f^c := \{1,\dots,n\}\backslash I_f$. Then for each $f\in [K]$, the machine learning estimator of the nuisance function are given by \begin{align*}
%		\hat{\eta}_{f} = \hat{\eta}(W_{i\in I_f^c}).
%	\end{align*}
%	Thus for any observation $i \in I_f$ the estimated score only uses the model for $\eta$ learned from the complementary folds $\psi_B(W_i,\hat{\eta}) =  \psi_B(W_i,\hat{\eta}_{f})$.
%\end{mydef}
The use of cross-fitting controls potential bias arising from over-fitting using flexible machine learning methods without the need to evaluate complexity/entropy conditions for the function class that contains true and estimated nuisance quantities with high probability. If finite dimensional parametric models such as linear or logistic are assumed and estimated for the nuisance quantities, the proposed methodology can be applied without the need for cross-fitting. %\paragraph{Remark:} These estimated bounds with cross-fitting are not sharp in the sense that they provide the smallest bound conditional on the given sample due to the data splitting, i.e.~they are constructed from averages over different samples. They are, however, sharp in these sense that they converge to sharp population bounds in probability.    
%	Under suitable assumptions, the predictions using the estimator in \eqref{eq_bhat1} are consistent for $g(z)$. Moreover, it is possible to conduct asymptotically valid inference around the  linear predictor, i.e.~for any $z_0 = z_{0,n}$ we can construct $(1-\alpha)\%$ confidence intervals for the true decomposition function as
%	\begin{align}
	%		CI_{1-\alpha}(g(z_0)) = \bigg[b(z_0)'\hat{\beta} \pm q_{1-\alpha/2}\sqrt{\frac{b(z_0)'\hat{\Omega}b(z_0)}{{n}}} \bigg]\label{eq_CI1}
	%	\end{align}
%	where $q_{1-\alpha/2}$ denotes the $(1-\alpha/2)$-quantile of the standard normal distribution and $\hat{\Omega}$ is a consistent sample estimator of the asymptotic variance $\Omega_0$[WHITE]. The estimator explicitly takes into account the additional uncertainty from estimating the unconditional version probabilities in the decomposition terms. 

Under suitable assumptions, the heterogeneous bound estimators are jointly asymptotically normal at each $z \in \mathcal{Z}$	\begin{align}
	\sqrt{n}&\hat{\Omega}_n^{-\frac{1}{2}}(z)\begin{pmatrix}
		\hat{\theta}_L(z) -	\theta_L(z) \\
		\hat{\theta}_U(z) -	\theta_U(z)
	\end{pmatrix} \overset{d}{\rightarrow} \mathcal{N}\left(\begin{pmatrix}
		0\\0
	\end{pmatrix}, \begin{pmatrix}
		1 &0 \\
		0 &1
	\end{pmatrix}\right), \notag \\ &\hat{\Omega}_n(z) =  \begin{pmatrix}
		\hat\sigma_L^2(z) &\hat\rho(z)\hat\sigma_L(z)\hat\sigma_U(z) \\
		\hat\rho(z)\hat\sigma_L(z)\hat\sigma_U(z) &\hat\sigma_U^2(z)
	\end{pmatrix}. \label{eq_asyN_intro}
\end{align}
For the complete definitions of the variance terms consider Appendix \ref{sec_variances1}. The variance can depend on the sample size and is generally increasing in norm if we allow the basis functions $b_B(z)$ to grow with the sample size. Thus, convergence is slower than the parametric rate equivalently to conventional nonparametric series regression \citep{belloni2015some}. If the approximation errors $r_{B}(z)$ are rather large, then this distributional result is still valid if centered around the linear predictors of the true bounds \eqref{eq_BLP} similar to standard regression estimation under misspecification.
\subsection{Inference}
Inverting the quantiles of the distribution in \eqref{eq_asyN_intro} could in principle be used to construct confidence intervals for the upper and lower bounds. However, this would be overly conservative for the actual effect of interest $\theta(z)$. Inference on this partially identified parameter should be adaptive to the underlying true width of the interval. %&, see \cite{stoye2009more} and \cite{andrews2010inference} for a thorough treatment and adaptive confidence intervals. 
In addition, if we allow for (local) misspecification, corresponding confidence regions could be empty or very narrow suggesting overly precise inference \citep{andrews2019inference}. Robustness to misspecification is important in our setup as, in contrast to unconditional bounds, the heterogeneous bound functions are estimated at potentially many points with different variances and varying strength of identification in the sense of different widths of the underlying true identified set. A flexible estimator that is chosen e.g.~by a global goodness-of-fit criterion for the effect bound curves could well be locally misspecified at some points. A limiting case of this type of misspecification would be to ``overfit'' the effect of a treatment that is fully independent of selection for some units instead of imposing local point identification. Corresponding confidence regions should be adaptive to such cases. 
To do so, we introduce the notion of a pseudo-true parameter $\theta^*(z)$ and its corresponding standard deviation $\sigma^*(z)$ that can be estimated as \begin{align}
	\hat{\theta}^*(z) = \frac{\hat\sigma_U(z)\hat{\theta}_L(z)+\hat\sigma_L(z)\hat{\theta}_U(z)}{\hat\sigma_L(z) + \hat\sigma_U(z)}, \qquad 
	\hat{\sigma}^*(z) = \frac{\hat\sigma_L(z)\hat\sigma_U(z)\sqrt{2(1+\hat{\rho}(z))}}{\hat\sigma_L(z) + \hat\sigma_U(z)}.
\end{align}
This is a variance weighted version of the upper and lower bound. 
The pointwise $(1-\alpha)\%$-confidence intervals for the true always-taker effect $\theta(z)$ can then be obtained by the union of two intervals \begin{align} \label{eq_CIhat1}
	CI_{\theta(z),1-\alpha} &= \big[\hat{\theta}_L(z) - \frac{\hat\sigma_L(z)}{\sqrt{n}}\hat{c}(z), \hat{\theta}_U(z) + \frac{\hat\sigma_U(z)}{\sqrt{n}}\hat{c}(z)\big] \cup \big[\hat{\theta}^*(z) \pm \frac{\hat{\sigma}^*(z)}{\sqrt{n}}\Phi^{-1}(1-\alpha/2)\big],
\end{align}
where the critical value $\hat{c}(z)$ uniquely solves \begin{align} \label{eq_CI_concentrate1}
	\inf_{\Delta \geq 0} P(u_1 - \Delta - c \leq 0 \leq u_2\ or\ |u_1 + u_2 - \Delta| \leq \sqrt{2(1+\hat{\rho}(z))}\Phi^{-1}(1-\alpha/2)) = 1 - \alpha,
\end{align}
with $(u_1,u_2)$ being jointly normal with unit variances and covariance $\hat{\rho}(z)$. 
This is an adaptation of the method by \cite{stoye2020simple} who considers simple parametric estimators for generic bounds without two-stage estimation, cross-fitting, or additional nuisance functions.	
Interval \eqref{eq_CIhat1} is robust against misspecification that yields reverse ordering of the bounds as it is never empty and guarantees at least nominal coverage over an extended parameters space. In particular, it has at least $(1-\alpha)\%$ asymptotic coverage uniformly for all widths $\theta_U(z) - \theta_L(z)$ pointwise at each $z\in\mathcal{Z}$. For uniform confidence bands consider Section \ref{sec_uniformbands}. 

In principle, population moments \eqref{eq_boundsX_DEFINITION} should not cross or be arbitrarily close to point identification under Assumptions 3.1 to 3.4. The same applies to the nonparametric projections at any $z$ under correct specification. However, when estimating the unknown conditional probabilities and quantile functions as well as the final heterogeneity projections, (local) misspecification is increasingly likely. Thus, this problem could also be analyzed from a model selection perspective. Our inference approach is agnostic whether narrow or empty identified sets are a result of a violation of the identification assumptions or of such model selection based misspecification. However, the pseudo-true parameter should be interpreted from that perspective. Note that relaxing the bounds pointwise could potentially be crude. Optimally, an inference procedure should incorporate common information across all bounds. Our misspecification robust intervals do this indirectly: When the density of $z$ and conditional moments are smooth, bounds and confidence intervals will be smooth as well. Thus, there is an implied smoothness on the relaxation parameter of the underlying conditional moment inequalities in the sense of \cite{andrews2019inference}. %, see Appendix D. However, how to do the joint relaxation or smoothing more systematically/optimally with respect to inference is an open question. 
For more details regarding the misspecification framework, consider Appendix \ref{sec_misspecFnE}.

\section{Large Sample Theory} \label{sec_LargeSample1}
\subsection{Assumptions and Pointwise Limiting Distribution}
In this section we provide the assumptions for asymptotic normality, validity of the confidence intervals \eqref{eq_CIhat1}, and some more technical discussion. We again present the case with known propensity scores.\footnote{With propensity scores estimated, one has to augment the bias-correction in the moment functions and the nuisance parameter space as in \cite{semenova2023generalized} as well as Assumption A.6 by additional terms that equivalently depend on the (squared) $L_p$ error rate of the propensity score estimator.} Denote $||\cdot||_p$ as the $L_p$ norm. Let $E[\psi_B(W_i,\eta_0)|Z_i=z] \in \mathcal{G}_B$ where $\mathcal{G}_B$ is a space of functions (possibly depending on $n$) that map from $\mathcal{Z}$ to the real line. Note that $E[\psi_B(W_i,\eta_0)|Z_i=z] = b_B(z)'\beta_{B,0} + r_{B}(z)$ with basis transformations $b_B(z) \in  \mathcal{S}^{k_B} := \{b \in \mathbb{R}^{k_B}: ||b|| = 1\}$ and $\beta_{B,0}$ being the parameter of the linear predictor defined as root of equation $E[b_B(Z_i)(\psi_B(W_i,\eta_0) - b_B(Z_i)'\beta_{B,0})] = 0$. Define basis bound $\xi_{k,B} = \sup_{z\in\mathcal{Z}}||b_{B}(z)||$. Let $\eta \in T$ where $T$ is a convex subset of some normed vector space. Denote the realization set $\mathcal{T}_n = (\mathcal{S}_{0,n}\times\mathcal{S}_{1,n}\times \mathcal{Q}_n) \subset T$ as the set that with high probability contains estimators $\hat{\eta} = \{\hat{s}(0,X_i), \hat{s}(1,X_i),\hat{q}_0(u,X_i),\hat{q}_1(u,X_i)\}$ for nuisance quantities  $\eta_0 = \{s(0,X_i),s(1,X_i),q_0(u,X_i),q_1(u,X_i)\}$. Let their corresponding $L_p$ error rates be
\begin{align*} 
	\lambda_{s,n,p} &= \sup_{d\in\{0,1\}}\sup_{\hat{s}(d) \in \mathcal{S}_{d,n}} E[|\hat{s}(d,X_i) - s(d,X_i)|^p]^{1/p}, \\
	\lambda_{q,n,p} &= \sup_{u \in \tilde{U}}\sup_{\hat{q}(u) \in \mathcal{Q}_{n}} E[|\hat{q}(u,X_i) - q(u,X_i)|^p]^{1/p}, 
\end{align*}
where $\tilde{U}$ is a compact subset of $(0,1)$ containing the relevant quantile trimming threshold support unions $([supp(p_0(X_i)) \cup supp(1-p_0(X_i))] \cap \mathcal{X}_{+}) \cup ([supp(1/p_0(X_i)) \cup supp(1-1/p_0(X_i))] \cap \mathcal{X}_{-})$. All of the following assumptions are uniformly over $n$ if not stated differently for the required elements $B \in \{L^+,U^+,L^-,U^-,S_0,S_1,S_0^+,S_1^-\}$: 
\textit{\begin{enumerate}[itemsep=0pt]
		\item[A.1)] (Identification) $Q_B = E[b_B(Z_i)b_B(Z_i)']$ has eigenvalues bounded above and away from zero. % uniformly over $n$. 
		\item[A.2)] (Regular outcome) The outcome has bounded conditional moments $E[Y_i^m|X_i=x,D_i=d,S_i=1]$ for some $m > 2$ and a continuous density  $f(y|X=x,D=d,S_i=1)$ that is bounded from above and away from zero with bounded first derivative for any $x\in \mathcal{X}$ and $d\in \{0,1\}$.
		\item[A.3)] (Strong multiple overlap) There exist constants $\underline{e},\underline{s} \in (0,1/2)$ such that \begin{align*}
			\underline{e} < \inf_{x\in\mathcal{X}} e(x) \leq \sup_{x\in\mathcal{X}} e(x) < 1- \underline{e}, 
		\end{align*}\begin{align*}
			\underline{s} < \inf_{x\in\mathcal{X},d\in{\{0,1\}}} s(d,x) \leq \sup_{x\in\mathcal{X},d\in{\{0,1\}}} s(d,x) < 1- \underline{s}. 
		\end{align*}
		\item[A.4)] (Approximation) For any $n$ and $k_B$, there are finite constants $c_{k,B}$ and $l_{k,B}$ such that for each $E[\psi_B(W_i,\eta_0)|Z_i=z] \in \mathcal{G}_B$ \begin{align*}
			||r_{B}||_{P,2} &:= \sqrt{\int_{z\in\mathcal{Z}}r_{B}^2(z)dP(z)} \leq c_{k,B},  \\
			||r_{B}||_{P,\infty} &:= \sup_{z\in\mathcal{Z}}|r_{B}(z)|  \leq l_{k,B}c_{k,B}.
		\end{align*}
		\item[A.5)] (Basis growth) Let $\sqrt{n}/\xi_{k,B} - l_{k,B}c_{k,B} \rightarrow \infty$ such that \begin{align*}
			\sqrt{\frac{\xi_{k,B}^2\log k_B}{n}}\bigg(1 + \sqrt{k_B}l_{k,B}c_{k,B} \bigg)&= o(1). 
		\end{align*}
		\item[A.6)]  (Machine learning bias) Let $e_n = o(1)$. For all folds, the nuisance parameters obtained via cross-fitting belong to a shrinking neighborhood $\mathcal{T}_n$ around $\eta_0$ with probability of at least $1-e_n$, such that \begin{align*}
			\xi_{k,B}(\lambda_{q,n,1} + \lambda_{s,n,1} +\lambda_{q,n,2} + \lambda_{s,n,2}) &= o(1)
		\end{align*}
		and (i) either
		\begin{align*}
			\sqrt{nk_B}(\lambda_{q,n,4}^2 + \lambda_{s,n,4}^2) &= o(1)
		\end{align*}
		or (ii) the basis is bounded  $\sup_{z\in\mathcal{Z}}||b(z)||_{\infty} < C$ and 	 \begin{align*}
			\sqrt{nk_B}(\lambda_{q,n,2}^2 + \lambda_{s,n,2}^2) &= o(1).
		\end{align*}
		 For $B\neq S_0,S_1$ under conditional monotonicity, we also require that \begin{align*}
			\sqrt{n\xi_{k_B}^2}\lambda_{s,n,2}^2 &= o(1).
		\end{align*}
		%		\item[A.7)] (Lipschitz quantile) On the realization set with probability of at least $1-e_n$, the conditional quantile estimator is Lipschitz continuous over $\tilde{U}$, i.e.~there exists a $C >0$ such that \begin{align*}
			%			\sup_{u,u'\in \tilde{U} }\sup_{\hat{q}(u),\hat{q}(u') \in \mathcal{Q}_n}|\hat{q}(u,x) - \hat{q}(u',x)| < C|u - u'|.
			%		\end{align*} 
		%		almost surely in $\mathcal{X}$. 
\end{enumerate} }

Assumption A.1 excludes collinearity of the basis transformations of the heterogeneity variables. A.2 puts restrictions on the tails and the smoothness of the distribution for the observed outcome distribution in different selection and treatment states. A.3 assures that there are comparable units between units of different selection and/or treatment status. A.2 and A.3 together imply a continuously differentiable conditional quantile function for the selected observed units that is almost surely bounded. This, together with the strong overlap for the treatment and selection probabilities, assures that the effect bounds are regularly identified \citep{khan2010irregular,HEILER2021valid}.

A.4 defines $L_2$ and uniform approximation error bounds for function class $\mathcal{G}_B$. This is a typical characterization in the literature on nonparametric series regression without nuisance functions \citep{belloni2015some}. We say the model is correctly specified if the basis is sufficiently rich to span $\mathcal{G}_B$, i.e.~$c_{k,B} \rightarrow 0$ as $k_B \rightarrow \infty$. However, the distributional theory also allows for the case of misspecification, i.e.~$c_{k,B} \not\rightarrow 0$. 
A.5 controls the approximation error from linearization of the estimator with unknown design matrix $Q_B$. This is equivalent to the condition required for localization in general least squares series regression \citep{belloni2015some}.\footnote{For more specific series methods such as splines \citep{huang2003local} or local partitioning estimators \citep{cattaneo2020large}, this rate can be improved to $\sqrt{\xi_{k,B}^2 \log k_B/n}(1+ \sqrt{\log k_B}l_{k,B}c_{k,B})$, see also \cite{belloni2015some}, Section 4 and \cite{cattaneo2020large}, Supplemental Appendix Remark SA-4 for a related discussion.}

A.6 is crucial: It says that the machine learning estimators for the conditional quantiles and selection probabilities have sufficiently good approximation qualities in an $L_p$ sense. In the case of a bounded finite basis, the conditions reduce to $L_1$ and $L_2$ consistency as well as the well-known requirement in the semiparametric/DML literature that the nuisance functions have root have mean squared error rates of order $o(n^{-1/4})$ \citep{chernozhukov2018double}. In the more general case, the higher $L_p$ rates are identical to the ones in \cite{semenova2021debiased} required for nonparametric estimation of conditional average treatment effects. The additional condition for conditional monotonicity is due to potential classification error. In the typical case of well-behaved basis functions such splines, wavelets, and local partitioning, we have that $\xi_{k,B} \lesssim \sqrt{k_B}$ \citep{belloni2015some} and thus this requirement is subsumed by the prior $\sqrt{nk_B}\lambda_{s,n,p}^2$ condition. When $B \in \{S_0,S_1\}$, the $L_1$ requirement is omitted as the AIPW moment functions do not contain any trimming indicator.

For the estimation of the asymptotic variance, we also assume that A.V holds:   
\textit{\begin{enumerate}[itemsep=0pt]
		\item[A.V)] (Asymptotic variance) The conditions in Appendix \ref{app_matrixEst2}, Assumption \ref{ass_AV} hold, i.e.\\ $||\hat{\Omega}_n(z) - \Omega(z)|| = o_p(1)$ pointwise at each $z \in \mathcal{Z}$.  
\end{enumerate}} \noindent
A.V can require somewhat stronger outcome moment/tail and basis growth conditions. The corresponding primitive assumptions and discussion can be found in Appendix \ref{app_matrixEst2} and are omitted for brevity.
We obtain the following Theorem:

%\subsection{Pointiwse Limiting Distributions}

\begin{thm}[Asymptotic Normality] \label{thm_AsyNor}
	Suppose Assumptions 3.1 - 3.4 and A.1 - A.6 hold and $\hat{\theta}_B(z_0)$  for $B = L, U$ and $\hat{\Omega}_n(z_0)$ are estimators according to \eqref{eq_theta_hat1} and \eqref{eq_variance_estimated2} respectively. Let $\theta_B^{LP}(z)$ be the the population predictor according to \eqref{eq_BLP}.
	Then, for any sequence $z_0 = z_{0,n}$,  \begin{align*}
		\sqrt{n}&\hat{\Omega}_n^{-\frac{1}{2}}(z_0)\begin{pmatrix}
			\hat{\theta}_L(z_0) -	\theta_L^{LP}(z_0)\ \\
			\hat{\theta}_U(z_0) -	\theta_U^{LP}(z_0)
		\end{pmatrix} \overset{d}{\rightarrow} \mathcal{N}\left(\begin{pmatrix}
			0\\0
		\end{pmatrix}, \begin{pmatrix}
			1 &0 \\
			0 &1
		\end{pmatrix}\right).
	\end{align*} Moreover if $\sup_{B}n^{1/2}k_B^{-1/2}l_{k,B}c_{k,B} = o(1)$, then \begin{align*}
		\sqrt{n}&\hat{\Omega}_n^{-\frac{1}{2}}(z_0)\begin{pmatrix}
			\hat{\theta}_L(z_0) -	\theta_L(z_0) \\
			\hat{\theta}_U(z_0) -	\theta_U(z_0)
		\end{pmatrix} \overset{d}{\rightarrow} \mathcal{N}\left(\begin{pmatrix}
			0\\0
		\end{pmatrix}, \begin{pmatrix}
			1 &0 \\
			0 &1
		\end{pmatrix}\right).
	\end{align*}
\end{thm}

Theorem \ref{thm_AsyNor} shows that nonparametric heterogeneous bounds using DML are jointly asymptotically normal. It allows for the case of misspecification when centered around the linear predictor. It is most useful under the additional undersmoothing condition that makes any misspecification bias vanish sufficiently fast.\footnote{For example, when $\mathcal{G}_B$ is in a $s$-dimensional ball on $\mathcal{Z}$ of finite diameter, then the condition simplifies to $n^{1/2}k_B^{-(\frac{1}{2} + \frac{s}{d})}\log(k_B) \rightarrow 0$. See \cite{belloni2015some}, Comment 4.3 for additional details. Note that undersmoothing does in general not admit mean-squared error optimal $k_B$ choices, see e.g.~\cite{cattaneo2020large} for multiple bias-correction alternatives methods for local partitioning estimators.}

\subsection{Pseudo Parameterization and Pointwise Inference}
Theorem \ref{thm_AsyNor} in principle is sufficient to construct confidence intervals for the heterogeneous effect bounds. They are, however, too wide for the actual effect parameter of interest $\theta(z)$ depending on the width of $\theta_U(z) - \theta_L(z)$ \citep{imbens2004confidence}. If the difference between upper and lower bound is large, the testing problem is essentially one-sided compared to the case of only having a small difference. This raises the question of how to conduct inference that is uniformly valid with respect to the underlying difference between upper and lower effect bound and has more power compared to using simple two-sided critical values. 
Moreover, there are multiple sources of misspecification that could invalidate or suggest spuriously precise inference when upper and lower bound estimates are close to each other or reverted. In particular, even when the population bounds do not intersect, their linear predictors still might.  
Handling such misspecification is particularly relevant for heterogeneous bounds as here we estimate two functions at potentially many points which makes (local) misspecification and/or reordering a much larger concern compared to a single partially identified parameter as in \cite{lee2009training} or \cite{semenova2023generalized}.\footnote{Note also that we allow for the use of different basis functions across when estimating upper and lower effect bounds. This seems reasonable as it could well be that lower and upper effect bounds are curves of e.g.~different smoothness and not equally difficult to approximate. Alternatively one could impose the largest complexity of any bound for both models, i.e.~paying the price of a potentially higher estimation variance in the effect bound with a higher degree of smoothness. While this will generally yield a better approximation for each separate bound, it could also contribute to a reverse ordering in the sense that $\hat{\theta}_L(z) \geq \hat{\theta}_U(z)$ at some $z$. % which requires an adaptive inference solution similar to partial identification problems resulting from misspecification \citep{andrews2019inference,stoye2020simple}.
} We relax the notion of coverage following \cite{andrews2019inference} to include an artificial pseudo-true target parameter that guarantees non-empty confidence intervals. For more details on misspecification and corresponding formalization consider Appendix \ref{sec_misspecFnE}. 
%This is  in a general moment inequality framework.

Our method is based on \cite{stoye2020simple} who demonstrates that, in the simple case of two regular, jointly asymptotically normally distributed parameters, uniformly valid intervals with good power properties can be obtained by concentrating out the unobserved true difference in effect bounds. We adapt his framework to the nonparametric heterogeneous effect bounds with estimated nuisances. For any $z \in \mathcal{Z}$ let the identified set be 
%\begin{align}
$\Theta_{z} = [\theta_L(z),\theta_U(z)]$
%\end{align}
%which contains the true causal effect $\theta(z)$.
Now denote the pseudo-true parameter $\theta^*(z)$ and its variance as  \begin{align}
	{\theta}^*(z) = \frac{\sigma_U(z){\theta}_L(z)+\sigma_L(z){\theta}_U(z)}{\sigma_L(z) + \sigma_U(z)}, \qquad 
	{\sigma}^*(z) = \frac{\sigma_L(z)\sigma_U(z)\sqrt{2(1+{\rho}(z))}}{\sigma_L(z) + \sigma_U(z)}. \label{eq_defPseudoParVar1}
\end{align}
The pseudo-true identified set is then given by %\begin{align}
$\Theta_{z}^* = \Theta_z \cup \{\theta^*(z)\}$.
%\end{align} 
This set is implicitly defined by an estimand corresponding to setting a test statistic for the heterogeneous effect at $z$ to (i) zero (meaning no rejection of the null) if the interval is nonempty and the null value is inside the estimated intervals and to (ii) the larger of the two $t$-statistics for upper and lower bound for the null of the bound being equal to the hypothesized value. 
In particular, it chooses the test statistic under the null as $\max\{(\theta(z)-\hat{\theta}_U(z))/\hat{\sigma}_U(z),(\hat{\theta}_L(z)-\theta(z))/\hat{\sigma}_L(z),0\}$. Case (ii) collapses to a standard test for the parameter lying in $\Theta_z$ in the case of a well-defined interval with $\theta_U(z) > \theta_L(z)$ and to a test on the pseudo-true parameter under misspecification $\theta_U(z) < \theta_L(z)$. This inference procedure can also be interpreted as resulting from a moment inequality problem that allows for misspecification by adding slacks/deviations to equations defining upper and lower effect bounds, see Appendix \ref{sec_misspecFnE}. If such slacks are large, $\theta_U(z) < \theta_L(z)$ can be admitted as solution. 
In principle, alternative definitions for the pseudo-true set $\Theta^*_z$ that use different weighting compared to \eqref{eq_defPseudoParVar1} could be considered as well. This would change the definition of the pseudo-true parameter. Not all choices of such pseudo-true parameter are of equal use. This is similar to e.g.~GMM models under misspecification where the pseudo-true parameter is defined implicitly as the maximizer of a GMM population criterion using a particular weighting matrix. Therefore, the choice of the pseudo-true parameter and its usefulness should be balanced versus the robustness against spurious precision under misspecification \citep{andrews2019inference}. The particular choice in \eqref{eq_defPseudoParVar1} leads to a convenient solution in terms of critical value adjustment due to the specific asymptotic bivariate normal distribution \citep{stoye2020simple}. We obtain the following Theorem:
\begin{thm}[Misspecification Robust Inference] \label{thm_Coverage1}
	Suppose Assumptions 3.1 - 3.4 and A.1 - A.6 hold and $\hat{\theta}_B(z_0)$  for $B = L, U$ and $\hat{\Omega}_n(z_0)$ are estimated according to \eqref{eq_theta_hat1} and \eqref{eq_variance_estimated2} respectively. Let the confidence interval be constructed according to \eqref{eq_CIhat1}.  Then, for any sequence $z_0 = z_{0,n}$,
	\begin{align*}
		\liminf_{n\rightarrow\infty}\inf_{{\theta}^{LP}(z_0)\in{\Theta}^*_{z_0}}\ P({\theta}^{LP}(z_0) \in CI_{{\theta(z_0)},1-\alpha}) \geq 1 - \alpha, 
	\end{align*}
	where ${\Theta}^*_{z_0} = [{\theta}_L^{LP}(z_0),{\theta}_U^{LP}(z_0)] \cup {\theta}^{LP,*}(z_0)$ where ${\theta}^{LP,*}(z)$ is defined analogously to \eqref{eq_defPseudoParVar1} with linear predictors instead of bounds. Furthermore if $\sup_{B}n^{1/2}k_B^{-1/2}l_{k,B}c_{k,B} = o(1)$, then  \begin{align*}
		\liminf_{n\rightarrow\infty}\inf_{\theta(z_0)\in{\Theta}^*_{z_0}}\ P(\theta(z_0) \in CI_{\theta(z_0),1-\alpha}) \geq 1 - \alpha, 
	\end{align*}
	where ${\Theta}^*_{z_0} = [\theta_L(z_0),\theta_U(z_0)] \cup {\theta^*}(z_0)$ as defined in \eqref{eq_defPseudoParVar1}.
\end{thm}	

Theorem \ref{thm_Coverage1} demonstrates the asymptotic validity of the confidence intervals proposed in \eqref{eq_CIhat1} for both the linear predictor as well as the correctly specified case. In particular, we achieve at least nominal coverage independently of the actual width of the true identified region for the heterogeneous effect bound. Coverage is uniform with respect to the width of the true identified region $\theta_U(z) - \theta_L(z)$ pointwise at each $z \in \mathcal{Z}$ or along sequences therein. The coverage notion over the augmented parameter set $\Theta_{z}^*$ assures non-emptiness of the interval and avoids spuriously precise inference in regions where the estimated lower bound might be too large relative to the estimated upper bound. Coverage will be closer to the nominal level when correlations between conditional mean errors are small. In particular, near \textit{one-sided} critical values apply for the usual levels of confidence when $\rho(z) = 0$. % (up to minor simulation inaccuracies of solving for $\hat{c}(z)$ in \eqref{eq_CI_concentrate1}). 
Note that $\rho(z)$ stems from the correlation between two linear combinations of residuals from up to four separate nonparametric regressions. Hence, $\rho(z)$ can generally vary over different values of $z$. Thus, power and size properties of the corresponding tests will depend on the location of the local effect bound. In particular, they are driven by the share of missing outcomes at given $z$. In the case of point identification (no missing outcomes), upper and lower bounds and pseudo true parameter are equivalent and $\rho(z) = 1$. Thus, the confidence interval collapses to one using standard \textit{two-sided} critical values.\footnote{Alternatively, one could employ an additional data splitting step for estimating the two bounds on different subsamples to assure that $\rho(z)$ is equal to zero. We refrain from this approach to avoid inaccuracies in finite samples as the first estimation step already requires cross-fitting and potential data splitting for tuning of the machine learning methods within folds.} 

\subsection{Uniform Inference}\label{sec_uniformbands}
We now consider inference for the whole process $\{\theta(z)\}_{z\in \mathcal{Z}}$ without misspecification. Under stronger approximation requirements, the series approach allows for a uniform linearization that can be used to obtain a process analogue of Theorem \ref{thm_AsyNor}. We exploit this to construct a simple multiplier bootstrap procedure that guarantees uniform coverage and avoids retraining the first-stage estimators. Inference is then based on a joint process results for the bounds. This does not exploit locally varying widths of the identified sets and convergence in distribution. %In particular, the critical values that are obtained by taking the supremum over $z$ along the whole process also use the corresponding worst-case $\Omega(z)$. 
As a result, inference will be generally be more conservative. 
The bootstrap works as follows: For each $b=1,2,\dots$, generate $n$ independent standard exponential random variables $h_1,\dots,h_n$. Then, for the relevant $B \in \{L^+,U^+,L^-,U^-,S_0,S_1,S_0^+,S_1^-\}$, use the same $h_i$ in multiple weighted regressions \begin{align}
	\sqrt{h_i}\psi_B(W_i,{\eta_0}) = \sqrt{h_i}b_B(Z_i)'\beta_{B,0} + \sqrt{h_i}r_{B}(Z_i) + \sqrt{h_i}\varepsilon_{i,B}, \label{eq_auxReg1_boot}
\end{align}
with estimated nuisance parameters to obtain bootstrap estimates \begin{align}
	\hat{\beta}^b_B = \bigg(\sum_{i=1}^nh_ib_B(Z_i)b_B(Z_i)'\bigg)^{-1}\sum_{i=1}^nh_ib_B(Z_i)\psi_B(W_i,\hat{\eta}). \label{eq_bhat1_boot}
\end{align}
Then, for all $z$, calculate the centered bootstrap $t$-statistic for  bounds $B = L,U$ as \begin{align}
	t_B^b(z) = \frac{\hat{\theta}_B^b(z) - \hat{\theta}_B(z)}{\hat{\sigma}_B^b(z)}.  %\frac{b_B(z)'(\hat{\beta}_B^b - \hat{\beta}_B)}{\hat{\sigma}^b_B(z)}.
\end{align}
Denote $\bar{t}_B^b = \sup_{z\in\mathcal{Z}} t_B^b(z)$ and $\underline{t}_B^b = \inf_{z\in\mathcal{Z}} t_B^b(z)$ and, for any $\alpha$, let $c_{n,\alpha}(t^b)$ be the $\alpha$-quantile of $t^b$ over the bootstrap samples $b=1,2,\dots$. The $(1-\alpha)$ confidence band for $\theta(z)$ for all $z \in \mathcal{Z}$ is then given by \begin{align}
	CB_{\theta(z),1-\alpha} = \bigg[\hat{\theta}_L(z) - c_{n,\alpha/2}(\underline{t}_L^b)\frac{\hat{\sigma}_L(z)}{\sqrt{n}}, \hat{\theta}_U(z) + c_{n,1-\alpha/2}(\bar{t}_U^b)\frac{\hat{\sigma}_U(z)}{\sqrt{n}}\bigg]. \label{eq_CBAND1}
\end{align}
Theorem \ref{thm_strongApprox1} contains the joint strong approximation result for both series and bootstrap processes. Theorem \ref{thm_uniformInf1} contains the uniform inference result. The assumptions are generally stronger with regards to the basis function growth and existence of moments compared to the pointwise case \citep{belloni2015some}.

\begin{thm}[Strong Approximation of Series and Bootstrap Process] \label{thm_strongApprox1}
	Let $k^*$ be defined according to Table \ref{tab_k_all}.
	Under the assumptions of Theorem \ref{thm_AsyNor} with (i) $m > 3$ and (ii) $\sup_B \sqrt{{\xi_{k,B}^2\log^3n}/{n}}(n^{1/m_B}\log^{1/2}n + \sqrt{k_B}l_{k,B}c_{k,B}) = o(a_n^{-1})$, (iii) $a_n^3n^{-1/2}\sup_Bk_B^2$ $\xi_{k,B}(\log n + \log^2 n /k_B) = o(1)$, %and (iii) $ \sup_Bl_{k,B}c_{k,B}\log n  = o(a_n^{-1})$, 
	and (iv) $\sup_{B}n^{1/2}$ $k_{B}l_{k,B}c_{k,B}\log^2n = o(a_n^{-1})$, there exists a random standard normal vector $\mathcal{N}_{k^*}$ of length $k^*$ such that \begin{align*}
		\sqrt{n}{\hat{\Omega}_n(z)^{-1/2}}\begin{pmatrix}
			{\hat{\theta}_L(z) - \theta_L(z)} \\
			{\hat{\theta}_U(z) - \theta_U(z)}
		\end{pmatrix} =_d {\Omega(z)^{1/2}}\mathcal{N}_{k^*} + o_P(a_n^{-1}) \textit{ in } \ell^{\infty}(\mathcal{Z}) 
	\end{align*} and \begin{align*}
		\sqrt{n}{\hat{\Omega}^b_n(z)^{-1/2}}\begin{pmatrix}
			{\hat{\theta}^b_L(z) - \hat{\theta}_L(z)} \\
			{\hat{\theta}^b_U(z) - \hat{\theta}_U(z)}
		\end{pmatrix} =_d {\Omega(z)^{1/2}}\mathcal{N}_{k^*} + o_{P^*}(a_n^{-1}) \textit{ in } \ell^{\infty}(\mathcal{Z}), 
	\end{align*}
	where $P^*$ is the conditional probability computed given the data $\{W_i\}_{i=1}^n$. 
\end{thm}

\begin{thm}[Uniform Inference] \label{thm_uniformInf1}
	Denote $\mathcal{P}$ as the set of probability measures obeying the assumptions of Theorem \ref{thm_strongApprox1}. The confidence bands \eqref{eq_CBAND1} have uniform coverage \begin{align*}
		\underset{n\rightarrow\infty}{\lim \inf} \inf_{P\in\mathcal{P}} P(\theta(z) \in CB_{\theta(z),1-\alpha}~\textit{ for all } z\in\mathcal{Z}) \geq 1 - \alpha. 
	\end{align*}
\end{thm}

\section{Monte Carlo Study} \label{sec_MC1}
In this section, we analyze size and power properties of the proposed misspecification robust confidence intervals in finite samples. In particular we look at the size along a grid of $z$-values that vary with respect to the share of always-takers and thus the width of the identified set in the population. We consider a generalized Roy model with a random binary treatment and missing responses: 

\begin{table}[!h] \centering 
	\caption{Monte Carlo Study: Generalized Roy Model} \label{tab_MCdesign1} \footnotesize
	\begin{minipage}{0.4\textwidth} 	\begin{align*}
			S_i &= \mathbbm{1}(X_i'\gamma + D_i - v_i \geq 0),  \\
			Y_i(0) &= \varepsilon_i(0), \\
			Y_i(1) &= \mu_1(X_{i,1}) + \varepsilon_i(1), \\
			Y_i^* &= D_iY_i(1) + (1-D_i)Y_i(0), \quad \\
			Y_i &= S_iY_i^*,
	\end{align*}\end{minipage} \begin{minipage}{0.4\textwidth} 
		\begin{align*}
			D_i &\sim binomial(0.5), \\
			X_{i,j} &\sim uniform(0,1) \quad j=1,\dots,p, \\
			\begin{pmatrix}
				\varepsilon_i(1) \\ \varepsilon_i(0) \\ v_i
			\end{pmatrix} &\sim \mathcal{N} \begin{pmatrix}
				\begin{bmatrix}
					0 \\ 0 \\ 0
				\end{bmatrix},\begin{bmatrix}
					&\sigma_1^2 \ &0 \ &\rho\sigma_1 \\ 
					&0 \ &\sigma_0^2 \ &0 \\ 
					&\rho\sigma_1 \ &0 \ &1 
				\end{bmatrix} 
		\end{pmatrix}, \end{align*}
	\end{minipage} \begin{justify}
		with $X_i = (X_{i,1},\dots,X_{i,p})'$, $\gamma = (\Phi^{-1}(0.99)-1,0,0,\dots,0)'$, $\mu_1(x) = 0.35 - 4x^2 + 4x^3$, $\sigma_1 = \sigma_0 = 0.2$, and $\rho = 0.5$. $\Phi^{-1}(\cdot)$ denotes the quantile function of the standard normal distribution. \end{justify}
\end{table} %\vspace{-18pt}
Similar designs have been considered for simulation for point-identified effects under exclusion restrictions, see e.g.~\cite{heckman2005structural} or \cite{heiler2022efficient}. Note that the true model here assumes strong monotonicity of the effect of the treatment on response. This knowledge, however, is not imposed onto the estimation procedures. We chose $Z_i = X_{i,1}$. Thus, the true response function for the always-takers is given by
\begin{align}
	\theta_{AT}(z) &= E[E[Y_i(1) - Y_i(0)|X_i, S_i(1) = S_i(0) = 1]|Z_i = z, S_i(1) = S_i(0) = 1] \notag \\
	&= \mu_1(z) - \rho \frac{\phi(z\gamma)}{\Phi(z\gamma)}. \label{eq_MC_thetaz}
\end{align}
The parameters are chosen such that non-response rates vary from 84.13\% to 99.00\% and are monotonically increasing in $z$. This will allow us to discover potential differences in coverage rates for varying setups including scenarios close to point identification. We consider both the continuous case, i.e.~the size of confidence intervals for the continuous $\theta(z)$ as well as power curves for a simple discretized version where $z$ is integrated from 0 to 0.5 and 0.5 to 1 respectively.
The nuisance quantities are estimated using honest probability random forests and honest quantile regression forests from the \verb|grf| package \citep{athey2019generalized} with default tuning parameters and two-fold cross-fitting. The design is sufficiently sparse for the forests to achieve the required convergence rates for estimating \eqref{eq_MC_thetaz} according to Assumption A.6. For the heterogeneity analysis, we use basis splines with nodes and order selected via leave-one-out cross-validation for the continuous case and indicator functions for the discrete case. We analyze the size and power of the misspecification robust confidence intervals \eqref{eq_CIhat1} at a 95\% confidence level.

\begin{figure}[!h]
	\caption{Coverage Rates for $\hat{\theta}(z)$}
	\label{fig_size_both}
	\begin{subfigure}{0.5\textwidth}
		\centering
		% include first image
		\includegraphics[width = \textwidth, trim = 0 100 0 100, clip]{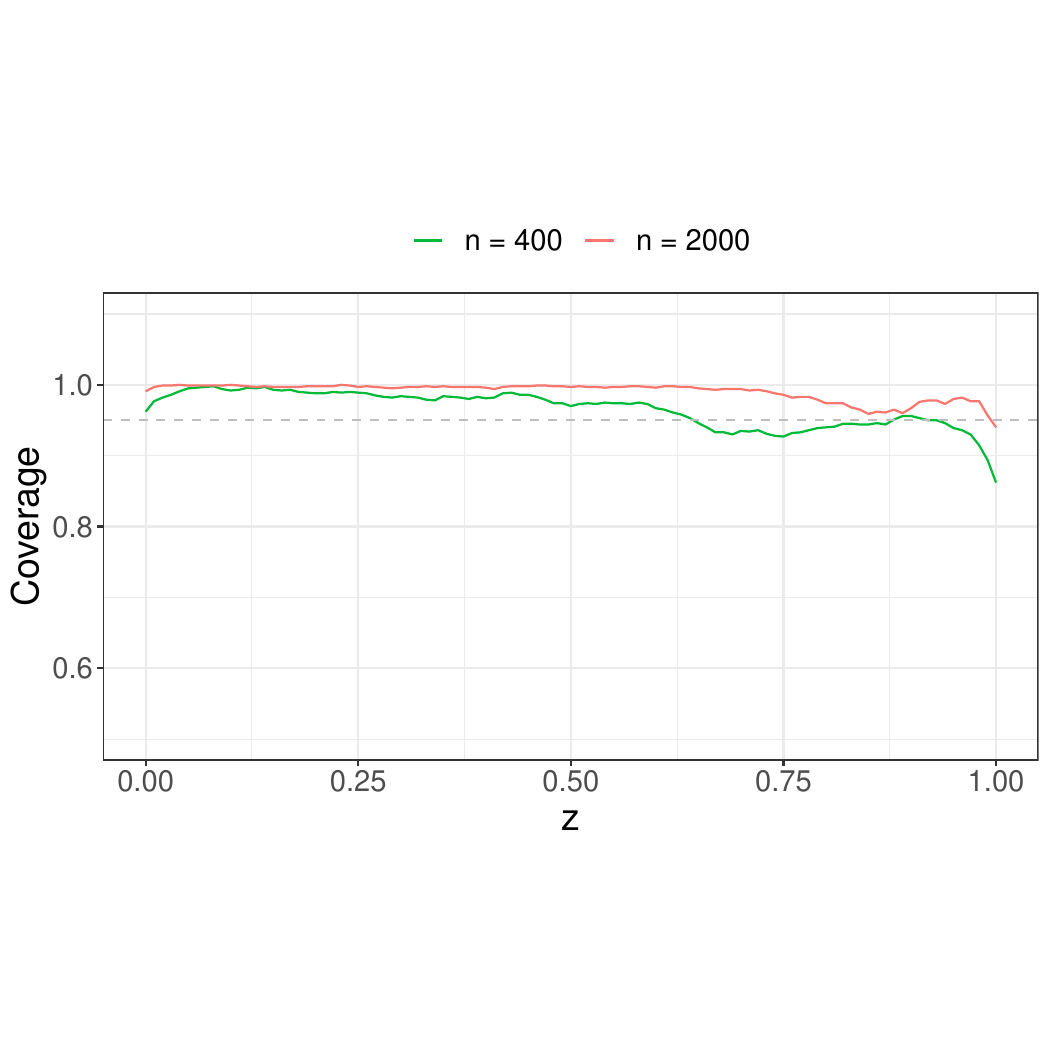}
		\caption{Design 1: $p = 10$}
		\label{fig_size_p10}
	\end{subfigure}
	\begin{subfigure}{0.5\textwidth}
		\centering
		% include second image
		\includegraphics[width = \textwidth, trim = 0 100 0 100, clip]{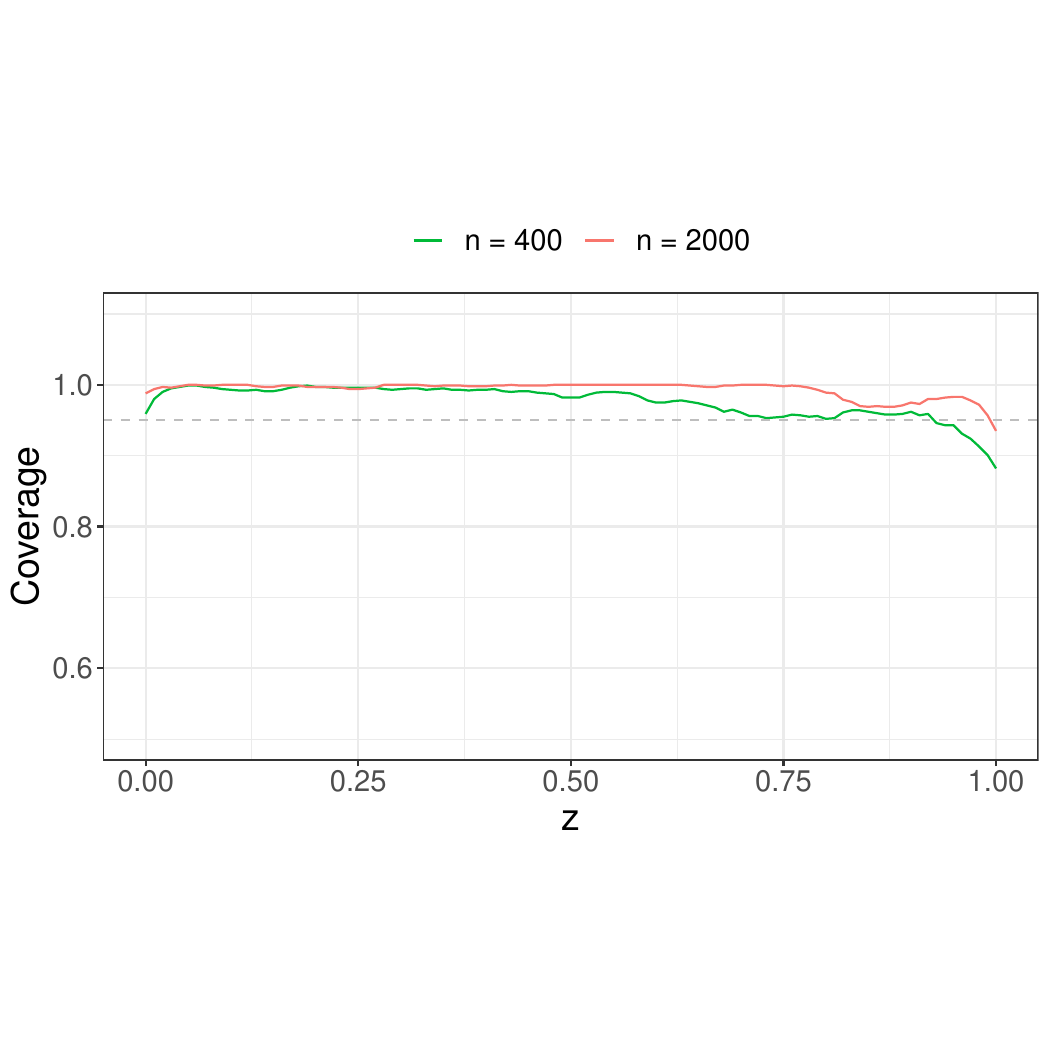}
		\caption{Design 2: $p = 100$}
		\label{fig_size_p100}
	\end{subfigure}	\footnotesize
	\begin{justify}
		Coverage rates for varying parameter and sample sizes with nominal level 95\%. Results are based on 1000 replications. 	
	\end{justify} \vspace{-12pt}
\end{figure}

Figure \ref{fig_size_both} depicts the simulated coverage rates in the case of $p=10$ and $p=100$ regressors for total sample sizes of $n=400$ and $n=2000$. The rates for $n=400$ can sometimes drop to around 90\% but overall coverage is still close to nominal. For $n=2000$, the confidence intervals have at least nominal coverage ranging from 95\% to 100\% depending on $z$. Thus, the theoretical large sample guarantee in Theorem \ref{thm_Coverage1} seems to approximate the finite sample behavior reasonably well in these designs. 

\begin{figure}[!h]
	\caption{Power Curves for $\hat{\theta}(z)$}
	\label{fig_power_all}
	\begin{subfigure}{0.5\textwidth}
		\centering
		\includegraphics[width=\textwidth, trim = 0 100 0 100, clip]{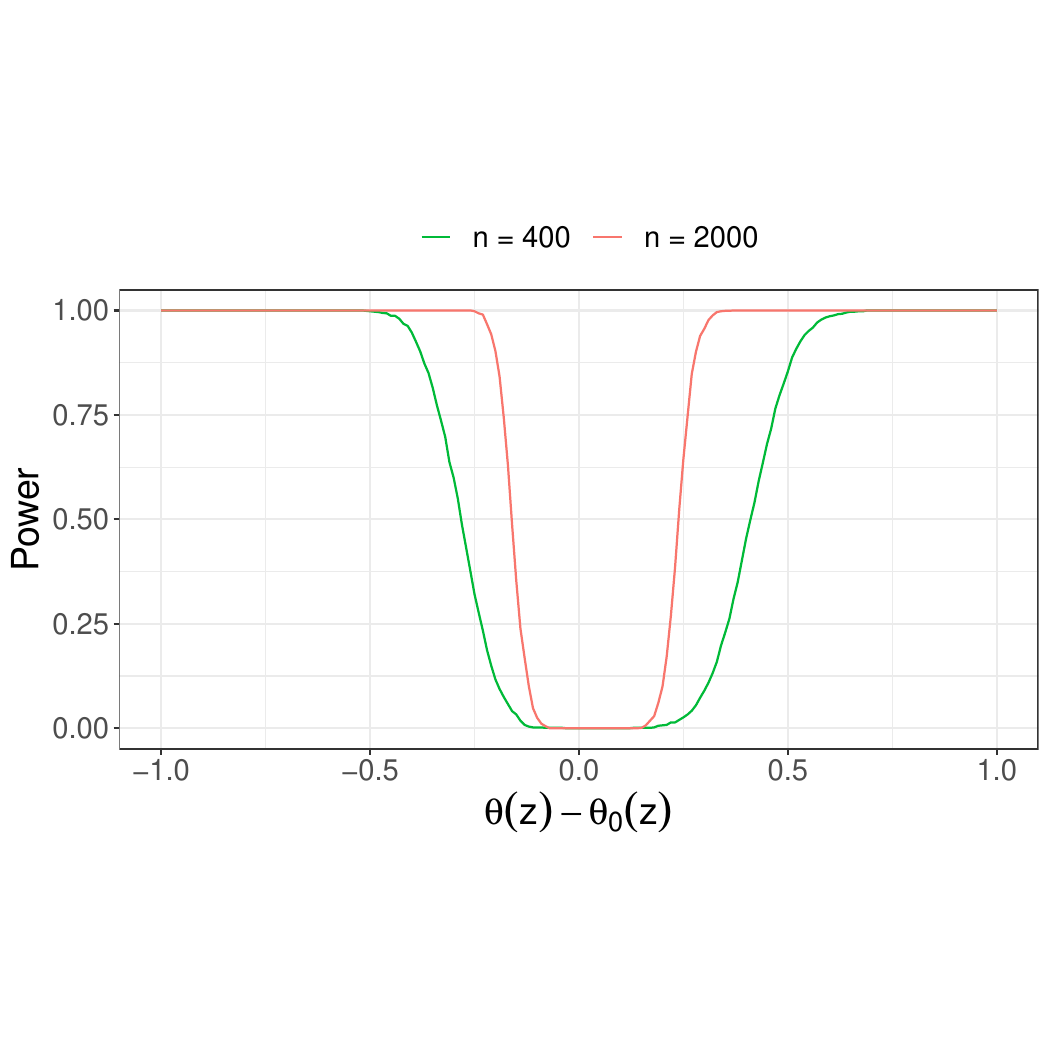}
		\caption{Design: $z_0$, $p = 10$}
		\label{fig_power1_p10}
	\end{subfigure}
	\begin{subfigure}{0.5\textwidth}
		\centering	
		\includegraphics[width=\textwidth, trim = 0 100 0 100, clip]{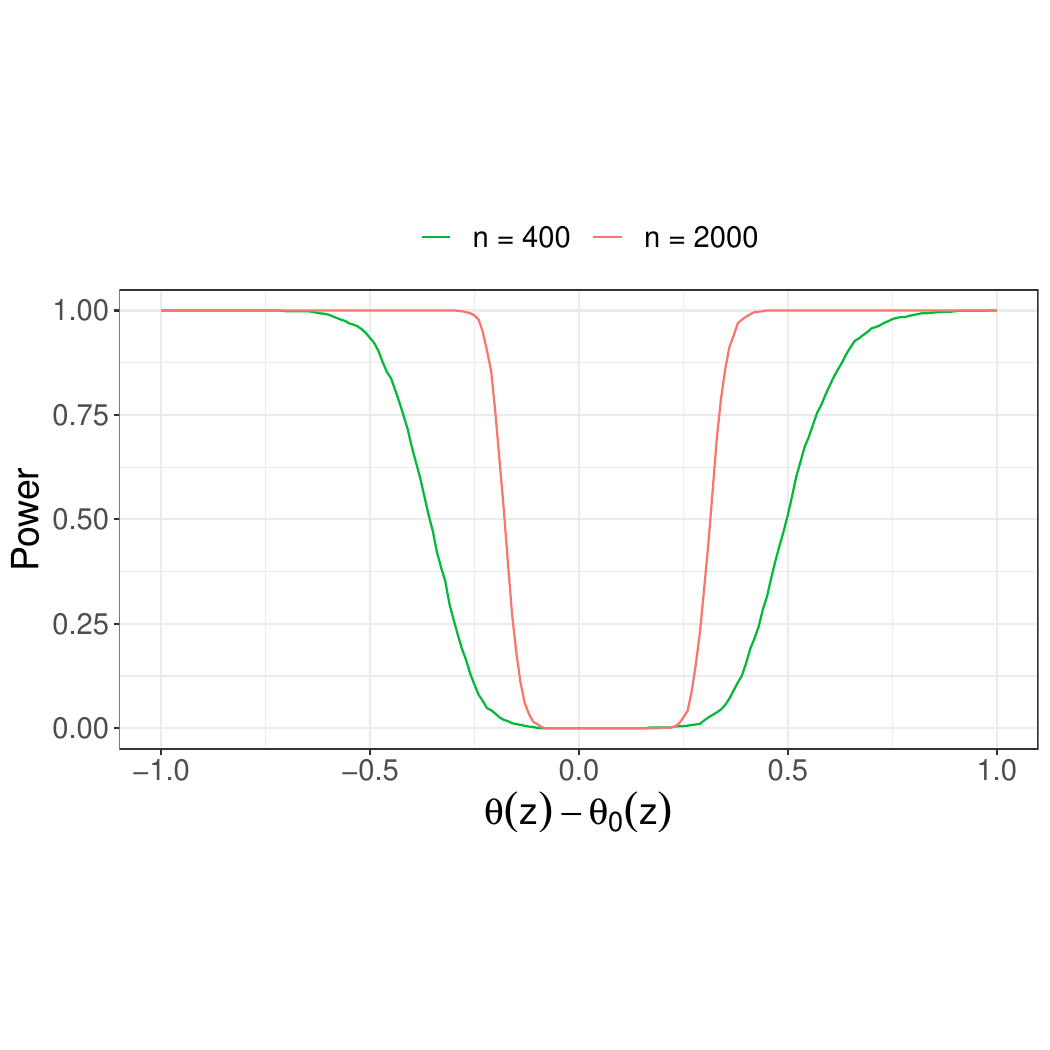}		
		\caption{Design: $z_0$, $p = 100$}
		\label{fig_power1_p100}
	\end{subfigure}
	\begin{subfigure}{0.5\textwidth}
		\centering
		\includegraphics[width=\textwidth, trim = 0 100 0 100, clip]{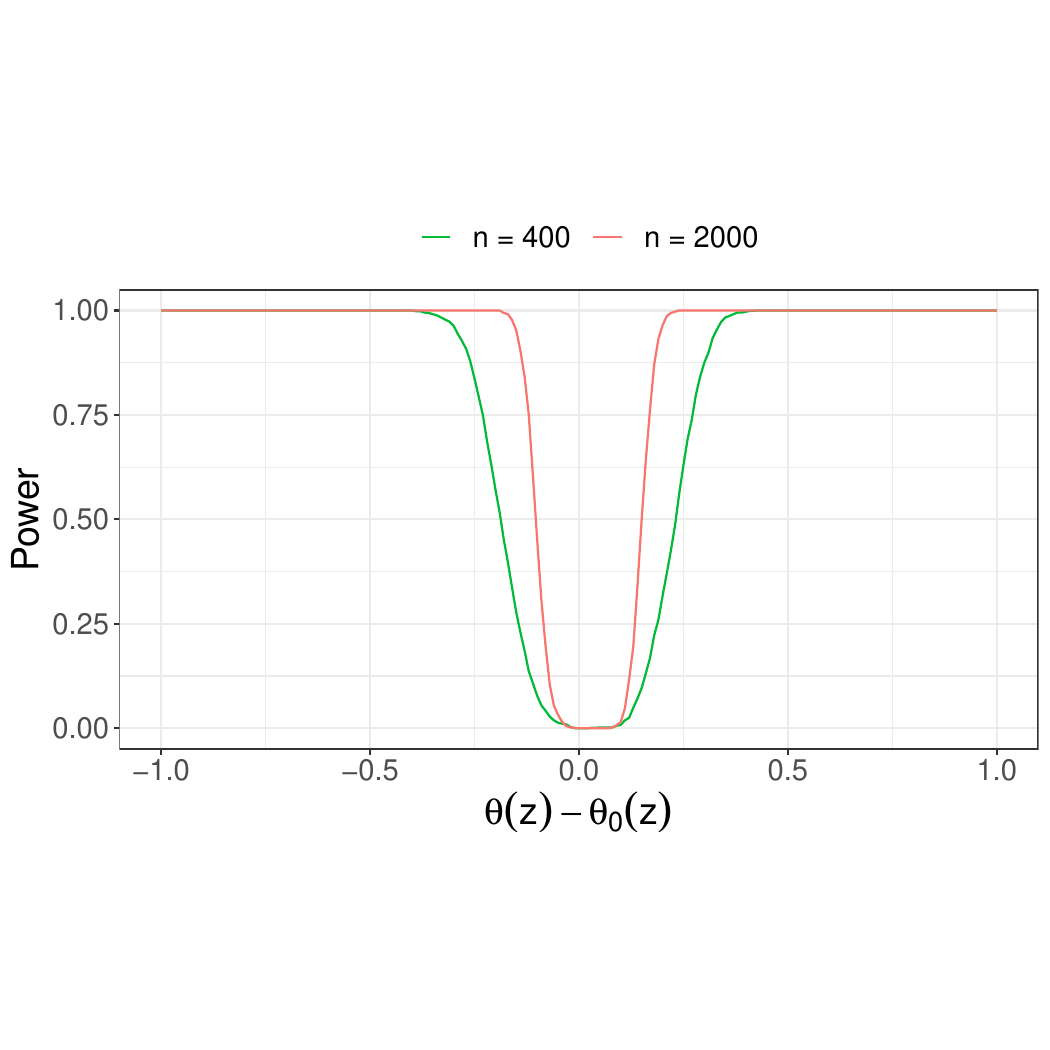}
		\caption{Design: $z_1$, $p = 10$}
		\label{fig_power2_p10}
	\end{subfigure}
	\begin{subfigure}{0.5\textwidth}
		\centering
		\includegraphics[width=\textwidth, trim = 0 100 0 100, clip]{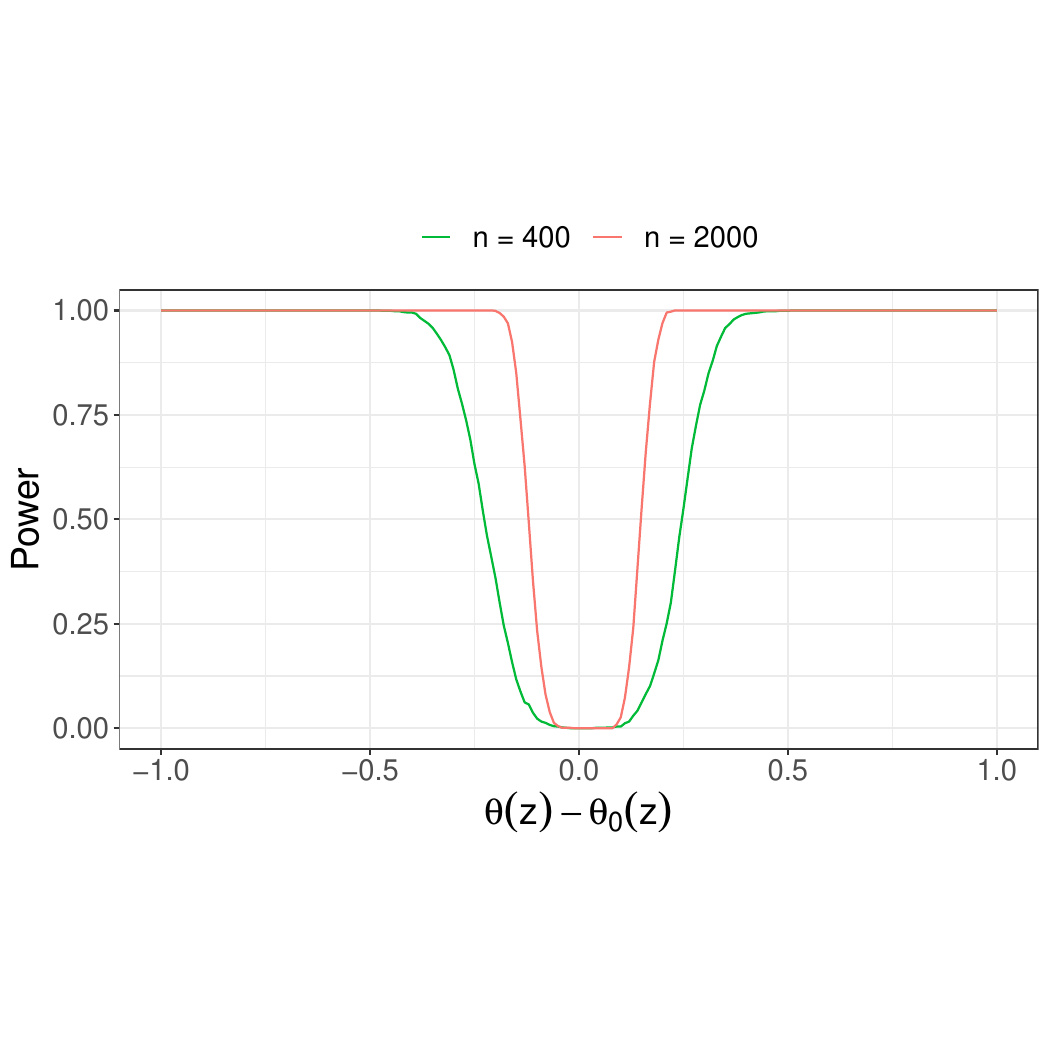}
		\caption{Design: $z_1$, $p = 100$}
		\label{fig_power2_p100}
	\end{subfigure}	\footnotesize
	\begin{justify}
		Designs with $z = z_0$ depict power curves as function of the deviation from the null for parameter $\theta(z_0) = \int_0^{1/2}\theta(z)dF(z)$. Designs with $z = z_1$ depict the power curves as a function of the deviation from the null for parameter $\theta(z_1) = \int_{1/2}^1\theta(z)dF(z)$. Results are based on 1000 replications. 
	\end{justify} \vspace{-12pt}
\end{figure}

Figure \ref{fig_power_all} depicts the power curves for two different heterogeneous effect parameters $\theta(z_0)$ and $\theta(z_1)$ for $p=10$ and $p=100$ at sample sizes $n=400$ and $n=2000$. $\theta(z_0)$ corresponds to an area with larger uncertainty regarding the partially identified parameter, i.e.~it is integrated over the range of the heterogeneity with the largest share of unobserved outcomes while $\theta(z_1)$ integrates over the range with the largest share of observed outcomes. The difference in uncertainty can also be seen in Figure \ref{fig_example1_both}. The results show that the power curves are close to zero around the null for all sample size and designs reflecting the conservativeness of the inferential method. However, power converges quickly to 100\% when moving away from the null. Power is lower for smaller sample sizes and a larger amount of possible confounding variables as expected. Moreover, in the $z=z_0$ case for which the share of missing outcome is larger, confidence intervals have lower power compared to the $z=z_1$ case that is closer to the case of point identification. Overall, intervals seem to perform reasonably well for different degrees of identification in the overall heterogeneous effect but being closer to point identification tends to yield more power in finite samples.

\section{Empirical Study} \label{sec_application}
\subsection{Literature on Social Media and Political Polarization}
There is a large developing literature regarding the effects of social media on political polarization.
However, the evidence on the effects of social media and news exposure via social media on polarization is very mixed, see \cite{haidtONGOINGsocialmediaPolitical} (ongoing) for a comprehensive collaborative review. Effects and channels are often context-dependent and can vary by time, country, platform, algorithm, research design, subgroup, outcome measure, and more. Here we focus on research regarding affective polarization, Facebook, online news, and questions of selection and heterogeneity.   

There are significant changes in measures of affective polarization in many countries since the 1980's with steepest increase in the US \citep{boxell2020cross}. The precise role of online news consumption and social media is still under debate: 
\cite{suhay2018polarizing} show that online news that contains partisan criticism that derogates political opponents increases affective polarization. 
\cite{munger2020null} use different Amazon Turk and Facebook Ad samples with clickbait and conventional news headline treatments. They find that older non-democrat users read clickbait articles more often but no effects on polarization. 
\cite{cho2020search} demonstrate that political videos selected via the YouTube algorithm increase affective polarization.
\cite{nordbrandt2021affective} uses Dutch Survey data to argue that higher levels of affective polarization lead to increased social media consumption but finds no evidence for the reverse channel. However, there seems to be significant user-level heterogeneity. 
\cite{beam2018facebook} show that, over the course of the US 2016 presidential election campaign, attitudes towards political opponents remained relatively stable. Moreover, they find that Facebook leads to modest depolarization due to an increase in counter-attitudinal news exposure in this period. However, they look at partisan measures based on identity formation and not at standard affective polarization metrics. 
\cite{bail2018exposure} study the effects of following counter-attitudinal Twitter bots. Their findings suggest a heterogeneous increase in polarization from counter-attitudinal exposure on Twitter. In particular, Republicans adopted more conservative attitudes after being exposed to a liberal bot while for Democrats the increase in liberal attitudes after following a conservative bot is insignificant.  
\cite{allcott2020welfare} show that deactivation of Facebook for one month in Fall 2018 decreased exposure to polarizing news and polarization of political views. Their point estimate on affective polarization is negative but insignificant. However, the study is underpowered to detect small effects and they use a consumption based metric of polarization instead of attitudinal measures.
\cite{feezell2021exploring}  do not find evidence that algorithmic or non-algorithmic news sources contribute to higher levels of partisan polarization using explorative survey data.
\cite{di2021does} study varying social media status treatments for inside and outside echo chamber units. They find that Twitter, in particular when allowing for interactions, increases polarization for groups which are already classified as being inside an echo chamber. Subjects in the outside echo chamber group show no significant increases. 
%The difference between platforms and changes in user base over seem to be important. 
\cite{yarchi2021political} argue that there are important cross-platform differences. %Their findings suggest that Twitter is polarizing while What's App is depolarizing. Moreover, 
They provide evidence that Facebook is the least homophilic social media network in terms of interactions, positions, and expressed emotions.
%\cite{waller2021quantifying} study polarization of Reddit in terms of behavioral outcomes over time. They argue that system level shifts in polarization are primarily driven by the arrival of new posting users after the 2016 presidential election in the US. 

\cite{levy2021social} employs a large field experiment regarding news consumption, polarization, and algorithmic news selection on Facebook. He shows that a counter-attitudinal nudge towards subscribing to an outlet with an opposing political ideology decreases affective polarization but does not affect political opinions. He argues that the small nudge setup reflects a realistic user experience on Facebook. A channel seems to be that a shock to the selection of news consumption has lasting effects as units do not re-optimize their feed much afterwards. However, \cite{levy2021social} also provides evidence that the Facebook algorithm limits exposure to precisely these counter-attitudinal news outlets and thus can increase polarization overall. 
This study has a clear stratified randomization design but suffers from large differential attrition rates in the endline survey, in particular when considering pro- and counter-attitudinal treatments separately. Conventional unconditional Lee bounds for the pro- and counter-attitudinal treatment effects are relatively wide and include zero as well as moderate polarization effects. 
\cite{levy2021social} does not find significant heterogeneity in treatment effects when using simple interacted linear regression models that ignore attrition. We re-analyze the specific question of the effect of a counter-attitudinal nudge on Facebook on affective polarization by \cite{levy2021social} using the refined DML based heterogeneous bounding method developed in this paper. 

\subsection{Experiment, Data, and Attrition}
%In this section, we re-analyze the data collected by \cite{levy2021social} in a large scale field experiment on Facebook that studies news exposure, effects of a subscription nudge on polarization and opinions, and algorithmic news supply. 
In \cite{levy2021social}, users were recruited via Facebook ads and filled out a baseline survey between February--March 2018. Units were then stratified by self-reported political ideology and randomly allocated into one of three different treatment arms 1) Liberal, 2) Conservative, or 3) Control. The treatments in 1) and 2) consisted of a nudge to subscribe to (``like'') a selection of four potential (liberal or conservative) outlets. It was explained that a subscription could provide new perspectives, but there were no other incentives or rewards offered. Likes on Facebook make posts from the corresponding outlet more likely to appear on the user's feed, thus exposes them to potentially new information and opinions. The liberal outlets were \textit{HuffPost}, \textit{MSNBC}, \textit{The New York Times}, and \textit{Slate}. The conservative outlets were \textit{Fox News}, \textit{The National Review}, \textit{The Wall Street Journal}, and \textit{The Washington Times}. Based on this, the \textit{counter-attitudinal treatment} is defined as nudge towards outlets with ideological leanings contrary to the leaning of the user.\footnote{Individual leanings are based on party affiliation. If units do not identify as Democrats or Republicans, it is according to self-reported ideology. If they neither identify as liberal nor conservative, support of the candidate in the 2016 elections is used. This excludes about 3\% of the total sample that provide no information on leaning.}  

Around two months after the baseline survey, participants were asked to fill out an endline survey where political opinions and measures of affective polarization were recorded. 
\cite{levy2021social} does not find effects from any treatment on political opinions, Without controlling for attrition, there is a significant decrease in affective polarization for the counter-attitudinal treatment group. %However, he detects significant effects for the counter-attitudinal treatment on affective polarization relative to the pro-attitudinal treatment. %The counterfactual comparison for affective polarization there is made between treatment arms and not relative to the control group due to differential attrition rates. 
%Our method can account for such differences and thus enables us to study whether the counter-attitudinal treatment itself drives the effect on polarization compared to no intervention.
%Our method can account for the high level of (differential) attrition. 

Thus, we consider the index for affective polarization constructed by \cite{levy2021social} as outcome in what follows. In particular, we analyze the causal effects of \textit{nudging} users towards counter-attitudinal subscriptions on affective polarization. The parameters can also be interpreted as intent-to-treat effects of \textit{subscription}. The outcome measure is standardized such that all coefficients are measured in terms of standard deviations in what follows. 

The covariates collected via the baseline survey and Facebook contain information on political ideology, party affiliation, voting behavior, approval of President Trump, baseline polarization, news consumption, and socio-demographic variables such as age and gender. For more information and descriptive statistics consider \cite{levy2021social}, Section II. 
The final sample (including missing endline survey units) consists of 24230 units of which 12126 are in the treatment and 12104 in the control group. 
\cite{levy2021social} estimates the effects of the intervention on affective polarization using only the units which replied to the endline survey, i.e.~for which the outcome variable is observed. He argues that the main estimates are likely to generalize beyond the selected population. 
%When considering the pro- versus counter-attitudinal treatments.  

\begin{table}\footnotesize \centering \caption{Attrition Rates for Endline Survey} \label{tab_attrition1}
	\begin{tabular}{lrccl}  \hline \hline \\[-1ex]
		&     $n$  & Treatment & Control & p-value \\[1ex] \hline \\[-1ex]
		1. Extremely liberal        &     3868  &  0.5031  &0.4623  &$0.0110^{**}$ \\
		2. Liberal                  &     7168  &  0.5221  &0.4959  &$0.0267^{**}$ \\
		3. Slightly liberal         &     3195  &  0.5516  &0.5287  &0.1933 \\
		4. Moderate					&	  2156  &  0.5944  &0.5660  &0.1808 \\
		5. Slightly conservative    &     2372  &  0.5687  &0.5595  &0.6512 \\
		6. Conservative             &     3990  &  0.5719  &0.5330  &$0.0136^{**}$ \\
		7. Extremely conservative   &     1236  &  0.5460  &0.5527  &0.8149 \\
		Haven't thought much    	&	   245  &  0.6535  &0.5932  &0.3325 \\[1ex] \hline \\[-1ex]
		Total                       &    24230  &  0.5447  &0.5173  &$0.0000^{***}$ \\[1ex] 
		 \hline \hline 
	\end{tabular} \vspace{-6pt}
	\begin{justify} \centering
		{The p-values are for tests of equal means. $^{***} = p < 0.01$, $^{**} = p < 0.05$.}
	\end{justify}
\end{table}

The experiment, however, suffers from large differential attrition rates: 
Table \ref{tab_attrition1} contains the differential attrition rates between treatment and control group stratified by political ideology. Attrition in the endline survey is large with rates between 46.23\% to 65.35\%. Moreover, there is significant heterogeneity when looking at the difference in attrition rates between treatment and control condition. In addition, the association between treatment and attrition is not homogeneous across all subgroups. %violating the validity of the basic Lee bounds used by \cite{levy2021social} for robustness checks. 
This indicates that there are potential interactions between treatment and baseline characteristics that can lead to heterogeneous response rates. In this case, looking only at unconditional attrition rates between treatment and control group provides a distorted view on the potential bias introduced by ignoring the selection into response. 
We provide a more thorough analysis of the effect bounds for the counter-attitudinal treatment fully accounting for heterogeneity in treatment effects and attrition rates and potential effects of the treatment on selection into the endline survey. In particular, we re-analyze the unconditional effect bounds using the method suggested in this paper as well as heterogeneous bounds in terms of relevant pre-treatment characteristics.   

\subsection{Parameter, Estimation, and Inference Methods}
Replication of unconditional point estimates and unconditional Lee bounds provided by \cite{levy2021social}, Table A.12(b) use the same methods. 
We bound the (un)conditional average treatment effect(s) for the always-takers using the methods developed in this paper. Always-takers in this experiment are units that would be taking part in the endline survey regardless of whether they have been nudged or not. They are estimated to make up around 46.18\% of the total study population. 

We employ two versions of the DML based generalized Lee bounds that differ in terms of nuisance parameter models: The first (DML parametric) uses logistic regression with all confounding variables interacted with the treatment for the response selection probabilities and linear quantile regression with all confounding variables for the conditional quantile of the selected treated and selected controls. These models are more likely to be misspecified. The second (DML forest) uses probability forests and quantile forests with 1000 trees and honest splitting \citep{athey2019generalized}.
For both specifications, all categorical variables are coded as flexible dummy variables leaving us with 36 confounders in total. Conditional quantile trimming levels are rounded towards the closest value on a grid from $(0.01, 0.02,\dots,0.99)$. Cross-fitting is based on 10 folds. For the heterogeneity analysis we use the estimated signals $\psi_B(W_i,\hat{\eta})$ provided by the forest-based DML specification and basis splines for the continuous variables with node and order selection via leave-one-out cross-validation. Confidence intervals are based on the misspecification robust method \eqref{eq_CIhat1}, confidence bands on the multiplier bootstrap \eqref{eq_CBAND1}. We report both at 90\% due to the convservativeness of the methods. 

\subsection{Results}
\subsubsection{Unconditional Effects}
In this subsection, we provide the unconditional effect analysis of the counter-attitudinal nudge on affective polarization.  
Table \ref{tab_uncondBounds1} contains the estimates by \cite{levy2021social}, the naive Lee bounds assuming (strong) monotonicity as well as the two DML-based methods.\footnote{Table A.12(b) by \cite{levy2021social} contains an error as his bounds $[-0.172, 0.060]$ (no CI provided) are calculated from $\theta_L(x) = E[Y_i|D_i=1,S_i=1, Y_i \leq q(p_0(X_i)),X_i=x] - E[Y_i|D_i=0,S_i=1, Y_i \leq q(p_0(X_i)),X_i=x]$ (and equivalently for $\theta_U(x)$) and not based on \eqref{eq_boundsX_DEFINITION}. Moreover, in the specifications with controls, he assumes constant effect bounds and a linear form of the truncated mean which is heavily restrictive and generally does not identify the true bounds. Columns (4) and (5) produce the correctly calculated bounds under the more credible assumptions allowing for non-linearity and weak monotonicity.} 
\begin{table}[!h] \footnotesize \centering \caption{Unconditional Counter-Attitudinal Treatment Effect on Affective Polarization} \label{tab_uncondBounds1}
	\begin{tabular}{lccccc} \hline \hline \\[-1ex]
		& Levy (2021)    &   Levy (2021)   &   Lee Bounds & DML Bounds & DML Bounds \\ 
		& (unconditional) & (conditional) &  (unconditional) & (parametric) & (forest) \\ 
		& (1) 	&(2) & (3) & (4) & (5) \\[1ex] \hline \\[-1ex]
		Estimate	&     $\mathbf{-0.055}^{***}$ 	& $\textbf{-0.028}^{**}$    &   [-0.161,~0.060] &   [-0.083,~0.010] &  [-0.080,~0.011] \\
		$90\%$-CI 	& (-0.086,~-0.024)				& (-0.048.~-0.008) 			& (-0.195,~0.094) 	&  (-0.109,~0.037)  & (-0.120,~0.043) \\[1ex] \hline \hline 
	\end{tabular} \begin{justify} \footnotesize
		Point and interval estimates of the counter-attitudinal treatment on affective polarization using 16896 (columns 1 and 2) and 24230 (columns 3 to 5) observations. (Un)conditional refers to the case of (not) using $X$ in the regression or bound analysis. Parametric and forest refers to the nuisance models for the DML bounds. $^{***} = p < 0.01$, $^{**} = p < 0.05$. %$^{*} = p < 0.10$.
	\end{justify} \vspace{-12pt}
\end{table}
The conventional Lee bounds (3) as well as the DML bounds (4) and (5) contain a null effect in the estimated identified set. All contain the point estimates provided by \cite{levy2021social}. 
Both DML bounds are much shorter than the Lee bounds, ruling out even small to moderate polarization effects. 
%From Theorem \ref{thm_Coverage1} we know that inference on the effect is conservative, thus we interpret this as weak evidence against a null effect of the counter-attitudinal treatment on affective polarization in line with the original study. 
The applicability of the conventional Lee bounds (3) under strong monotonicity is questionable. In particular, our estimates for the conditional selection probabilities based on the specification for Column (5) suggest that the effect of the treatment on attrition is negative ($\hat{p}_0(x) > 1$) for 65.42\% and positive ($\hat{p}_0(x) < 1)$ for 34.58\% of the sample indicating violation of strong monotonicity. This is in line with the heterogeneous attrition rates observed in Table \ref{tab_attrition1}. 

\subsubsection{Heterogeneous Effects}

%%%%%%%%%%%%%%%%%%%%%
%The relative size of the identified set benefits from the inclusion of covariates even under randomized treatment assignment. In our empirical application, depending on nuisance models, our identified set are between 33.4\% to 48.9\% shorter compared to conventional unconditional Lee-type bounds. 
%
%For some subgroups we are getting close to point identification, e.g.~for conservatives and 18-year olds, the identified sets are 
%cons -0.076794583 -0.038488497
%18 -0.08830682 -0.070940052

%> (-0.09768692 - 0.01533579)/(-0.1608569 - 0.06014163)
%[1] 0.5114184
%> (-0.1053947 - 0.03737267)/(-0.1608569 - 0.06014163)
%[1] 0.6460105
In this subsection, we analyze the effect bounds of the counter-attitudinal nudge on affective polarization as functions of heterogeneity variables. In particular, we look at political ideology (categorical) and age (continuous). We select political ideology from the baseline survey as it was used for block-randomization of the original experiment and could provide insight regarding potential asymmetries in the effect of counter-attitudinal nudges in terms of partisanship. Age can easily be used for targeting of such an intervention based on social media information only (no survey required) and has been shown to be an important determinant of aggregate affective polarization levels in the US \citep{phillips2022affective}. These variables were also suggested by \cite{levy2021social} for heterogeneity analysis.  

\begin{figure}[!h]
	\centering \caption{Counter-Attitudinal Treatment Effect Bounds for Ideology} \label{fig_boundsIdeology}
	\includegraphics[width=\textwidth, trim = 0 170 0 170, clip]{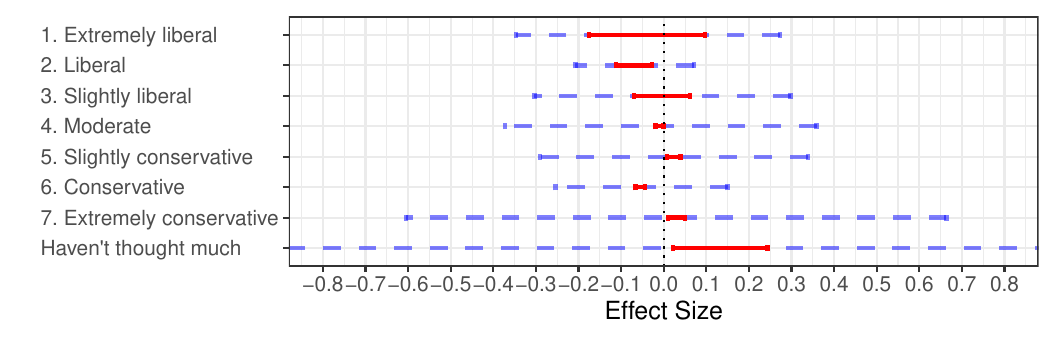}
	\begin{justify}\vspace{-12pt} \footnotesize
		This Figure contains the effect bounds (in standard deviations) of the counter-attitudinal treatment on affective polarization stratified by ideology. The red intervals are the estimated identified sets. The dashed blue lines are the misspecification robust 90\%-confidence intervals for the heterogeneous effects. 
	\end{justify} \vspace{-12pt}
\end{figure}

Figure \ref{fig_boundsIdeology} contains the heterogeneous bounds sorted by ideology plus $90\%$ confidence intervals. The identified sets for \textit{Liberal}, \textit{Moderate}, \textit{Conservative}, \textit{Extremely Conservative} and \textit{Haven't thought much} do not include zero. However, even for these groups, the statistical uncertainty dominates and we cannot rule out a null effect with precision. This reflects the fact that the experiment is not powered enough to detect small effects within smaller subgroups with high probability. 

%The effects are also significant for moderates with set estimate $[-0.159, -0.160]$ and for conservatives with $[-0.167, -0.091]$. This suggests that a nudge on Facebook for these groups can significantly decrease their affective polarization. It is noteworthy that these estimated effect bounds are about 2-3 times as large as the unconditional estimates suggested by \cite{levy2021social} that do not correct for attrition. 

\begin{figure}[!h]
	\centering \caption{Counter-Attitudinal Treatment Effect Bounds for Age} \label{fig_boundsAGE}
	\includegraphics[width=\textwidth, trim = 0 160 0 160, clip]{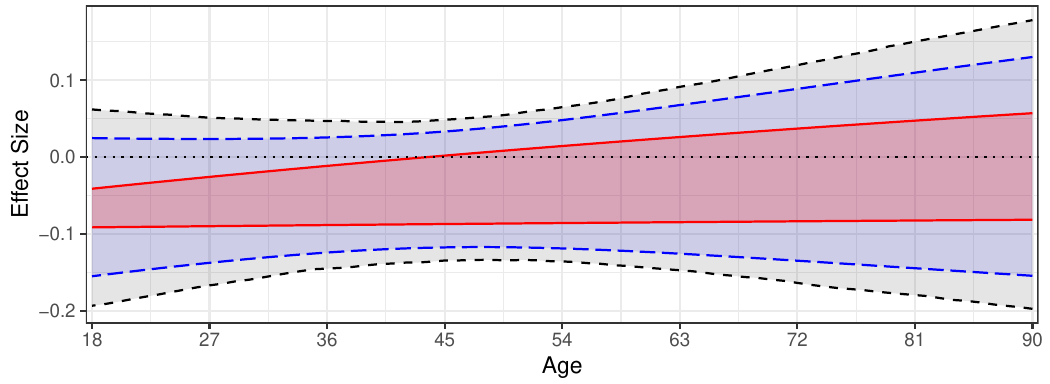}
	\begin{justify}\vspace{-12pt} \footnotesize
		This Figure contains the effect bounds (in standard deviations) of the counter-attitudinal treatment on affective polarization as a function of age. The red band is the estimated identified set. The dashed blue lines are the misspecification robust 90\%-confidence intervals. The black area is the 90\% uniform confidence band using the multiplier bootstrap with 999 replications. The dotted black line indicates zeroes. 
	\end{justify}\vspace{-12pt}
\end{figure}

Figure \ref{fig_boundsAGE} contains the estimated effect bounds and confidence intervals as functions of age.\footnote{Note that here we have excluded the units for which age information was missing ($n = 791$). Thus, the estimate for the lower bound differs slightly from the unconditional estimate in Table \ref{tab_uncondBounds1}, Column (5). Estimating the bounds separately for the omitted category yields interval $[0.096,~0.167]$ with $90\%$-CI $(-0.779,~1.007)$.} We can see that the identified set is widening monotonically in age reflecting a larger sample selection problem with older users. The identified set suggest a depolarization effect for 18-43 year old users ranging from $[-0.091,-0.041]$ to $[-0.087,-0.001]$. The estimates rule out moderate to large polarization effects for these younger users. The negative bounds, however, cannot statistically reject any positive effects due to relevant standard errors. As we know that inference is conservative, we interpret this as weak evidence in favor of a depolarization effect on young users with magnitudes close to the unconditional point estimate suggested by \cite{levy2021social} that does not correct for attrition. 

%We can see that the estimated lower bound is constant and very close to the unconditional lower bound provided in Table \ref{tab_uncondBounds1}, Column (5). The upper bound, however, is monotonically increasing with age. In particular, the identified set is very narrow at age 18, suggesting a significant effect with estimated set $[-0.160, -0.121]$. From there, the identified set gets progressively wider. The last age of significance is at 44 years with estimated set $[-0.166, -0.041]$. From 57 years on, the estimated set does no longer exclude zero. It is noteworthy that both effect bounds for the lowest age groups are much larger in magnitude compared to the unconditional point estimates that do not correct for attrition by, again, a rough factor of 2-3.

%AGE01 $
%z    zb_lower   zb_upper   ci_lower   ci_upper
%FALSE FALSE  0.04435922 0.22871395 -0.7599550 1.04145218
%TRUE   TRUE -0.11095240 0.03031928 -0.1409145 0.06054039$
\subsection{Discussion}	
Overall, the findings do not contradict the conclusion by \cite{levy2021social} that a counter-attitudinal nudge can decrease affective polarization. There is weak evidence in favor of effect heterogeneity. Tight depolarization bounds can be obtained for users from various political ideologies but they are not statistically significant.
% With regards to political ideology, the fact that effects are only significant for conservatives and moderates could be driven by measurement/low power given the amount of attrition considering that the identified set is negative for liberals as well. Note that, as the treatment varies depending on political ideology, we cannot distinguish heterogeneity in user response from heterogeneity in news outlets. It could well be that conservative or liberal leaning outlets offer counter-attitudinal viewpoints at smaller or larger doses. Assessing the differences in partisan attitudes by outlets and the consequences on polarization is an interesting question. However, from a policy assignment perspective, the total effects presented here are most relevant, not hypothetical channels mediated through treatment intensity as the latter can reasonably be considered fixed. Similarly, the baseline of news consumption on Facebook could differ between users with different political ideology and thus being exposed to the same dose of nudges has a different effect on the \textit{relative} amount of counter-attitudinal news which could potentially be more important for affective polarization than the absolute level. 
Regarding age differences, the identified set excludes non-negative values for young users. This could be due to measurement or, if accurate, dose-related as very young users spend significantly more time online. Their overall activity is also likely contributing to lower attrition rates which leads to the tighter bounds reported. Alternatively or additionally, affective polarization is usually understood through the lens of social identity theory: People internalize partisan affiliation as part of their sense of self. The latter tends to be more malleable for younger people, in particular in their formative years \citep{phillips2022affective}, which could explain larger effects. %The last age at which we detect significant effects (44 years) roughly corresponds to the median age of Facebook users (43 years). Thus, the evidence suggests that targeting users with counter-attitudinal subscription nudges on Facebook could decrease affective polarization for a substantial share of the population.

To put the size of identified sets into perspective, we compare the estimates to experimental estimate by \cite{allcott2020welfare}. The bounds for the youngest age levels suggest that, for these groups, the impact is around 0.4 to 0.9 times as large as the effect of deactivating Facebook for a whole month. As a limitation, note that we are comparing to an unconditional baseline from \cite{allcott2020welfare}. The effect of deactivation could potentially be much larger (or smaller) for these groups as well.
For further research, targeting interventions at particular young age and particular ideological groups could provide more definitive evidence.

%affective polarization in line with \cite{beam2018facebook} and \cite{yarchi2021political} Further research it would be interesting to see whether typically encountered subscription advertisements run by news outlets themselves have comparable intent-to-treat effects as nudges provided by outside sources. 

%All these results have to be taken with some caution given the finding by \cite{levy2021social} that the Facebook algorithm itself does not favor counter-attitudinal outlets. However, they suggest that Facebook has the potential to decrease or at least mitigate its effects on affective polarization in line with \cite{beam2018facebook} and \cite{yarchi2021political}. For further research it would be interesting to see whether typically encountered subscription advertisements run by news outlets themselves have comparable intent-to-treat effects as nudges provided by outside sources. 

\section{Concluding Remarks} \label{sec_conclusion1}
This paper provides a method for estimation and inference for bounds for  heterogeneous treatment effects under sample selection.
We make the general point that heterogeneity in partially identified problems requires special attention as both effect parameters as well as identified sets can be subject to heterogeneity. Exploiting the latter can yield more precise inference in empirical applications compared to crude, unconditional approaches. 
There are also multiple extensions possible: In many applications where the method could be useful, the i.i.d.~assumption is overly restrictive. In particular, in social experiments units are often clustered within groups such as schools or regions. It would also be interesting to see under which conditions the methodology in this paper can be extended to more general moment inequality problems.

\addcontentsline{toc}{section}{References}	
{\setstretch{1}
	\bibliography{bounds2,bounds1}
	\bibliographystyle{apa}	
}

%\newpage 
\begin{appendices}

\section{Definition of Variances} \label{sec_variances1} %sec_app_AV1
%The true variance is given by {\footnotesize\begin{align}
%	\Omega(z) = B_{\omega}(z)'E[b_i(\varepsilon_i + r_i)(\varepsilon_i + r_i)'b_i']B_{\omega}(z).
%\end{align}}
%with definitions given in Appendix \ref{sec_App_proof_prelims}. Its consistent estimator is {\footnotesize\begin{align}
%	\hat{\Omega}_n(z) = B_{\hat{\omega}}(z)'\frac{1}{n}\sum_{i=1}^n[b_ie_ie_i'b_i']B_{\hat{\omega}}(z), \label{eq_variance_estimated1}
%\end{align}} where
% {\footnotesize\begin{align*}
%	e_i = (e_{i,L},\ e_{i,S_0},\ e_{i,U},\ e_{i,S_0})'
%\end{align*}}
%with $e_{i,B} = \psi_B(W_i,\hat{\eta}) - b_B(Z_i)'\hat{\beta}_B$ being the residuals of the nonparametric regression for $B= L,U,S_0$ and
%{\footnotesize\begin{align*}
%		{B}_{\hat{\omega}}(z) &= \begin{pmatrix}
%			b_L(z)\hat{\omega}_{1,L}(z) & 0 \\
%			b_{S_0}(z)\hat{\omega}_{2,L}(z) & 0 \\
%			0 & b_U(z)\hat{\omega}_{1,U}(z) \\
%			0 & b_{S_0}(z)\hat{\omega}_{2,U}(z) 
%		\end{pmatrix}
%\end{align*}}
%with weights {\footnotesize\begin{align*}
%	\hat{w}_{1,B} = \frac{1}{b_{S_0}(z)'\hat{\beta}_{S_0}}, \quad 
%	\hat{w}_{2,B} = \frac{b_{B}(z)'\hat{\beta}_{B}}{(b_{S_0}(z)'\hat{\beta}_{S_0})^2}
%\end{align*}}
%for $B = L,U$. The bootstrapped $\hat{\Omega}_n^b(z)$ is defined analogously to \eqref{eq_variance_estimated1} with bootstrap weights $h_i$ applied to each regression first which produces different residuals and weights. 
%\subsection{Variance Estimation}
For any $z$, define the total or stacked basis matrix as {\footnotesize\begin{align*}
b(z) &= \begin{pmatrix}
		b_{L^+}(z) & 0& 0& 0& 0& 0& 0& 0 \\
		0 & b_{S_0^+}(z) & 0& 0& 0& 0& 0& 0 \\
		0 &0 &b_{L^-}(z) & 0& 0& 0& 0& 0 \\
		0 &0 &0 &b_{S_1^-}(z) & 0& 0& 0& 0 \\
		0 & 0& 0& 0 & b_{U^+}(z) & 0& 0& 0 \\
		0 & 0& 0& 0 & 0 & b_{S_0^+}(z) & 0& 0 \\
		0 & 0& 0& 0 & 0& 0 & b_{U^-}(z) & 0&  \\
		0 & 0& 0& 0 & 0 & 0& 0 & b_{S_1^-}(z)  \\ 
	\end{pmatrix}_{[k^*\times 8]}
\end{align*} } and {\footnotesize\begin{align*}
{B}_{{\omega}}(z) &= \begin{pmatrix}
	b_{L^+}(z){\omega}_{1,L}(z) & 0 \\
	b_{S_0^+}(z){\omega}_{2,L}(z) & 0 \\
	b_{L^-}(z){\omega}_{1,L}(z) & 0 \\
	b_{S_1^-}(z){\omega}_{2,L}(z) & 0 \\
	0 & b_{U^+}(z){\omega}_{1,U}(z) \\
	0 & b_{S_0^+}(z){\omega}_{2,U}(z) \\
	0 & b_{U^-}(z){\omega}_{1,U}(z) \\
	0 & b_{S_1^-}(z){\omega}_{2,U}(z)  
\end{pmatrix},
\end{align*}}
with weights 
{\footnotesize\begin{align*}
		\omega_{1,B} &= \frac{1}{b_{S_0^+}(z)'\beta_{S_0^+,0} + b_{S_1^-}(z)'\beta_{S_1^-,0}}, \\
		\omega_{2,B} &= \frac{b_{B^-}(z)'\beta_{B^-,0} + b_{B^+}(z)'\beta_{B^+,0}}{(b_{S_1^+}(z)'\beta_{S_1^+,0} + b_{S_1^-}(z)'\beta_{S_1^-,0})^2},	\end{align*}}
%
%{\footnotesize\begin{align*}
%{\omega}_{1,B^+} &= \frac{1}{b_{S_0}(z)'{\beta}_{S_0,0}}, \quad 
%{\omega}_{2,B^+} = \frac{b_{B^+}(z)'\beta_{B^+,0}}{(b_{S_0}(z)'{\beta}_{S_0,0})^2} \\	
%{\omega}_{1,B^-} &= \frac{1}{b_{S_1}(z)'{\beta}_{S_1,0}}, \quad  	
%{\omega}_{2,B^-} = \frac{b_{B^-}(z)'\beta_{B^-,0}}{(b_{S_1}(z)'{\beta}_{S_1,0})^2}	\end{align*}}
for $B=L,U$. For the $\theta_{AT}(z)$ bounds, the true variance is then given by {\footnotesize\begin{align*}
		\Omega(z) = B_{\omega}(z)'Q^{-1}E[b(Z_i)(\varepsilon_i + r_i)(\varepsilon_i + r_i)'b(Z_i)']Q^{-1}B_{\omega}(z),
\end{align*}}
where $Q = E[b(Z_i)b(Z_i)']$ is the total design matrix. The consistent estimator for $\Omega(z)$ is given by {\footnotesize\begin{align}
		\hat{\Omega}_n(z) = {B}_{\hat{\omega}}(z)'\hat{Q}^{-1}\frac{1}{n}\sum_{i=1}^n[b(Z_i)e_ie_i'b(Z_i)']\hat{Q}^{-1}{B}_{\hat{\omega}}(z), \label{eq_variance_estimated2}
\end{align}}
with $\hat{Q} = \frac{1}{n}\sum_{i=1}^nb(Z_i)b(Z_i)'$ and {\footnotesize\begin{align*}
		e_i = (
		e_{i,L^+} ,\ %\omega_{1,L^+} \\
		e_{i,S_0^+},\ %\omega_{2,L}^+ \\
		e_{i,L^-},\ %\omega_{1,L^-} \\
		e_{i,S_1^-},\ %\omega_{2,L}^- \\
		e_{i,U^+},\ %\omega_{1,U^+} \\
		e_{i,S_0^+},\ %\omega_{2,U}^+ \\
		e_{i,U^-},\ %\omega_{1,U^-} \\
		e_{i,S_1^-} %\omega_{2,U}^- \\
		)',
\end{align*}}
%		\begin{align*}
	%			e_i = \begin{pmatrix}
		%				e_{i,L}^+\hat{w}_{1,L} + e_{i,S_0}\hat{w}^+_{2,L} + 	e_{i,L}^-\hat{w}_{1,L} + e_{i,S_1}\hat{w}^-_{2,L} \\
		%				e_{i,U}^+\hat{w}_{1,L} + e_{i,S_0}\hat{w}^+_{2,L} + 	e_{i,U}^-\hat{w}_{1,L} + e_{i,S_1}\hat{w}^-_{2,L}
		%			\end{pmatrix}
	%		\end{align*}
with residuals
{\footnotesize\begin{align*}
		e_{i,S_0^+} &= \psi_{S_0^+}(W_i,\hat{\eta}) - b_{S_0^+}(Z_i)'\hat{\beta}_{S_0^+},\\
		e_{i,S_1^-} &= \psi_{S_1^-}(W_i,\hat{\eta}) - b_{S_1^-}(Z_i)'\hat{\beta}_{S_1^-},\\
		e_{i,B^+} &= \psi_{B^+}(W_i,\hat{\eta}) - b_{B^+}(Z_i)'\hat{\beta}_{B^+}, \\
		e_{i,B^-} &= \psi_{B^-}(W_i,\hat{\eta}) - b_{B^-}(Z_i)'\hat{\beta}_{B^-},
\end{align*}}
for $B= L,U$ and
{\footnotesize\begin{align*}
		{B}_{\hat{\omega}}(z) &= \begin{pmatrix}
			b_{L^+}(z)\hat{\omega}_{1,L}(z) & 0 \\
			b_{S_0}(z)\hat{\omega}_{2,L}(z) & 0 \\
			b_{L^-}(z)\hat{\omega}_{1,L}(z) & 0 \\
			b_{S_1}(z)\hat{\omega}_{2,L}(z) & 0 \\
			0 & b_{U^+}(z)\hat{\omega}_{1,U}(z) \\
			0 & b_{S_0}(z)\hat{\omega}_{2,U}(z) \\
			0 & b_{U^-}(z)\hat{\omega}_{1,U}(z) \\
			0 & b_{S_1}(z)\hat{\omega}_{2,U}(z)  
		\end{pmatrix},
\end{align*}}
with estimated weights 
{\footnotesize\begin{align*}
	\hat{\omega}_{1,B} &= \frac{1}{b_{S_0^+}(z)'\hat{\beta}_{S_0^+,0} + b_{S_1^-}(z)'\hat{\beta}_{S_1^-,0}}, \\
	\hat{\omega}_{2,B} &= \frac{b_{B^-}(z)'\hat{\beta}_{B^-,0} + b_{B^+}(z)'\hat{\beta}_{B^+,0}}{(b_{S_1^+}(z)'\hat{\beta}_{S_1^+,0} + b_{S_1^-}(z)'\hat{\beta}_{S_1^-,0})^2}.	\end{align*}}
%{\footnotesize\begin{align*}
%		\hat{\omega}_{1,B^+} &= \frac{1}{b_{S_0}(z)'\hat{\beta}_{S_0}}, \quad 
%		\hat{\omega}_{2,B^+} = \frac{b_{B^+}(z)'\hat{\beta}_{B^+}}{(b_{S_0}(z)'\hat{\beta}_{S_0})^2} \\	
%		\hat{\omega}_{1,B^-} &= \frac{1}{b_{S_1}(z)'\hat{\beta}_{S_1}}, \quad  	
%		\hat{\omega}_{2,B^-} = \frac{b_{B^-}(z)'\hat{\beta}_{B^-}}{(b_{S_1}(z)'\hat{\beta}_{S_1})^2}	\end{align*}}
for $B = L,U$. For the other parameters, formula are analogous. Under strong monotonicity, all minus-superscript terms can be completely omitted and $\psi_B^+ = \psi_{B^+}$ as well as $\psi_{S_0} = \psi_{S_0^+}$ for all observations.
The bootstrapped $\hat{\Omega}_n^b(z)$ is defined equivalently to \eqref{eq_variance_estimated2} with bootstrap weights $h_i$ applied to in each regression first which produce bootstrapped residuals and weights.

\section{Large Sample Properties and Lemmas} 
\subsection{Preliminaries}
First, we introduce notation and establish auxiliary results in this subsection. We then show that our assumptions imply small machine learning bias in Section \ref{app_mlbias} and provide a linearization result in Section \ref{app_linearization}. Section \ref{app_theorem12} verifies a sufficient condition for asymptotic normality for Theorem \ref{thm_AsyNor}, which, together with a matrix convergence, serves as main input for the assumptions required for adaptation of the method by \cite{stoye2020simple} that provides the coverage result in Theorem \ref{thm_Coverage1}. We then consider the strong approximation and bootstrap for Theorem in Section \ref{app_uniform1}. The extension of all results to conditional monotonicity can be found in Section \ref{app_conditional_ext}. 

We use $x \lesssim y$ whenever $x = O(y)$ and $x\lesssim_P y$ when $x = O_p(y)$. Statements about random variables are almost surely if not stated differently. We refer to \cite{semenova2021debiased} and \cite{belloni2015some} as SC and BCCK respectively. 

%\subsection{Auxiliary Results} \label{app_aux1}
\paragraph{Auxiliary $p$-rate:}
By definition $p_0(x) = s(0,x)/s(1,x)$ and equivalently for $\hat{p}(x)$, thus, on $\mathcal{T}_n$, we have that {\footnotesize\begin{align*}
	 \hat{p}(x) - p_0(x) &= \frac{\hat{s}(0,x)}{\hat{s}(1,x)}  - \frac{s(0,x)}{s(1,x)} 
	 = \frac{\hat{s}(0,x) - s(0,x)}{s(1,x)}\frac{s(1,x)}{\hat{s}(0,x)} - \frac{\hat{s}(1,x) - s(1,x)}{s(1,x)}\frac{s(0,x)}{s(1,x)}\frac{s(1,x)}{\hat{s}(1,x)} \\
	 &\lesssim \sup_d|\hat{s}(d,x) - s(d,x)|  
\end{align*}}
almost surely in $\mathcal{X}$ due to A.3 and A.6.
\subsection{Machine Learning Bias and Linearization} \label{app_mlbias}
We now verify that our assumptions are sufficient for Assumption 3.5 in SC and use this to provide a linearization result. %Throughout let $b_B(Z_i) = b_i$ whenever it does not cause confusion. 
Assumption 3.5 in SC requires that for $B \in \{L,U,S_0\}$
\vspace{-4pt}
 {\footnotesize\begin{align*}
	B_n &:= \sqrt{n}\sup_{\eta \in \mathcal{T}_n}||E[b_B(Z_i)(\psi_B(W_i,\eta) - \psi_B(W_i,\eta_0))] || = o(1), \\
	\Lambda_n &:= \sup_{\eta \in \mathcal{T}_n}E[||b_B(Z_i)(\psi_B(W_i,\eta) - \psi_B(W_i,\eta_0))||^2]^{1/2}= o(1). 
\end{align*}}%\vspace{-12pt}
 First consider $B \in \{L,U\}$.
%Define the moment function for the lower bound with bias correction as $
% %{\footnotesize\begin{align*}
%	\psi(W,\eta) = \sum_{j=1}^4 \psi^{[j]}(W,\eta)$ 
%%\end{align*}}
%with 
%
%{\footnotesize\begin{align*}
%	\psi^{[1]}(W,\eta) &:= \frac{D}{e(X)}SY\IQ{p_0} - \frac{1-D}{1-e(X)}SY \\
%	\psi^{[2]}(W,\eta) &:= \Q{p_0}\bigg(\frac{1-D}{1-e(X)}S - s(0,X)\bigg) \\
%	\psi^{[3]}(W,\eta) &:= -\Q{p_0}p_0(X)\bigg(\frac{D}{e(X)}S - s(1,X)\bigg) \\
%	\psi^{[4]}(W,\eta) &:= -\Q{p_0}s(1,X)\bigg(\frac{DS}{e(X)s(1,X)}\IQ{p_0} - p_0(X)\bigg).
%\end{align*}}
For $B_n$ note that, conditional on the cross-fitted model, the following expansion applies for any $\eta \in \mathcal{T}_n$ {\footnotesize\begin{align*}
	\frac{1}{\sqrt{k}}E[b_B(Z_i)\psi_B(W_i,\eta)] &= \frac{1}{\sqrt{k}}E[b_B(Z_i)\psi_B(W_i,\eta_0)] +  \frac{1}{\sqrt{k}}\partial_rE[b_B(Z_i)\psi_B(W_i,\eta_0 + r(\eta - \eta_0))] \\ &\quad + \frac{1}{\sqrt{k}}\partial^2_rE[b_B(Z_i)\psi_B(W_i,\tilde{\eta})], \\
\end{align*}}
where $\partial_r$ denotes the Gateaux-derivative operator and $\tilde{\eta}$ is on the line segment between $\eta_0$ and $\eta$. Define $\bar{Q}:= \sup_{x\in\mathcal{X}}\sup_{u\in\tilde{U}}Q(u,x)$ which is bounded as the boundaries of $\tilde{U}$ are bounded away from $0$ and $1$ due to A.3. This, for any $\eta \in \mathcal{T}_n$, yields 

{\footnotesize\begin{align*}
	E[|\psi_B(W_i,\eta)|~|X_i=x] &\leq \bigg|\frac{1}{\underline{e}}E[|Y|~|X_i=x]\bigg| + \bigg|\frac{1}{1-\bar{e}}E[|Y(1)|~|X_i=x]\bigg| + \bigg|\frac{\bar{Q}}{\underline{e}}\bigg| + |\bar{Q}| \\
	&\quad + \bigg|\frac{\bar{Q}}{\underline{s}}\bigg(\frac{1}{\underline{e}} + \frac{1}{\underline{s}}\bigg)\bigg| + \bigg|\bar{Q}\bigg(\frac{1}{\underline{e}} + \frac{1}{\underline{s}}\bigg)\bigg| \\
	&\lesssim E[|Y_i|~|X_i=x] \\
	&\lesssim 1
\end{align*}}
almost surely in $\mathcal{X}$ due to A.2 and A.3. 
Now note that $b_B(Z_i)$ is $\mathcal{X}$-measurable and thus we can apply the following bound using the Cauchy-Schwarz inequality: {\footnotesize\begin{align*}
	\frac{1}{\sqrt{k}}E[|b_B(Z_i)\psi_B(W,\eta)|] \lesssim \frac{1}{\sqrt{k}}E[||b_B(Z_i)||~|E[|Y_i|~|X_i]|] 
	\leq \frac{E[b_B(Z_i)'b_B(Z_i)]^{1/2}}{\sqrt{k}}E[E[Y_i^2|X_i]]^{1/2} 
	\lesssim 1,
\end{align*}}
where the last inequality comes from A.1 and A.2. This bounds allows us to apply dominated convergence {\footnotesize\begin{align*}
	\frac{1}{\sqrt{k}}\partial_rE[b_B(Z_i)\psi_B(W_i,\eta_0 + r(\eta - \eta_0))] &=\frac{1}{\sqrt{k}}E[b_B(Z_i)\partial_rE[\psi_B(W_i,\eta_0 + r(\eta - \eta_0))|X]] = 0,
\end{align*}}
where the last step follows from Neyman-orthogonality of $\psi_B(W_i,\eta)$ around $\eta_0$, see \cite{semenova2023generalized}, Section 4.2. A uniform bound for the second derivative of $E[\psi_B(W_i,\eta)|X=x]$ has been provided in \cite{semenova2023generalized}, Proof of Lemma A.7. Thus, we obtain  {\footnotesize\begin{align*}
	||\frac{1}{\sqrt{k}}E[b_B(Z_i)\psi_B(W_i,\eta)] - \frac{1}{\sqrt{k}}E[b_B(Z_i)\psi_B(W_i,\eta_0)]||   &\lesssim \frac{1}{\sqrt{k}}E[||b_B(Z_i)||~||\eta - \eta_0||^2].  
\end{align*}}
Together this implies that, as $E[||b_B(z)||] = k_B$, {\footnotesize\begin{align*}
	\sqrt{n}\sup_{\eta \in \mathcal{T}_n}||E[b_B(Z_i)(\psi_B(W_i,\eta) - \psi_B(W_i,\eta_0))] 
	&\lesssim \sqrt{nk_B}(\lambda_{s,n,4}^2 + \lambda_{q,n,4}^2) = o(1) 
\end{align*}}
by the Cauchy-Schwarz and triangle inequality. In the case of a bounded basis similarly {\footnotesize\begin{align*}
\sqrt{n}\sup_{\eta \in \mathcal{T}_n}||E[b_B(Z_i)(\psi_B(W_i,\eta) - \psi_B(W_i,\eta_0))] 
&\lesssim \sqrt{nk_B}(\lambda_{s,n,2}^2 + \lambda_{q,n,2}^2) = o(1)
\end{align*}}
by A.6 and Lemma 6.1 in \cite{chernozhukov2018double}.
 Now consider the variance term $\Lambda_n$. Note that our assumptions imply the conditions in Assumption 3 in Semenova (2023), version 3. Thus, by her Lemma (A.7), Equation (A.36) we have that, for $B \in \{L,U\}$,
 {\footnotesize\begin{align*}
 		\sup_{\eta \in \mathcal{T}_n} E[(\psi_B(W_i,{\eta}) - \psi_B(W_i,\eta_0))^2]^{1/2} \lesssim \lambda_{q,n,1} + \lambda_{s,n,1} + \lambda_{q,n,2} + \lambda_{s,n,2}.
 	\end{align*}}
 Hence, we can bound our variance term analogously by 
 {\footnotesize\begin{align*}
	\sup_{\eta \in \mathcal{T}_n} E[||b_B(Z_i)(\psi_B(W_i,{\eta}) - \psi_B(W_i,\eta_0))||^2]^{1/2} 
	&\leq \xi_{k,L} 	\sup_{\eta \in \mathcal{T}_n} E[(\psi_B(W_i,{\eta}) - \psi_B(W_i,\eta_0))^2]^{1/2} \\
	&\lesssim \xi_{k,L}(\lambda_{q,n,1} + \lambda_{s,n,1} + \lambda_{q,n,2} + \lambda_{s,n,2})  \\
	&= o(1)
\end{align*}}
using A.6. Therefore Assumption 3.5 in SC applies. This together with A.1, A.2, A.4, and A.5 then covers all conditions required for Lemma 3.1(b) in SC. 
Thus, for any $z \in \mathcal{Z}$  and $B \in \{L,U\}$, the estimators are asymptotically linear {\footnotesize\begin{align*}
	\sqrt{n}b_B(z)'(\hat{\beta}_{B} - \beta_{B,0}) = b_B(z)'E[b_B(Z_i)b_B(Z_i)']^{-1}\frac{1}{\sqrt{n}}\sum_{i=1}^n[b_B(Z_i)(\varepsilon_{i,B} + r_{B}(Z_i))] + o_p(1),
\end{align*}}
where the remainder convergence follows from A.5.

For $B = S_0$ and the AIPW moment function $\psi_{S_0}(W_i,\eta)$, note that our assumptions are equivalent to the ones in Assumption 4.10 and 4.11 in SC for the conditional average treatment effect using AIPW moment functions. Thus, their Lemma 3.1 applies directly which yields an equivalent linearization to the one above for $B = S_0$ under the convergence rate requirements of Assumption 4.11 in SC. 

\subsection{Linearization of Single Bound} \label{app_linearization}
We can linearize the difference between estimated and true bounds similarly to $p_0(x)$. For any $B \in \{L,U\}$ note that
{\footnotesize\begin{align*}
	&\frac{b_B(z)'\hat{\beta}_B}{b_{S_0}(z)'\hat{\beta}_{S_0}} - \frac{b_B(z)'{\beta}_{B,0}}{b_{S_0}(z)'{\beta}_{S_0,0}} \\
	&\quad = \frac{b_{S_0}(z)'{\beta}_{S_0,0}}{b_{S_0}(z)'\hat{\beta}_{S_0}}\frac{1}{b_{S_0}(z)'{\beta}_{S_0,0}}b_B(z)'(\hat{\beta}_B - \beta_{B,0}) + \frac{b_{B}(z)'{\beta}_{B,0}}{b_{S_0}(z)'{\beta}_{S_0,0}b_{S_0}(z)'\hat{\beta}_{S_0}}b_{S_0}(z)'(\hat{\beta}_{S_0} - \beta_{S_0,0}) \\
	&\quad =: \hat{\omega}_{1,B}(z)b_B(z)'(\hat{\beta}_B - \beta_{B,0}) + \hat{\omega}_{2,B}(z)b_{S_0}(z)'(\hat{\beta}_{S_0} - \beta_{S_0,0})
\end{align*}}
%where (A.5) implies that uniformly $\hat{\omega}_{1,B} -  \omega_{1,B} \overset{p}{\rightarrow} 0$ and  $\hat{\omega}_{2,B} -\omega_{2,B} \overset{p}{\rightarrow} 0$ with
with estimates weights. We denote the corresponding estimands as $\omega_{1,B}(z) = 1/b_{S_0}(z)'{\beta}_{S_0,0}$,  $\omega_{2,B}(z) = b_{B}(z)'{\beta}_{B,0}/(b_{S_0}(z)'{\beta}_{S_0,0})^2$, and $\omega = (\omega_{1,L}(z), \omega_{2,L}(z),$ $ \omega_{1,U}(z), \omega_{2,U}(z))'$.

\subsection{Proof of Theorem \ref{thm_AsyNor} and Theorem \ref{thm_Coverage1}} \label{app_theorem12}
%We now consider the case of strong monotonicity first to simplify notation. The extension to conditional monotonicity is in Appendix 
\subsubsection{Joint Linearization} \label{sec_App_proof_prelims}
Without loss of generality, instead of A.1, assume basis functions are orthogonalized, i.e.~$E[b_B(Z_i)b_B(Z_i)'] = I_{k_B}$ for $B \in \{L,U,S_0\}$. Define now for $B \in \{L,U\}$ {\footnotesize\begin{align*}
	\hat{\theta}_B(z) = \frac{b_B(z)'\hat{\beta}_B}{b_{S_0}(z)'\hat{\beta}_{S_0}}, \quad 	{\theta}_B^{LP}(z) = \frac{b_B(z)'{\beta}_{B,0}}{b_{S_0}(z)'{\beta}_{S_0,0}}.
\end{align*}}
Stacking the linearization for both bounds then yields
{\footnotesize\begin{align*}
		\sqrt{n}\begin{pmatrix}
			\hat{\theta}_L(z) - {\theta}_L^{LP}(z) \\
			\hat{\theta}_U(z) - {\theta}_U^{LP}(z)
		\end{pmatrix} = B_{\omega}(z)'\frac{1}{\sqrt{n}}\sum_{i=1}^n[b(Z_i)(\varepsilon_i + r_i)] \bigg(1 + O(||\hat{\omega} - \omega ||)\bigg),
\end{align*}}
where {\footnotesize\begin{align*}
		B_{\omega}(z) &= \begin{pmatrix}
			b_L(z)w_{1,L}(z) & 0 \\
			b_{S_0}(z)w_{2,L}(z) & 0 \\
			0 & b_U(z)w_{1,U}(z) \\
			0 & b_{S_0}(z)w_{1,U}(z)  
		\end{pmatrix}_{[(k_L+k_U+2k_{S_0})\times 2]}, \qquad \varepsilon_i = \begin{pmatrix}
		\varepsilon_{i,L} \\ %\omega_{1,L} \\
		\varepsilon_{i,S_0}\\ %\omega_{2,L} \\
		\varepsilon_{i,U}\\ %\omega_{1,U} \\
		\varepsilon_{i,S_0}\\ %\omega_{2,U} \\
		\end{pmatrix}_{[4\times 1]},  \\[1ex]  b(z) &= \begin{pmatrix}
		b_L(z) & 0& 0& 0 \\
		0 & b_{S_0}(z) & 0& 0 \\
		0 & 0& b_U(z)& 0 \\
		0 & 0& 0& b_{S_0}(z)  
		\end{pmatrix}_{[(k_L+k_U+2k_{S_0})\times 4]},  \qquad r_i = \begin{pmatrix}
		r_{L}(Z_i)\\ %\omega_{1,L} \\
		r_{S_0}(Z_i)\\ %\omega_{2,L} \\
		r_{U}(Z_i)\\ %\omega_{1,U} \\
		r_{S_0}(Z_i)\\ %\omega_{2,U} \\
		\end{pmatrix}_{[4\times 1]}. 
\end{align*}}
Define {\footnotesize\begin{align*}
	\Omega(z) = B_{\omega}(z)'E[b(Z_i)(\varepsilon_i + r_i)(\varepsilon_i + r_i)'b(Z_i)']B_{\omega}(z)
\end{align*}}
and  {\footnotesize\begin{align*}
	\sqrt{n}\Omega(z)^{-1/2}B_{\omega}(z)\frac{1}{\sqrt{n}}\sum_{i=1}^n[b(Z_i)(\varepsilon_i + r_i)] = \sum_{i=1}^{n}w_{i,n}(z)(\varepsilon_i + r_i),
\end{align*}}
with {\footnotesize\begin{align*}
	w_{i,n}(z) = \frac{\Omega(z)^{-1/2}}{\sqrt{n}}B_{\omega}(z)'b(Z_i).
\end{align*}}
Note that invertibility of $\Omega(z)$ is guaranteed by Assumption 3.3 as this excludes the case of perfectly correlated conditional mean errors between upper and lower bound. %The remainder in the linearization can be ignored in what follows due to the results in Section \ref{app_linearization}.

\subsubsection{Auxiliary Results}
\paragraph{$(an.i)$ Matrices and Weights}
For simplification, we use notation $B = B_{\omega}(z)$ and $b_i = b(Z_i)$ whenever it does not cause confusion. We make use of the fact that for block matrices the induced $L_2$ norm $||\cdot||$ is given by {\footnotesize\begin{align*}
M = 	\begin{pmatrix}
		M_1 & 0  \\ 0 & M_2
	\end{pmatrix} \Rightarrow ||M|| = \max\{||M_1||,||M_2||\}.
\end{align*}} 
and that {\footnotesize\begin{align*}
	B'B = \begin{pmatrix}
		b_L(z)'b_L(z)w_{1,L}^2(z) + b_{S_0}(z)'b_{S_0}(z)w_{2,L}^2(z) & 0 \\
		0 & 	b_U(z)'b_U(z)w_{1,U}^2(z) +	b_{S_0}(z)'b_{S_0}(z)w_{2,U}^2(z)
	\end{pmatrix},
\end{align*}}
which implies that $||(B'B)^{-1}|| \lesssim 1$ due to the bounded weights. Moreover note that, due to A.2 and A.3, the population predictions are bounded from above and bounded away from zero for $B = S_0$. Thus %$\underline{s} < \inf_{z}b_{S_0}(z)'\beta_{S_0,0} \leq \sup_{z}b_{S_0}(z)'\beta_{S_0,0} < 1 - \underline{s} $ and thus 
$c < \inf_{B}|\omega_{1,B}| + |\omega_{2,B}| < \sup_{B}|\omega_{1,B}| + |\omega_{2,B}| \lesssim 1$ for some $c > 0$. 

\paragraph{$(an.ii)$ Expected norm of weights $w_{i,n}(z)$}
Note that Assumption 3.3 implies that the conditional mean errors have correlation bounded away from one. Thus, their conditional variance-covariance matrix is strictly p.d.~with $\inf_{z\in\mathcal{Z}} ||E[(\varepsilon_i + r_i)(\varepsilon_i + r_i)'|Z_i = z]|| > \lambda_{min} > 0$ and hence $||E[b_i(\varepsilon_i + r_i)(\varepsilon_i + r_i)'b_i']|| \geq ||\lambda_{\min}^{-1}E[b_ib_i']||$.
This implies that

{\footnotesize\begin{align*}
	nE[||w_{i,n}(z)||^2] &\leq ||(B'E[b_i(\varepsilon_i + r_i)(\varepsilon_i + r_i)'b_i']B)^{-1/2}||^2 E[||B'b_i||^2] \\
	&= ||(B'E[b_iE[(\varepsilon_i + r_i)(\varepsilon_i + r_i)'|Z_i]b_i']B)^{-1}|| E[||B'b_i||^2] \\
	&\leq \frac{1}{\lambda_{\min}}||(B'E[b_ib_i']B)^{-1}||\sup_B E[(b_B(z)'b_B(Z_i))^2] \\
	&= \frac{1}{\lambda_{\min}}||(B'B)^{-1}||\sup_B b_B(z)'b_B(z) \\
	&\lesssim 1,
\end{align*}}
as $||b_B(z)|| = 1$ and $E[b_B(Z_i)b_B(Z_i)'] = I_{k_B}$ for $B=L,U,S_0$. 

\paragraph{$(an.iii)$ Norm of weights $w_{i,n}(z)$}
{\footnotesize\begin{align*}
	||w_{i,n}|| & = \frac{1}{\sqrt{n}}||(B'E[b_i(\varepsilon_i + r_i)(\varepsilon_i + r_i)'b_i']B)^{-1/2}B'b_i|| \\
	&\leq \frac{1}{\sqrt{n}}||(B'E[b_iE[(\varepsilon_i + r_i)(\varepsilon_i + r_i)|Z_i]'b_i']B)^{-1/2}||\ ||B||\ ||b_i|| \\
	&\leq \frac{1}{\sqrt{n \lambda_{\min}}}||(B'E[b_ib_i']B)^{-1/2}||\ ||B|| \sup_{z\in\mathcal{Z}} ||B_{\omega}(z)|| \\
	&\leq \sup_{B}\sup_{z\in\mathcal{Z}}\frac{||b_B(z)||}{\sqrt{n}} \\
	&\leq \sup_{B}\frac{\xi_{k,B}}{\sqrt{n}}, \\
\end{align*}}
which is $o(1)$ by A.5. 

\subsubsection{Multivariate Lindeberg Condition} 
In the following denote $\bar\xi = \sup_{B}\xi_{k,B}$ and $\bar{c} = \sup_{B}c_{k,B}(1+\xi_{k,B})$ and note that by A.5 we have that for any constant $C_1,C_2 > 0$ 

{\footnotesize\begin{align*}
	\frac{C_1\sqrt{n}}{\bar{\xi}} - C_2\bar{c} \rightarrow \infty.
\end{align*}}
Now consider the weighted sum of the asymptotically linear representation. First note that by definition of the variance we have that {\footnotesize\begin{align*}
	V\bigg[\sum_{i=1}^nw_{i,n}(z)(\varepsilon_i + r_i)\bigg] = I_2.
\end{align*}}
Now we verify the Lindeberg condition. Note that for any $\delta > 0$,
{\footnotesize\begin{align*}
	\sum_{i=1}^n &E[||w_{i,n}(z)(\varepsilon_i + r_i)||^2\mathbbm{1}(||w_{i,n}(\varepsilon_i + r_i))||> \delta] \\&= n E[||w_{i,n}(z)||^2||\varepsilon_i + r_i||^2\mathbbm{1}(||\varepsilon_i + r_i||> \delta/||w_{i,n}(z)||)]\\
	&\leq 2n E[||w_{i,n}(z)||^2||\varepsilon_i||^2\mathbbm{1}(||\varepsilon_i|| + ||r_i||> \delta/||w_{i,n}(z)||)] \\
	&\qquad + 2n E[||w_{i,n}(z)||^2||r_i||^2\mathbbm{1}(||\varepsilon_i|| + ||r_i||> \delta/||w_{i,n}(z)||)] \\
	&= (E.1) + (E.2).
\end{align*}}
Now note that by the law of iterated expectations for some $C > 0$ {\footnotesize\begin{align*}
	(E.1) &\lesssim 2n E[||w_{i,n}(z)||^2E[||\varepsilon_i||^2\mathbbm{1}(||\varepsilon_i|| + ||r_i||> \delta/||w_{i,n}(z)||)]] \\
	&\leq 2n E[||w_{i,n}(z)||^2\sup_{z\in\mathcal{Z}}E[||\varepsilon_i||^2\mathbbm{1}(||\varepsilon_i|| + ||r_i||> \delta/||w_{i,n}(z)||)|Z_i=z]] \\
	&\leq 2n E[||w_{i,n}(z)||^2\sup_{z\in\mathcal{Z}}E[||\varepsilon_i||^2\mathbbm{1}(||\varepsilon_i|| + 2\bar{\xi}> C\delta/\bar{c})|Z_i=z]] \\
	&= o(1)
\end{align*}}
by $(an.ii)$ and A.5 implying $C\delta \sqrt{n}/\bar{\xi} - 2\bar{c} \rightarrow \infty$ together with the fact that the higher moment in A.2 implies uniform integrability of $||\varepsilon_i||^2$ as the moment function error tails are driven only by the (conditional) distribution of $Y_i$. Moreover,  {\footnotesize\begin{align*}
	(E.2) &\leq E[||w_{i,n}(z)||^22\sup_{B}\sup_{z\in\mathcal{Z}}||r_B(z)||^2E[\mathbbm{1}(||\varepsilon_i|| + ||r_i||> \delta/||w_{i,n}(z)||)|Z_i]] \\
	&\leq nE[||w_{i,n}(z)||^2]2\bar{c}^2\sup_{z\in\mathcal{Z}}P(||\varepsilon_i|| > C\delta\sqrt{n}/\bar{\xi} - 2\bar{c}|Z_i=z) \\
	&\lesssim \bar{c}^2 \frac{2\sup_{B}\sup_{z\in\mathcal{Z}}E[\varepsilon_{i,B}^2|Z_i=z]}{[C\delta\sqrt{n}/\bar{\xi} - 2\bar{c}]^2} \\
	&= o(1), 
\end{align*}}
where the second to last step follows from the bounded weights and Chebyshev's inequality. The divergence of the denominator in the last step is a consequence from A.5. The numerator is again bounded due to A.2. Thus, overall we have that {\footnotesize\begin{align*}
	\lim_{n\rightarrow\infty}	\sum_{i=1}^n &E[||w_{i,n}(z)(\varepsilon_i + r_i)||^2\mathbbm{1}(||w_{i,n}(\varepsilon_i + r_i))||> \delta] = 0,
\end{align*}}
which implies asymptotic normality around the linear predictor. For the case with small approximation error, note that 	{\footnotesize\begin{align*}
	\sqrt{n}&\Omega(z)^{-1/2}\begin{pmatrix}
		\hat{\theta}_L(z) - {\theta}_L^{LP}(z) \\
		\hat{\theta}_U(z) - {\theta}_U^{LP}(z)
	\end{pmatrix} \\
		&= \sqrt{n}\Omega(z)^{-1/2}\begin{pmatrix}
			\hat{\theta}_L(z) - {\theta}_L(z) \\
			\hat{\theta}_U(z) - {\theta}_U(z)
		\end{pmatrix} + \sqrt{n}\Omega(z)^{-1/2}\begin{pmatrix}
			r_{L}(z)w_{1,L} + r_{S_0}(z)w_{2,L}  \\
			r_{U}(z)w_{1,U} + r_{S_0}(z)w_{2,U}
		\end{pmatrix} + o_p(1) \\
		&= \sqrt{n}\Omega(z)^{-1/2}\begin{pmatrix}
			\hat{\theta}_L(z) - {\theta}_L(z) \\
			\hat{\theta}_U(z) - {\theta}_U(z)
		\end{pmatrix} + o_p(1),
\end{align*}}
where the last step follows from the bounded weights and {\footnotesize\begin{align*}
		||\sqrt{n}\Omega(z)^{-1/2}(r_L(z) + r_U(z) + r_{S_0}(z))'|| \leq \sqrt{n}||\Omega(z)^{-1/2}||\sup_{B,z\in\mathcal{Z}}|r_B(z)| 
		\lesssim \sqrt{n}\sup_{B} {k_B}^{-1/2}l_{k,B}c_{k,B},
\end{align*}}
which is $o(1)$ by the additional assumption in Theorem \ref{thm_AsyNor}, Part 2. 

\subsubsection{Matrix Estimation} \label{app_matrixEst2}
Now we consider difference between $\Omega(z)$ and $\hat{\Omega}_n(z)$. Let $a(z) := b(z)/||b(z)||$ and denote	%{\footnotesize\begin{align*}
		$	\xi_{k,B}^L:= \sup_{z,z'\in\mathcal{Z},z\neq z'} {||a(z) - a(z')||}/{||z-z'||} $.
		%	\end{align*}}
Assume the following conditions hold for $B=L,U,S_0$: \begin{ass} \label{ass_AV}
	There exists an $m_B > 2$ such that (i) $\sup_{x\in\mathcal{X}}E[|Y_i|^{m_B}|X_i=x] \lesssim 1$, (ii) $(\xi_{k,B}^L)^{2m/(m-2)}\log k_B/n \lesssim 1$, and (iii) $\xi_{k,B}^L \lesssim \log k_B$. Moreover assume that uniformly over $\mathcal{T}_n$ {\footnotesize\begin{align*}
			\kappa_n^1 &:=	\underset{\eta \in \mathcal{T}_n}{\sup}E[\underset{1\leq i \leq n}{\max}|\psi_B(W_i,\eta) - \psi_B(W_i,\eta_0)|] =  o(n^{-\frac{1}{m_B}}), \\
			\kappa_n &:=	\underset{\eta \in \mathcal{T}_n}{\sup}E[\underset{1\leq i \leq n}{\max}(\psi_B(W_i,\eta) - \psi_B(W_i,\eta_0))^2]^{1/2} =  o(1).
	\end{align*}} 
\end{ass} 
These assumptions imply Assumptions 3.6 - 3.8 in SC and the additional conditions (i) and (ii) in their Theorem 3.3. Thus, pointwise consistency of $\hat{\Omega}_n(z)$ follows directly for any $z \in \mathcal{Z}$. 
Note that in their Assumption 3.8 they first assume the weaker $\kappa_n^1 = o(1)$ for the asymptotic normality with true covariance. For estimating the latter consistently, however, they also require that $n^{1/m}\kappa_n^1 = o(1)$, see Assumption (ii) in their Theorem 3.3. Thus, there is no qualitative difference to the assumptions imposed here, see also \cite{heiler2021effect} for a similar argument in a modified setup.
\subsubsection{Proof of Thereom \ref{thm_Coverage1}}
For each $z$, the asymptotic distribution in Theorem \ref{thm_AsyNor} matches the assumptions behind \cite{stoye2020simple}, Theorem 1 (his Assumption 1) with adapted parameter spaces as in the Theorem \ref{thm_Coverage1}. As all statements are pointwise in $z$, they follow directly. 

\subsection{Proof of Theorem \ref{thm_strongApprox1} and Theorem \ref{thm_uniformInf1}} \label{app_uniform1}
First, we introduce some auxiliary definitions. Second, we provide a joint uniform linearization result for the series process for the bounds and its remainder. Third, we derive a strong approximation result for the whole series process under correct specification. Fourth, we then provide an equivalent result for the weighted bootstrap up to some logarithmic terms. At last, we prove the uniform asymptotic size control of the suggested confidence bands for the treatment effect under the correct specification.
\subsubsection{Preliminaries}
Let	$\hat{Q}_B =\frac{1}{n}\sum_{i=1}^nb_B(Z_i)b_B(Z_i)'$ and define matrix 
{\footnotesize\begin{align*}
	\hat{Q} := \begin{pmatrix}
		\hat{Q}_L & 0 & 0 &0 \\
		0&\hat{Q}_{S_0} & 0 & 0 \\
		0&0 & \hat{Q}_U & 0\\
		0& 0& 0 & \hat{Q}_{S_0}
	\end{pmatrix} =   \frac{1}{n}\sum_{i=1}^nb_ib_i' 
\end{align*}}
and quantities {\footnotesize\begin{align*}
	{\theta}_{LU}^{}(z) = \begin{pmatrix}
	{\theta}_{L}^{}(z) \\
	{\theta}_{U}^{}(z) 
\end{pmatrix}, \quad	{\theta}_{LU}^{LP}(z) = \begin{pmatrix}
		{\theta}_{L}^{LP}(z) \\
		{\theta}_{U}^{LP}(z) 
	\end{pmatrix}, \quad \hat{\theta}_{LU}(z) = \begin{pmatrix}
		\hat{\theta}_L(z) \\
		\hat{\theta}_U(z) 
	\end{pmatrix}, \quad \Sigma(z) = E[b_i\varepsilon_i\varepsilon_i'b_i'].
\end{align*}}

\subsubsection{Uniform Linearization and Remainder}
The uniform linearization follows exactly the same steps as the Proof of Lemma 4.2 in BCCK. Due to the multi-dimensional problem, however, we have to use the following modified bounds
{\footnotesize\begin{align*}
	\max_{1\leq i\leq n}||\varepsilon_i|| &\lesssim_P E[\max_{1\leq i\leq n}|\varepsilon_{i,L}| + |\varepsilon_{i,U}| + |\varepsilon_{i,S_0}|] \lesssim n^{1/m_L} + n^{1/m_U} + n^{1/m_{S_0}} \lesssim \sup_Bn^{1/m_{B}},  \\
	||\hat{Q} - I_{[k_L + k_U + 2k_{S_0}]}|| &\lesssim_P \sup_B ||\hat{Q}_B - I_{k_B}|| \lesssim_P \sup_B \sqrt{\frac{k_B^2 \log k_B}{n}}, \\
	||\hat{Q}|| &\lesssim_P \sup_B||\hat{Q}_B|| \lesssim_P 1 \\
	||B_{\omega}(z) - B_{\omega}(\tilde{z})|| &\lesssim \sup_B ||b_B(z) - b_B(\tilde{z})|| \leq \sup_B \xi_{k,B}^L||z - \tilde{z}||, \\
	||G_n[b_ir_i]|| &\lesssim_P \sup_B ||G_n[b_B(Z_i)r_{B}(Z_i)]|| \lesssim_P \sup_B l_{k,B}c_{k,B}\sqrt{k_B}
\end{align*}}
by the block-diagonality and definition of the underlying components, Assumption A.V and the proof of the previous theorems. 
We decompose {\footnotesize\begin{align*}
	\sqrt{n}(\hat{\theta}_{LU}(z) - \theta_{LU}^{LP}(z)) &= B_{\omega}(z)'G_n[b_i\varepsilon_i] + B_{\omega}(z)'G_n[b_ir_i] + B_{\omega}(z)'[\hat{Q}^{-1} - I]G_n[b_i(\varepsilon_i + r_i)] + o_p(1)
\end{align*}}
uniformly over $z$. Using the modified bounds above within the Proof of Lemma 4.2 in BCCK, Step 1 and Step 2 yields
{\footnotesize\begin{align*}
	\sup_{z\in\mathcal{Z}} B_{\omega}(z)'[\hat{Q}^{-1} - I]G_n[b_i(\varepsilon_i + r_i)] \lesssim_P \sup_B \sqrt{\frac{\xi_{k,B}^2\log k_B}{n}}\bigg(n^{1/m_B}\log^{1/2}k_B \sqrt{k_B}l_{k,B}c_{k,B} \bigg),
\end{align*}}
while Step 3 implies
{\footnotesize\begin{align*}
	\sup_{z\in\mathcal{Z}} ||B_{\omega}(z)'G_n[b_ir_i]|| \lesssim_P \sup_B l_{k,B}c_{k,B}\log^{1/2}k_B.
\end{align*}}
Plugging this back into the decomposition, we obtain a uniform linearization 
{\footnotesize\begin{align*}
	\sqrt{n}(\hat{\theta}_{LU}(z) - \theta^{LP}_{LU}(z)) &= B_{\omega}(z)'G_n[b_i(\varepsilon_i + r_i)] + R_{1n}(z) \\
	&= B_{\omega}(z)'G_n[b_i\varepsilon_i] + R_{2n}(z) + R_{1n}(z)
\end{align*}}
with remainder {\footnotesize\begin{align*}
	\sup_{z\in\mathcal{Z}}|| R_{1n}(z)|| &\lesssim_P \sup_B \sqrt{\frac{\xi_{k,B}^2\log k}{n}}\bigg(n^{1/m_B}\log^{1/2}k_B \sqrt{k_B}l_{k,B}c_{k,B} \bigg) = o(a_n^{-1}), \\
	\sup_{z\in\mathcal{Z}} ||R_{2n}(z)|| &\lesssim_P \sup_B l_{k,B}c_{k,B}\log^{1/2}k_B = o(a_n^{-1})
\end{align*}}
under the conditions of Theorem \ref{thm_strongApprox1}.

\subsubsection{Strong Approximation}
Now we apply Yurinskii's Coupling to the first-order process from above when $\sup_{z\in\mathcal{Z}} ||R_{2n}(z)|| = o_p(a_n^{-1})$. Consider the following copy of the first-order term above {\footnotesize\begin{align*}
	\frac{1}{\sqrt{n}}\sum_{i=1}^n\zeta_i \text{ with } \zeta_i = \Sigma(z)^{-1/2}b_i\varepsilon_i.
\end{align*}} 
Note that $E[\zeta_i] = 0$ and $V[\zeta_i] = I_{k_L+k_U+2k_{S_0}}$ by definition. To apply the coupling, we need have to bound the (sum of the) third moments $\sum_{i=1}^n E[||\zeta_i||^3]$. Note that {\footnotesize\begin{align*}
	E[||\zeta_i||^3] &\leq \frac{1}{\lambda_{min}(\Sigma(z))^{3/2}}E[||b_i\varepsilon_i||^3]\\
	&\lesssim E[\sup_B||b_B(Z_i)||^3\sup_{z\in\mathcal{Z}}E[|\varepsilon_{i,B}|^3|Z_i=z]] \\
	&\lesssim \sup_B \xi_{k,B} E[||b_B(Z_i)||^2] \\
	&\lesssim \sup_B \xi_{k,B}k_B,
\end{align*}}
where the bounded eigenvalues follows from Assumption A.1 and Assumption 3.3 that ensures invertibility. The third inequality follows from the conditional moment bound in Theorem \ref{thm_strongApprox1}. Thus, Yurinskii's coupling says implies that, for each $\delta > 0$, there exist a random standard normal vector of length $k_L + k_B + 2k_{S_0}$, $\mathcal{N}_{[k_L+k_u+ 2k_{S_0}]}$, such that {\footnotesize\begin{align*}
	P\bigg(\bigg|\bigg|\frac{1}{\sqrt{n}}\sum_{i=1}^n\zeta_i - &\mathcal{N}_{[k_L+k_u+ 2k_{S_0}]} \bigg|\bigg| \geq 3\delta a_n^{-1}\bigg) \\
	%	&\lesssim \Delta [k_L+k_U]\delta^{-3}\bigg(1+ \frac{|\log(1/\Delta[k_L+k_B]\delta^{-3})}{[k_L+k_B]}\bigg) \\
	&\lesssim \frac{n[k_L+k_U+ 2k_{S_0}]\sup_B k_b \xi_{k,B} a_n^3}{(\delta \sqrt{n})^3}\bigg(1 + \frac{|\log(\delta^3/\sup_B\xi_{k,B}k_B[k_L+k_U+ 2k_{S_0}])|}{[k_L + k_U+ 2k_{S_0}]}\bigg) \\
	&\lesssim n^{-1/2}a_n^3[k_L + k_U+ 2k_{S_0}]\sup_B k_b \xi_{k,B}\bigg(1 + \frac{\log n}{[k_L + k_U+ 2k_{S_0}]}\bigg) \\
	&\lesssim n^{-1/2}a_n^3\sup_B k_B^2 \xi_{k,B}\bigg(1 + \frac{\log n}{k_B}\bigg),
\end{align*}}
where the second inequality follows from $\delta$ being fixed and $\sup_B\xi_{k,B}k_B^2/n = o(1)$. Now note that, under correct specification $\sup_{B}n^{1/2}k_{B}l_{k,B}c_{k,B}\log^2n = o(a_n^{-1})$, we have that {\footnotesize\begin{align*}
	\sqrt{n}\Omega(z)^{-1/2}(\hat{\theta}_{LU}(z) - \theta_{LU}^{LP}(z)) - \sqrt{n}\Omega(z)^{-1/2}(\hat{\theta}_{LU}(z) - \theta_{LU}(z)) = o_p(a_n^{-1}), 
\end{align*}}
where $\Omega(z) = B_{\omega}(z)'\Sigma(z)B_{\omega}(z)$. Thus, overall we have that, on $\ell^{\infty}(\mathcal{Z})$, {\footnotesize\begin{align*}
	\sqrt{n}\Omega(z)^{-1/2}(\hat{\theta}_{LU}(z) - \theta_{LU}(z)) =_d  \Omega(z)^{1/2}\mathcal{N}_{[k_L + k_u + 2k_{S_0}]} + o_p(a_n^{-1}).
\end{align*}} %\vspace{-18pt}
\subsubsection{Bootstrap: Strong Approximation}
Define the bootstrap estimator $\hat{\beta}^b_B$ analogously to first least squares problem \eqref{eq_auxReg1} but with random independent exponential weights for $B = L,U,S_0$ as {\footnotesize\begin{align*}
	\hat{\beta}_B^b = \arg\underset{\beta}{\min}~E_n[h_i(\psi_B(W_i,\hat{\eta}) - b_B(Z_i)'\beta)^2].
\end{align*}}
For $B=L,U$ and any $b$ we then obtain estimator {\footnotesize\begin{align*}
	\hat{\theta}^b_B(z) = \frac{b_B(z)'\hat{\beta}_B^b}{b_{S_0}(z)'\hat{\beta}_{S_0}^b}.
\end{align*}}
Note that the weights $h_1,\dots,h_n$ have to be the same for both bounds and the conditional selection probability regressions. Else, it would alter the dependency between estimators. Note that $E[(h_i - 1)] = 0$, $E[h_i^{m_B/2}] \lesssim 1$, $\max_{1\leq i\leq n}h_i \lesssim_P \log n$, and $E[\max_{1\leq i\leq n}|\sqrt{h_i}\varepsilon_{i,B}|] \lesssim_P n^{1/m_B}\log n$ by the properties of the standard exponential. The least squares problem above can be considered as the regression \eqref{eq_auxReg1} with both sides multiplied by $\sqrt{h}_i$. Thus, all results from the previous sections apply with $\xi_{k,B}^b = \xi_{k,B} \log^{1/2}n$ and $l^b_{k,B}c^b_{k,B} = l_{k,B}c_{k,B}\log^{1/2}n$, see also Proof of Theorem 4.5 in BCCK. 
Adding and subtracting then yields the following centered decomposition {\footnotesize\begin{align*}
	\sqrt{n}(\hat{\theta}_{LU}^b(z) -	\hat{\theta}_{LU}(z)) &= \sqrt{n}(	\hat{\theta}_{LU}^b(z) - {\theta}_{LU}(z)) -  \sqrt{n}(\hat{\theta}_{LU}(z) -{\theta}_{LU}(z)) \\
	&= B_{\omega}(z)G_n[b_i\varepsilon_i(h_i -1)] + R_{n1}^b(z) + R_{n2}^b(z),
\end{align*}}
with {\footnotesize\begin{align*}
	\sup_{z\in\mathcal{Z}}||R_{n1}^b(z)|| &\lesssim_P \sup_B \sqrt{\frac{\xi_k^2\log^3 n}{n}}(n^{1/m_B}\log^{1/2}n + \sqrt{k_B}l_{k,B}c_{k,B}) = o(a_n^{-1}), \\
	\sup_{z\in\mathcal{Z}}||R_{n2}^b(z)|| &\lesssim_P \sup_B l_{k,B}c_{k,B}\log n = o(a_n^{-1}). 
\end{align*}}
Then, due to existence of the higher moments of the (weighted) process, Yurinkii's coupling can be applied analogously. Thus, on $\ell^{\infty}(\mathcal{Z})$, {\footnotesize\begin{align*}
	\sqrt{n}\Omega(z)^{-1/2}(\hat{\theta}_{LU}^b(z) - \theta_{LU}(z)) =_d  \Omega(z)^{1/2}\mathcal{N}_{[k_L + k_u + 2k_{S_0}]} + o_p(a_n^{-1}),
\end{align*}}
which gives the desired result in Theorem \ref{thm_strongApprox1} by adding and subtracting the estimated parameter vector $\hat{\theta}_{LU}(z)$ and using the triangle inequality. 

\subsubsection{Uniform Confidence Band Coverage}
Under correct specification, uniformly over $\mathcal{Z}$, $\theta_L(z) \leq \theta(z) \leq \theta_U(z)$ and thus
{\footnotesize\begin{align*}
	P\bigg(\theta(z) &\in CB_{\theta(z),1-\alpha}\ \text{ for all } z \in \mathcal{Z}\bigg) \\
	&\geq P\bigg(\sup_{z\in\mathcal{Z}}\theta(z) \leq c_{n,1-\alpha/2}(\bar{t}^b_U)\hat{\sigma}_U(z)\bigg) - P\bigg(\inf_{z\in\mathcal{Z}}\theta(z) \leq c_{n,\alpha/2}(\underline{t}^b_L)\hat{\sigma}_L(z)\bigg) \\
	&\geq P\bigg(\sup_{z\in\mathcal{Z}}\theta_U(z) \leq c_{n,1-\alpha/2}(\bar{t}^b_U){\sigma}_U(z)\bigg) - P\bigg(\inf_{z\in\mathcal{Z}}\theta_L(z) \leq c_{n,\alpha/2}(\underline{t}^b_L){\sigma}_L(z)\bigg) + o(1)\\
	&\geq 1 - \alpha/2 - \alpha/2 + o(1),
\end{align*}}
where the second inequality is due to the correct specification and consistency of the asymptotic variance. The third inequality follows from the strong approximation result for the centered bootstrap process.

\subsection{Extension of Results to Conditional Monotonicity} \label{app_conditional_ext}
For the $\theta_{AT}(z)$ bounds, estimators and estimands under conditional monotonicity are given by {\footnotesize\begin{align*}
	\hat{\theta}_B(z) = \frac{b_{B^+}(z)'\hat{\beta}_{B^+} + b_{B^-}(z)'\hat{\beta}_{B^-}}{b_{S_0^+}(z)'\hat{\beta}_{S_0^+} + b_{S_1^-}(z)'\hat{\beta}_{S_1^-}}, \quad 
	{\theta}_B^{LP}(z) = \frac{b_{B^+}(z)'{\beta}_{B^+,0} + b_{B^-}(z)'{\beta}_{B^-,0}}{b_{S_0^+}(z)'{\beta}_{S_0^+,0} + b_{S_1^-}(z)'{\beta}_{S_1^-,0}}. 
\end{align*}}
Following the linearization as in Appendix \ref{app_linearization} twice, we obtain {\footnotesize\begin{align*}
	\hat{\theta}_B(z)- {\theta}_B^{LP}(z) 
	&= \hat{\omega}_{1,B}b_{B^+}(z)'(\hat{\beta}_{B^+} - \beta_{B^+,0}) + \hat{\omega}_{1,B}b_{B^-}(z)'(\hat{\beta}_{B^-} - \beta_{B^-,0}) \\
	&\quad + \hat{\omega}_{2,B}b_{S_0}(z)'(\hat{\beta}_{S_0} - \beta_{S_0,0}) + 
	\hat{\omega}_{1,B}b_{S_1}(z)'(\hat{\beta}_{S_1} - \beta_{S_1,0}),
\end{align*}}
where {\footnotesize\begin{align*}
	\omega_{1,B} &= \frac{1}{b_{S_0^+}(z)'\beta_{S_0^+,0} + b_{S_1^-}(z)'\beta_{S_1^-,0}} \\
	\omega_{2,B} &= \frac{b_{B^-}(z)'\beta_{B^-,0} + b_{B^+}(z)'\beta_{B^+,0}}{(b_{S_1^+}(z)'\beta_{S_1^+,0} + b_{S_1^-}(z)'\beta_{S_1^-,0})^2}	\end{align*}}
and equivalently for each estimated $\omega$ with $\beta$ replaced by their sample analogues. Therefore stacking everything yields the following linearization for both bounds {\footnotesize\begin{align*}
	\sqrt{n}\begin{pmatrix}
		\hat{\theta}_L(z)- {\theta}_L^{LP}(z) \\
		\hat{\theta}_U(z)- {\theta}_U^{LP}(z) 
	\end{pmatrix} = B_{\omega}(z)'\frac{1}{\sqrt{n}}\sum_{i=1}^nb(Z_i)(\varepsilon_i + r_i)\bigg(1 + O(||\hat{\omega} - \omega||)\bigg),
\end{align*}}
with {\footnotesize\begin{align*}
		B_{\omega}(z) &= \begin{pmatrix}
			b_{L^+}(z)\omega_{1,L}(z) & 0 \\
			b_{S_0^+}(z)\omega_{2,L}(z) & 0 \\
			b_{L^-}(z)\omega_{1,L}(z) & 0 \\
			b_{S_1^-}(z)\omega_{2,L}(z) & 0 \\
			0 & b_{U^+}(z)\omega_{1,U}(z) \\
			0 & b_{S_0^+}(z)\omega_{2,U}(z) \\
			0 & b_{U^-}(z)\omega_{1,U}(z) \\
			0 & b_{S_1^-}(z)\omega_{2,U}(z)  
		\end{pmatrix}_{[k^*\times 2]}, \qquad \varepsilon_i = \begin{pmatrix}
			\varepsilon_{i,L^+} \\ %\omega_{1,L^+} \\
			\varepsilon_{i,S_0^+}\\ %\omega_{2,L}^+ \\
			\varepsilon_{i,L^-}\\ %\omega_{1,L^-} \\
			\varepsilon_{i,S_1^-}\\ %\omega_{2,L}^- \\
			\varepsilon_{i,U^+}\\ %\omega_{1,U^+} \\
			\varepsilon_{i,S_0^+}\\ %\omega_{2,U}^+ \\
			\varepsilon_{i,U^-}\\ %\omega_{1,U^-} \\
			\varepsilon_{i,S_1^-}\\ %\omega_{2,U}^- \\
		\end{pmatrix}_{[8\times 1]},  \qquad r_i = \begin{pmatrix}
		r_{L^+}(Z_i)\\ %\omega_{1,L^+} \\
		r_{S_0^+}(Z_i)\\ %\omega_{2,L}^+ \\
		r_{L^-}(Z_i)\\ %\omega_{1,L^-} \\
		r_{S_1^-}(Z_i)\\ %\omega_{2,L}^- \\
		r_{L^+}(Z_i)\\ %\omega_{1,U^+} \\
		r_{S_0^+}(Z_i)\\ %\omega_{2,U}^+ \\
		r_{L^-}(Z_i)\\ %\omega_{1,U^-} \\
		r_{S_1^-}(Z_i)\\ %\omega_{2,U}^- \\
		\end{pmatrix}_{[8\times 1]},\end{align*} \begin{align*}
		b(z) = \begin{pmatrix}
			b_{L^+}(z) & 0& 0& 0  & 0& 0& 0 & 0\\
			0 & b_{S_0^+}(z) & 0& 0  & 0& 0& 0 & 0 \\
			0 & 0 & b_{L^-}(z) & 0  & 0& 0& 0 & 0 \\
			0 & 0 & 0 & b_{S_1^-}(z) & 0& 0& 0 & 0 \\
			0 & 0 & 0 & 0 & b_{U^+}(z)& 0& 0 & 0 \\
			0 & 0 & 0 & 0& 0& b_{S_0^+}(z)& 0 & 0 \\
			0 & 0 & 0 & 0& 0& 0 &b_{U^-}(z)& 0 \\
			0 & 0 & 0 & 0& 0& 0 & 0 &  b_{S_1^-}(z)  
		\end{pmatrix}_{[k^*\times 8]}, 
\end{align*}}
where $k^* = (k_L^++k_U^++k_L^- +k_U^- + 2(k_{S_0^+} + k_{S_1^+}))$.
%\begin{align*}
%	w_{1,B}^+ &= \frac{1}{b_{S_0}(z)'}
%\end{align*}}
%Define {\footnotesize\begin{align*}
%		\Omega(z) = B(z)'E[b(Z_i)(\varepsilon_i + r_i)(\varepsilon_i + r_i)'b(Z_i)']B(z)
%\end{align*}}
For the proof of Theorem \ref{thm_strongApprox1} and Theorem \ref{thm_uniformInf1} we can assume that the basis functions are separately normalized (wlog), i.e.~$E[b_{B^+}(Z_i)b_{B^+}(Z_i)'] = I_{k_{B^+}}$, $E[b_{B^-}(Z_i)b_{B^-}(Z_i)'] = I_{k_{B^-}}$, and equivalently for the other bases as before. 
For any signed $B^+$ and $B^-$ bounds we now also have an indicator function with estimated threshold inside. Consider the $B^+$ terms. We have to additionally show that {\footnotesize\begin{align*}
	B_{n,2} &:= \sup_{\eta \in \mathcal{T}_n}\sqrt{n}||E[b_i\psi(W_i,\eta_0)(\mathbbm{1}(\hat{p}_0(X_i) < 1) - \mathbbm{1}({p}_0(X_i) < 1))]|| = o(1), \\
	\Lambda_{n,2} &:= \sup_{\eta \in \mathcal{T}_n}E[||b_i\psi(W_i,\eta_0)(\mathbbm{1}(\hat{p}_0(X_i) < 1) - \mathbbm{1}({p}_0(X_i) < 1))||^2]^{1/2} = o(1).
\end{align*}}
The process with corresponding cross terms  $(\psi(W_i,\hat{\eta}) - \psi(W_i,\eta_0))(\mathbbm{1}(\hat{p}_0(X_i) < 1) - \mathbbm{1}({p}_0(X_i) < 1))$ will be dominated by these plus the bounds derived in Appendix \ref{app_mlbias}.
First note that, by Assumption A.2 and A.3, we have bounded conditional moments $\sup_{x\in\mathcal{X}}E[\psi(W_i,\eta_0)^{m_B}|X_i=x] \lesssim 1$. Now consider the term $E[|(\mathbbm{1}(\hat{p}_0(X_i) < 1) - \mathbbm{1}({p}_0(X_i) < 1))|]$. This can be interpreted as the excess classification risk of a Bayes plug-in classifier for the sign of a random variable that uses $\hat{p}_0(x) - 1$ instead of $p_0(x) - 1$. The risk of such plug-in classifiers has been studied in \cite{audibert2007fast}. Note that our Assumption 3.3 (Margin) corresponds to \cite{audibert2007fast}, Assumption (MA) with $\alpha = \infty$ (see their equation 3.9). Thus, we can achieve a fast rate for the classification error using \cite{audibert2007fast}, Lemma 3.6. Formally, for the given $c > 0$ in Assumption 2.3, we have that 
	{\footnotesize\begin{align*}
	E[|(\mathbbm{1}(\hat{p}_0(X_i) < 1) - \mathbbm{1}({p}_0(X_i) < 1))|]
	\leq P(|\hat{p}_0(X_i) - {p}_0(X_i)| > c) 
	\lesssim \lambda_{s,n,2}^2
\end{align*}}
by Chebyshev's inequality. Thus we have that {\footnotesize\begin{align*}
	B_{n,2} &\leq \sqrt{n}E[||b_i||~|E[\psi(W_i,\eta_0)|X=x]|~|(\mathbbm{1}(\hat{p}_0(X_i) < 1) - \mathbbm{1}({p}_0(X_i) < 1))|] \\
	&\lesssim \sqrt{n}\xi_{k,B}E[|(\mathbbm{1}(\hat{p}_0(X_i) < 1) - \mathbbm{1}({p}_0(X_i) < 1))|] \\
	&\lesssim \sqrt{n\xi_{k,B}^2}\lambda_{s,n,2}^2,
\end{align*}}
where the last step follows from the $p_0(x)$ linearization. Similarly, we obtain {\footnotesize\begin{align*}
	\Lambda_{n,2} &\lesssim \xi_{k,B}E[|(\mathbbm{1}(\hat{p}_0(X_i) < 1) - \mathbbm{1}({p}_0(X_i) < 1))|]^{1/2} 
	\lesssim \xi_{k,B}\lambda_{s,n,2}.
\end{align*}}
For the bounds, all theorems then follow exactly analogously to Appendix \ref{app_theorem12} and \ref{app_uniform1} by invoking Assumption A.1 to A.6 for both regressions functions \ref{eq_bounds_ATplmin1} and taking suprema not just over $B$ and $S_0$ but over $B$ for both the ``+'' and ``-'' models in \eqref{eq_auxReg1} and $S_0^+$ and $S_1^+$ respectively. The proof for $\theta_{AT}^+(z)$ and $\theta_{AT}^-(z)$ follows analogously with modified moments and bases with $k^*$ according to Table \ref{tab_momentsAll1} and Table \ref{tab_k_all}.

\section{Misspecification Framework and Examples} \label{sec_misspecFnE}
In this section, we provide a more explicit discussion of misspecification using a moment inequality framework as well as two specific examples. In particular, we consider (i) global misspecification of the conditional moment or signal arising from misspecified nuisance functions and (ii) local misspecification of the heterogeneous effect bounds themselves. In practice, of course, these phenomena can occur simultaneously. First, we review the minimal relaxation framework by \cite{andrews2019inference} and compare it to the heterogeneity setup considered in this paper. In essence, our formalization is a conditional version where the target parameter is nonparametric/semiparametric instead of a finite dimensional vector. \cite{andrews2019inference} consider minimally relaxed moment inequalities such that the identified set is non-empty under correct specification and misspecification. For a generic $k$-dimensional moment function $m(\cdot)$, they define {\footnotesize\begin{align}
	\delta(\theta) &:= \inf\{\delta \geq 0: E[m(W_i,\theta)] \geq -\delta 1_k \}, \notag \\
	\delta^{\inf} &:= \inf_{\theta \in \Theta} \delta(\theta).
\end{align} }
The corresponding misspecification robust identified set is then given by {\footnotesize\begin{align}
	\Theta_I := \{\theta \in \Theta: E[m(W_i,\theta)] \geq -\delta^{\inf}1_k\}.
\end{align}}
In our framework, the moment condition behind auxiliary regression model \eqref{eq_auxReg1}, can be rephrased as conditional versions at $Z_i = z$. In particular, for all $z\in\mathcal{Z}$, we consider pointwise deviations {\footnotesize\begin{align}
	\delta(\theta(z)) &:= \inf\{\delta \geq 0: E[m(W_i,\theta(Z_i))|Z_i=z] \geq -\delta 1_k \}, \notag \\
	\delta_z^{\inf} &:= \inf_{\theta(z) \in \Theta(z)} \delta(\theta(z)),
\end{align}}
and corresponding misspecification robust identified sets {\footnotesize\begin{align}
	\Theta_I(z) := \{\theta(z) \in \Theta(z): E[m(W_i,\theta(Z_i))|Z_i=z] \geq -\delta_z^{\inf}1_k\}.
\end{align} }

For simplicity, we restrict the basis function to be non-negative $b_B(z)\geq 0$ in what follows. This is without loss of generality and all derivations apply with reversed signs as well. Consider case (i): Under correct specification in the heterogeneity step, $\theta_B(z) = b_B(z)'\beta_{B,0}$ or $\psi_B(W_i,\eta_0) = \theta_B(Z_i) + \varepsilon_{i,B}$. The moment function for the bound with estimated and potentially misspecified nuisance parameters then yields {\footnotesize\begin{align}
	E[b_B(Z_i)(\psi_B(W_i,\hat{\eta}) - \theta_B(Z_i))|Z_i=z] = 0 
\end{align}}
for all $z\in\mathcal{Z}$. Neyman-orthogonality of the upper bound signal $\psi_U(W_i,\eta)$ then yields {\footnotesize\begin{align}
	E[b_U(Z_i)&(\psi_U(W_i,\hat{\eta}) - \psi_U(W_i,{\eta_0}) + \varepsilon_{i,U})|Z_i=z] \notag \\
	&= b_U(z)E[(\psi_U(W_i,\hat{\eta}) - \psi_U(W_i,{\eta_0}))|Z_i=z] \notag \\
	&= b_U(z)\partial^2_rE[\psi_U(W_i,\eta_0 + r(\tilde{\eta}-\eta_0))|Z_i=z],
\end{align}}
where $\tilde{\eta}$ is on the line-segment between $\eta_0$ and $\hat{\eta}$. Thus, overall we have that {\footnotesize\begin{align}
	E[b_U(Z_i)(\psi_U(W_i,\hat{\eta}) - \theta_U(Z_i))|Z_i=z] = b_U(z)g_{r}(z) =: -\delta_U(\theta(z)), \label{eq_deltamiss_case1}
\end{align}}
where $\sup_{r\in[0,1)}|g_{r}(z)| \lesssim_P \lambda_{q,n,2}^2 + \lambda_{p,n,2}^2$, see Proof of Theorem \ref{thm_AsyNor}. Now note that, for an upper bound $\theta_U(z)$, inference is potentially invalid if there is underestimation. Under overestimation, confidence intervals based on misspecified moments are still valid albeit more conservative. Thus, $\delta_U(\theta(z))$ can be treated as a positive function in what follows. As $\theta_U(z) \geq \theta(z)$ it also follows that {\footnotesize\begin{align}
	E[b_U(Z_i)(\psi_U(W_i,\hat{\eta}) - \theta(Z_i))|Z_i=z] %\notag \\ %&=	 E[b_U(Z)(\psi_U(W,\hat{\eta}) - \theta_U(Z))|Z=z] - E[b_U(Z)(\theta_U(Z) - \theta(Z))|Z=z] \notag  \\
	&\geq E[b_U(Z_i)(\psi_U(W_i,\hat{\eta}) - \theta_U(Z_i))|Z_i=z].
\end{align}}
Applying the same with reverted signs and ordering for the bounds then yields the following moment inequalities for the effect parameter {\footnotesize\begin{align}
	E[b_U(Z_i)(\psi_U(W_i,\hat{\eta}) - \theta(Z_i))|Z_i=z] \geq -\delta_U(\theta(z)), \notag \\
	E[b_L(Z_i)(\theta(Z_i) - \psi_L(W_i,\hat{\eta}))|Z_i=z] \geq -\delta_L(\theta(z)), 
\end{align}}
where $\delta_B(\theta(z))$ are (vectors of) positive functions. This exactly corresponds to the conditional relaxed moment inequality framework outlined above with $k= k_U + k_L$ and $\delta_z^{\inf} = \sup_B \inf_{\theta(z) \in \Theta(z)} \delta_{B}(\theta(z))$.
Under correct specification and convergence of the nuisance functions as in Assumption $A.6$, naturally $\delta_B(\theta(z)) = o(1)$, so the uniform deviation would be $\delta_z^{\inf} = 0$. However, without correct specification, $\delta_z^{\inf} > 0$ can occur if we overestimate the lower bound and/or underestimate the upper bound. The suggested confidence interval with the pseudo-parameterization is then robust to such deviations from Assumption $A.6$. 

This formalization also provides a bridge between a population misspecification and a finite sample misspecification. In particular, $\delta_z^{\inf} > 0$ arguably provides a better approximation to cases where, asymptotically, there might be a correct specification but, for a given $n$, the $L_p$ errors are relatively large. In finite samples, point estimates after orthogonalization can also be reversed simply because they are not population means. This is also more likely to occur if nuisance parameters are estimated imprecisely.

Now consider case (ii): Local misspecification in the heterogeneity step but correct specification of the nuisance functions. This is the linear predictor case. If one or both are misspecified, $b_B(z)'\beta_{B,0} \neq \theta_B(z)$ for some $B \in \{L,U\}$, then there is no guarantee that these linear predictors do not intersect or reverse ordering for some $z$. 
%Figure [REF] depicts a simple example where the upper bound is increasing quadratically in $z$ while the lower bound is constant (both dashed-black).
%
%\section{Refereepoints}
%\begin{figure}[!h]
%	\centering \caption{Best Linear Predictor: Example} %\label{fig_boundsAGE}
%	\includegraphics[width=\textwidth, trim = 0 160 0 160, clip]{plot_BLPintersect2}\begin{justify}\vspace{-12pt} \footnotesize
	%		This Figure contains 
	%	\end{justify}\vspace{-12pt}
%\end{figure}
%
%The plot shows the identified set (red) and related non-robust confidence intervals (blue, after sorting for the empty case). One can see that, due to the approximation error for the upper bound $r_{\theta_U}(z)$, the bounds cross. The associated identified sets and confidence intervals are empty. Even with reordering as depicted, the intervals would suggest spuriously precise inference around and below the point of intersection.
The relaxed moment inequality representation for such linear predictor cases follows by definition: Recall that $E[b_B(Z_i)(\psi_B(W_i,\eta_0) - b_B(Z_i)'\beta_{B,0})|Z_i=z] = 0$ and $\theta_U(z) \geq \theta(z)$ and thus %\begin{align}

{\footnotesize\begin{align}E[b_U(Z_i)(\psi_U(W_i,&\eta_0) -  b_U(Z_i)'\beta_{B,0})|Z_i=z] + E[b_U(Z_i)( b_U(Z_i)'\beta_{B,0} - \theta_U(Z_i))|Z_i=z] \notag  \\  
&\quad + E[b_U(Z_i)(\theta_U(Z_i)-\theta(Z_i))|Z_i=z] \geq b_U(z)r_U(z)\end{align}}
%\end{align}

and therefore

{\footnotesize\begin{align}
	E[b_U(Z_i)(\psi_U(W_i,\eta_0) - \theta(Z_i))|Z_i=z] &\geq b_U(z)r_U(z) =: -\delta_U(\theta(z)),
\end{align}}
where again $\delta_U(\theta(z))$ can be treated as positive as we are only concerned about underestimation of the upper bound. Applying the same with reverted sign and ordering for the lower bound then again yields 
{\footnotesize\begin{align}
	E[b_U(Z_i)(\psi_U(W_i,{\eta}_0) - \theta(Z_i))|Z_i=z] &\geq -\delta_U(\theta(z)), \notag \\
	E[b_L(Z_i)(\theta(Z_i) - \psi_L(W_i,{\eta}_0))|Z_i=z] &\geq -\delta_L(\theta(z)),
\end{align}}
where $\delta_B(\theta(z)) = b_B(z)r_B(z)$. %[EQREFabov] is again identical to the conditional [AKWONREF] in [EQREF]. 
For the joint problem of upper and lower bound using two moment functions at a given $z$, the minimal deviation $\delta_z^{\inf}$ is then chosen simultaneously such that the bounds are allowed to intersect $b_L(z)'\beta_{L,0} > b_U(z)'\beta_{U,0}$. Thus, even when the conditional moment functions $\psi_B(W_i,\eta_0)$ are correctly specified, their linear predictors for $\theta_B(z)$ can intersect for some $z \in \mathcal{Z}$ if there is local misspecification $r_B(z) \neq 0$. Our confidence intervals are adaptive to these situations whenever such misspecification would lead to undercoverage and guarantee nominal coverage over the augmented parameter space.

\end{appendices}

\end{document}